\documentclass[10pt]{article}%
\usepackage{amsmath}
\usepackage{amsfonts}
\usepackage{amssymb}
\usepackage{array}
\usepackage{graphicx}
\usepackage{ragged2e}
\usepackage[colorlinks=true,linkcolor=blue,urlcolor=blue,citecolor=blue,anchorcolor=blue]%
{hyperref}
\usepackage{natbib}
\usepackage{geometry}
\usepackage[flushleft]{threeparttable}
\usepackage{subcaption}
\usepackage[autostyle,english=american]{csquotes}
\usepackage{graphicx}
\usepackage{booktabs}
\usepackage{tikz}
\usepackage[capposition=top]{floatrow}
\usepackage{rotating}
\usepackage{appendix}
\usepackage{bookmark}
\usepackage{setspace}
\usepackage{enumitem}
\usepackage[yyyymmdd]{datetime}
\usepackage{dsfont}
\usepackage{varwidth}
\usepackage{xr}
\usepackage{filecontents}%
\providecommand{\U}[1]{\protect\rule{.1in}{.1in}}
\newtheorem{theorem}{Theorem}
\newtheorem{example}{Example}

\newtheorem{lemma}{Lemma}
\newtheorem{corollary}{Corollary}
\newtheorem{remark}{Remark}
\newenvironment{proof}[1][Proof]{\noindent\textbf{#1.} }{\ \rule{0.5em}{0.5em}}
\numberwithin{equation}{section}
\allowdisplaybreaks

\makeatletter
\newcommand*{\sectionbookmark}
[1][]{
\bookmark[level=section,dest=\@currentHref,#1]
}
\makeatother

\newtheorem{asu}{Assumption}
\newcounter{subassumption}[asu]
\renewcommand{\thesubassumption}{(\textit{\roman{subassumption}})}
\makeatletter
\renewcommand{\p@subassumption}{\theasu}
\makeatother

\begin{document}
%
\title
{Debiased Inference for Dynamic Nonlinear Panels with Multi-dimensional Heterogeneities}
\date{\today}%
\author{
Xuan Leng
\thanks
{Department of Statistics and Data Science at School of Economics, Xiamen University,
Xiamen (361000), China. Email: \href{mailto:xleng@xmu.edu.cn}%
{xleng@xmu.edu.cn}.}
\and Jiaming Mao
\thanks{Wang Yanan Institute for Studies in Economics, Xiamen University,
Xiamen (361000), China. Email: \href{mailto:jmao@xmu.edu.cn}{jmao@xmu.edu.cn}%
.}
\and Yutao Sun
\thanks
{Institute for Advanced Economic Research, Dongbei University of Finance and Economics,
Dalian (116025), China. Email: \href{mailto:yutao.sun@iaer.dev}%
{yutao.sun@iaer.dev}.}
\thanks{Corresponding Author.}
}%
\maketitle
%

\begin{abstract}%

We introduce a generic class of dynamic nonlinear heterogeneous parameter
models that incorporate individual and time fixed effects in both the
intercept and slope. These models are subject to the incidental parameter
problem, in that the limiting distribution of the point estimator is not
centered at zero, and that test statistics do not follow their standard
asymptotic distributions as in the absence of the fixed effects. To address
the problem, we develop an analytical bias correction procedure to construct a
bias-corrected likelihood. The resulting estimator follows an asymptotic
normal distribution with mean zero. Moreover, likelihood-based test
statistics---including likelihood-ratio, Lagrange-multiplier, and Wald
tests---follow the limiting chi-squared distribution under the null
hypothesis. Simulations demonstrate the effectiveness of the proposed
correction method, and an empirical application on the labor force
participation of single mothers underscores its practical importance.

\begin{description}
\item[\textbf{Keywords:}] incidental parameter problem; bias correction; fixed
effects; panel data models

\item[\textbf{JEL Classification}:] C23
\end{description}

%

\end{abstract}%

\section{Introduction}

\label{Section.Introduction}

Panel data are common in empirical research. Panel data models with two-way
fixed effects are widely used to control for unobserved heterogeneity that may
correlate with covariates. Individual effects capture time-invariant,
individual-specific characteristics, while time effects account for shifts
that are common across individuals but vary over time. Traditionally, these
models address \emph{level heterogeneity} by incorporating fixed effects only
in the intercept, adjusting for baseline outcome differences across
individuals and time periods. However, both economic theory and empirical
evidence indicate substantial variation in how outcomes respond to covariates
across individuals and over time \citep{bc2007}. Cross-sectional response
heterogeneity reflects individual-specific factors such as preferences,
productivity, and access to resources \citep{heckman2001}, whereas
time-varying response heterogeneity captures changes in behavior driven by
evolving circumstances---including policy reforms, business cycles, and
technological progress \citep{ow2021}. When such \emph{response
heterogeneities} are correlated with covariates, failing to account for them
can lead to biased estimates and invalid inference \citep{sc2013}.

To illustrate, consider the classic problem of estimating the impact of
children on a mother's labor force participation. Research indicates that
mothers with more children may differ systematically from those with fewer or
none \citep{br1988, ae1998, kleven2019}. First, mothers who choose to have
more children might be inherently less inclined to work. Second, mothers with
more children may be those whose labor supply is less sensitive to family
size, perhaps due to better financial resources or childcare access. Moreover,
over time, declining fertility and rising female labor force participation
suggest a shifting baseline, while advances in home technology and welfare
reforms may lessen the burden of additional children, enhancing labor supply
responsiveness. In a panel dataset tracking households over time, we can
control for the first type of heterogeneity---level differences in baseline
participation---by including fixed effects in the intercept. However, to
address the second type of heterogeneity---variations in response to children
across individuals and over time---we must include fixed effects in the slope.
Given that labor force participation is a binary outcome and typically
exhibits serial correlation, this example underscores the need for a dynamic
nonlinear panel data model with fixed effects in both the intercept and slope.

\label{moremotivation}In this paper, we introduce a general class of
\emph{dynamic nonlinear heterogeneous-parameter} (DN-HP) models that
incorporate individual and time fixed effects in both the intercept and the
slope. This flexible framework accommodates multi-dimensional heterogeneities
and encompasses many commonly used panel-data models, including both static
and dynamic specifications for linear and limited-dependent-variable outcomes.
It nests individual-specific slope models that capture heterogeneous responses
across individuals, as well as time-varying-coefficient models that allow
response parameters to evolve over time due to shifts in the underlying
economic or structural environment. Within this class of models, we focus on
estimation and inference for average (common) slope coefficients under a
large-$N,T$ framework. In many empirical applications, slope coefficients are
primary objects of interest because they summarize economically meaningful
responsiveness---often in elasticity or semi-elasticity form---and are the
quantities most directly reported and compared across studies. For example, in
international trade, gravity models interpret coefficients on distance and
policy variables as elasticities or percentage effects on bilateral flows
using long panels of trading partners \citep{silva2006}. In innovation
economics, nonlinear count models for patenting rely on slope coefficients to
measure the responsiveness of innovative activity to market structure or
policy incentives \citep{aghion2005}. In differentiated product demand
estimation, discrete-choice models focus on slope coefficients that capture
consumers' price sensitivity and valuation of product characteristics, and are
often estimated using product--market panel data with a large number of
products or markets observed repeatedly over an extended period of time \citep{nevo2001}.

\begin{center}
\begin{figure}[H]%
\caption
{Comparison of Uncorrected and Corrected Maximum Likelihood Estimators}%
\label{Result.EstLogitPreview}%

\includegraphics[width=0.35\linewidth]{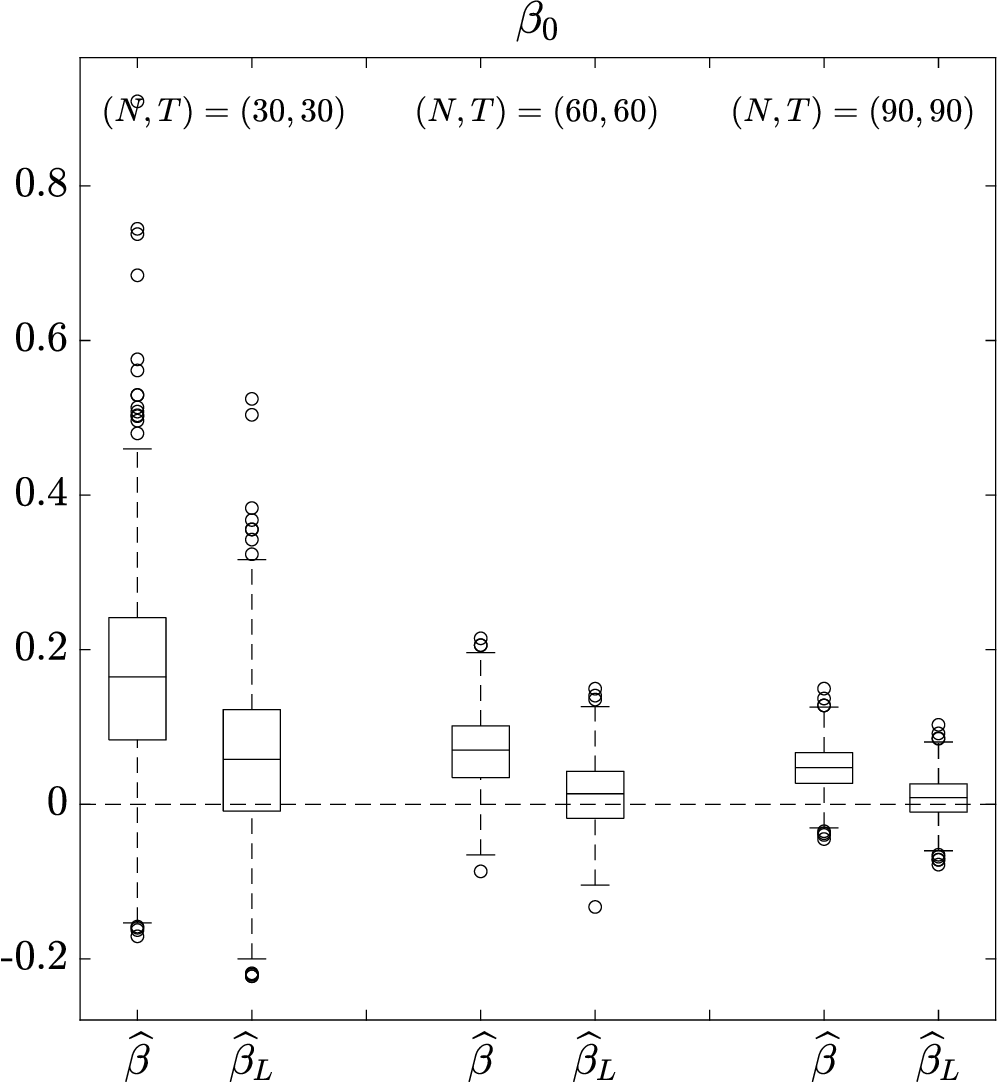}%
\qquad%
\includegraphics[width=0.35\linewidth]{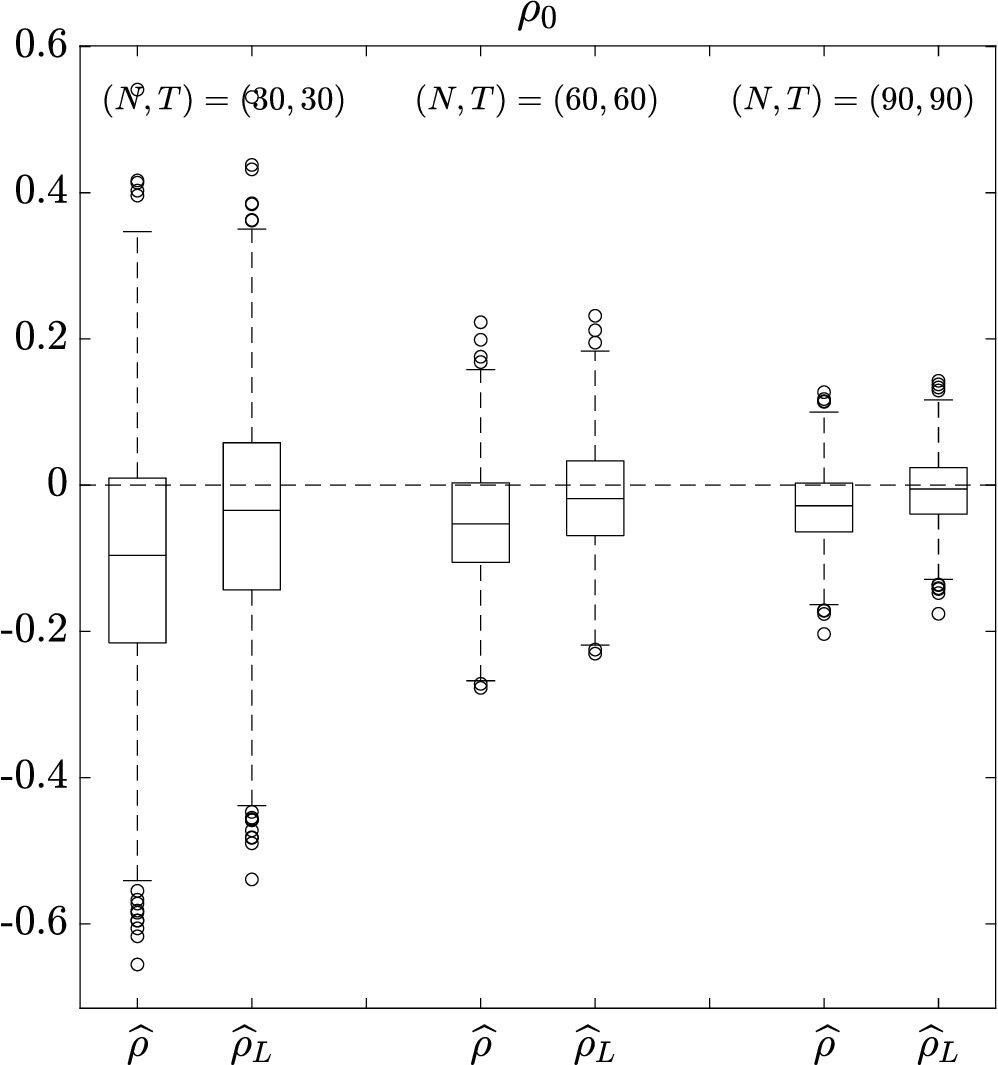}%
%

\begin{flushleft}
\begin{justify}
\begin{footnotesize}%

\noindent\textit{Notes}: Model. $Y_{it}=%
\mathds{1}%
\{\rho Y_{it-1}+(\beta+\alpha_{1,i}+\gamma_{1,t})Z_{it}+\alpha_{2,i}%
+\gamma_{2,t}+\varepsilon_{it}>0\}$ where $%
\mathds{1}%
\{\cdot\}$ denotes the indicator function, $\varepsilon_{it}$ is standard
logistic independent of the exogenous regressor $Z_{it}$, and the true values
$\rho_{0}=\beta_{0}=0.5$. $\widehat{\beta}$ and $\widehat{\rho}$ are the
uncorrected estimators while $\widehat{\beta}_{L}$ and $\widehat{\rho}_{L}$
are the corrected estimators. $1000$ replications. The detailed
data-generating process is given in Section \ref{Section.Simulation}.%

\end{footnotesize}
\end{justify}
\end{flushleft}%
%

\end{figure}
\end{center}

In such settings, DN-HP models are generally subject to the
incidental-parameter problem: the limiting distribution of the estimator is
not centered at zero, and standard test statistics (e.g., Wald, LM, and LR) no
longer follow their conventional asymptotic $\chi^{2}$ distributions. To
address this, we propose an analytical bias-correction procedure that restores
valid large-sample inference for both parameter estimates and test statistics.
As a preview, Figure \ref{Result.EstLogitPreview} presents simulated boxplots
of the estimation bias in a two-way heterogeneous parameter logit model,
comparing estimators from our bias correction procedure to the uncorrected
ones. The results reveal that our bias correction procedure significantly
reduces the estimation bias while maintaining comparable mean squared errors.%

\emph{Literature Review}. The estimation and inference of fixed-effects models
in the presence of the incidental-parameter problem have been extensively
studied. Early work, such as \citet{c1980}, \citet{ab1991}, and \citet{l2002},
focused on frameworks with only individual effects in the intercept,
establishing fixed-$T$ consistency (short panels) for structural parameter
estimators in specific models. In recent decades, the availability of long
panel datasets has motivated a large--$N,T$ framework, where the
cross-sectional size $N$ and the time dimension $T$ grow at similar rates
under rectangular asymptotics. In this setting, estimators remain consistent
but exhibit a non-negligible bias of order $O\left(  1/T\right)  $ because
each individual effect is estimated from only $T$ observations. This
incidental-parameter bias is especially pronounced in nonlinear or dynamic
models. To address it, researchers have developed a variety of bias correction
techniques within a maximum likelihood framework. \citet{hn2004} and
\citet{hk2011} propose \textquotedblleft parameter-based" corrections that
remove the bias from the likelihood estimator; \citet{w2002} and
\citet{llw2003} propose \textquotedblleft score-based" methods that modify the
profiled score; while \textquotedblleft likelihood-based" approaches such as
\citet{bh2009} and \citet{ah2016} adjust the log-likelihood directly. Most of
these procedures are analytical, relying on closed-form approximations to the
bias. Alternatively, numerical corrections estimate the bias through
resampling or re-estimation, including the jackknife \citep{dj2015}, the
bootstrap \citep{ks2016,bkss2020,hj2024}, and integrated-likelihood methods \citep{ab2009}.

For two-way fixed effects models, incorporating time effects in the intercept
generates an additional bias of order $O\left(  1/N\right)  $ on top of the
$O\left(  1/T\right)  $ term from individual effects. Several studies extend
bias-correction methods to this setting. \citet{mw2015a} develop a
parameter-based analytical correction for dynamic linear models with
interactive effects. For nonlinear models, \citet{fw2016} propose
parameter-based techniques for additive fixed effects, while \citet{cfw2014}
study interactive effects in static models. Alternatively, \citet{ko2018}
provide a likelihood-based approach for static models with an arbitrary but
known fixed-effect structure. Other model-specific contributions include
\citet{b2009} and \citet{c2017}. For a comprehensive overview of these
developments, see \citet{fw2017}. Despite this progress, the two-way
literature primarily addresses heterogeneity in levels---that is, intercept
effects---rather than heterogeneity in responses (slopes). Extending
bias-correction methods to settings with slope heterogeneity remains largely
unexplored and is the focus of our contribution.

A substantial related literature explores heterogeneity in slope coefficients
within panel models, emphasizing individual-specific parameters
\citep{robertson1992, ps1995} and time-varying coefficients
\citep{robinson1989,sw1996}, most often in linear or static contexts
\citep{h2014, pesaran2015}. Recent research in dynamic nonlinear panels has
advanced these ideas further. \citet{fl2013} study linear and nonlinear panel
data models with individual-heterogeneous coefficients and endogenous
regressors, and propose a bias correction method based on the generalized
method of moments. See also \citet{fglv2025} for a panel distribution
regression with individual-heterogeneous coefficient. \citet{cfhn2013} derive
partial identification with uniform inference for nonseparable panels under
time-homogeneity. \citet{bc2014} provide mixture-based point identification
for dynamic binary models with maximal cross-sectional heterogeneity, while
\citet{bc2010} analyze short-panel dynamic binary models with heterogeneous
transition probabilities and propose a mean-integrated-MSE (MIMSE) estimator
that better balances bias--variance trade-offs in small--$T$ settings than
analytical bias correction. While these studies focus on identification or
fixed--$T$ settings within semiparametric or nonparametric frameworks, we
adopt a parametric, likelihood-based approach that achieves point
identification and enables bias-corrected inference for dynamic nonlinear
models with two-way slope and intercept effects under large--$N,T$ asymptotics.

Two recent studies are closely aligned with our framework. \citet{kn2020}
examine linear models with two-way slope and intercept effects, employing an
iterative ``mean-observation OLS''\ estimator to address bias. Their analysis
of U.S. agricultural data reveals pronounced regional differences in
heat-yield sensitivity, alongside temporal adaptation as farmers adopt new
technologies over time. \citet{ls2023} study a similar two-way
heterogeneous-slope specification and implement a parameter-based jackknife
estimator that enables uniform inference for structural parameters. Their
cross-country analysis of the Feldstein--Horioka relation uncovers variation
across nations and periods, driven by financial integration and evolving
policy regimes. Both studies underscore the importance of slope heterogeneity
across individuals and time but remain confined to linear settings. Our study
advances this literature by extending the framework to a dynamic nonlinear
context and developing a likelihood-based analytical bias-correction procedure
that ensures valid estimation and inference. To our knowledge, neither our
DN-HP model nor its associated bias correction has been previously explored.
Together, the model and method provide a unified approach to estimation and
inference in dynamic panels with multi-dimensional heterogeneity.

\emph{This paper}. We employ the maximum likelihood framework and focus on
additive multi-dimensional two-way fixed effects, whose number grows with the
sample size. We consider an arbitrary DN-HP model whose log-likelihood
function is specified up to the unknown parameters. We construct a modified,
or corrected, log-likelihood function by adding two bias correction terms to
the original one. These two bias terms are analytically derived by combining
i) the Taylor expansion of the original log-likelihood function, in terms of
the fixed effects, and ii) an asymptotic expansion of the fixed effects
themselves. We show that the corrected log-likelihood function does not suffer
from the incidental parameter problem, under appropriate regularity
conditions. Our method can be viewed as an extension of \cite{ah2016} and
\cite{bh2009} to the two-way DN-HP models.

One key benefit of our procedure is that we achieve bias corrections for the
point estimators and the test statistics by modifying a single object, the
log-likelihood function. We rigorously show that the estimators obtained by
maximizing the corrected log-likelihood function retain a limiting normal
distribution with zero mean, consistent with the asymptotic theory of the
classical maximum likelihood estimators (MLE) in the absence of incidental
parameters. Beyond the estimators, we also show that the likelihood ratio
(LR), Lagrange-multiplier (LM), and Wald statistics, derived from the same
corrected log-likelihood function, are asymptotically equivalent, sharing the
same asymptotic $\chi^{2}$ distribution under the null hypothesis.

We demonstrate the finite-sample performance of our method through Monte Carlo
simulations, comparing it to the original likelihood and potential bias
correction devices of the bootstrap and the jackknife. We show that our
correction procedure reduces the bias significantly, without increasing the
root mean squared errors, and restores the test sizes of the LR test to its
nominal level. We find that our procedure outperforms the jackknife when it
comes to heterogeneous slope coefficients. We also find that the LR test based
on our approach has a stronger power near the true hypothesis than the LR test
based on the bootstrap. In the empirical application, we apply our approach
with a probit model to study the aforementioned question of how the number of
children affects a mother's labor force participation, with a focus on
single-mother households. Our analysis reveals that this impact varies
significantly across individuals and over time. Neglecting such variation can
lead to biased estimates and markedly different conclusions, highlighting the
need to address both level and response heterogeneities, as well as the value
of our framework in empirical research.

The remainder of the paper is organized as follows. Section \ref{Section.IPP}
explains our settings, the identification restriction, and the incidental
parameter problem of DN-HP models. Section \ref{Section.Asymptotics} describes
our bias correction procedure and provides relevant statistical properties.
Section \ref{Section.Simulation} presents simulation studies to demonstrate
the performance of the method. Section \ref{Section.Empirical} applies our
method to study single mothers' labor force participation. Finally in Section
\ref{Section.Conclusion}, we leave some closing remarks. Proofs, technical
details, additional results, elaborated discussions, etc. are provided in the
appendix. These items have references starting with letters.

\emph{Notation}. We denote $\mathbb{I}_{n}$ to be the $n\times n$ identity
matrix and $\iota_{n}$ the $n\times1$ vector of ones. $\otimes$ denotes the
Kronecker product and $%
\mathds{1}%
\{\cdot\}$ is the indicator function. Next, for a sequence of square matrices
$A_{i}$, $i=1,\ldots,n$, $\operatorname*{diag}\{A_{1},\ldots,A_{n}\}$
represents a block diagonal matrix where each $A_{i}$ is the $i$-th diagonal
element. For a vector $v$ and a function $f(v)$, we write $\partial_{v}$ and
$\partial_{vv^{\prime}}$ to represent the first and second partial derivatives
of $f(v)$ with respect to (w.r.t.) $v$.

\section{Bias in Heterogeneous Parameter Models}

\label{Section.IPP}

\subsection{Model and Estimation}

\label{ME}

In this section, we explain our models and the estimation procedure, leaving
two motivating examples in Section
\ref{Section.TechDetails.ModelAndEstimation} of the supplementary appendix.
Let $\{(Y_{it},X_{it}^{\prime})^{\prime}:i=1,\ldots,N;t=1,\ldots,T\}$ be a
panel data set of $N$ individuals and $T$ time periods, where, $i$ represents
the individual and $t$ represents the time period. Here $Y_{it}$ is a scalar
response variable, and $X_{it}$ is a $K\times1$ vector of regressors, with $K$
known and fixed. For each individual $i$, the response $Y_{it}$ is generated
sequentially over $t$, by%
\begin{equation}
Y_{it}|X_{it},\ldots,X_{i1},\phi^{0}\sim f(y|X_{it},\theta_{0}+\alpha_{i}%
^{0}+\gamma_{t}^{0}), \label{Eq.IPP.DGP}%
\end{equation}
where $\theta_{0}$ is a $K\times1$ vector of (non-random) unknown structural
parameters of interests, $\phi^{0}:=(\alpha_{1}^{0\prime},\ldots,\alpha
_{N}^{0\prime},\gamma_{1}^{0\prime},\ldots,\gamma_{T}^{0\prime})^{\prime}$
consists of the $K\times1$ random vectors $\alpha_{i}^{0}$ and $\gamma_{t}%
^{0}$, and $f(y|\cdot)$ is a density known up to $\theta_{0}$, $\alpha_{i}%
^{0}$, and $\gamma_{t}^{0}$. We restrict our attention to the case where
$f\left(  \cdot\right)  $ depends on $\theta_{0}$, $\alpha_{i}^{0}$, and
$\gamma_{t}^{0}$ through the additive structure $\theta_{0}+\alpha_{i}%
^{0}+\gamma_{t}^{0}$, the \textquotedblleft heterogeneous parameter". The
vectors $\alpha_{i}^{0}$ and $\gamma_{t}^{0}$ capture the individual-specific
and time-specific heterogeneities, respectively. Generally, $\alpha_{i}^{0}$
and $\gamma_{t}^{0}$ may be correlated with the regressor $X_{it}$. Throughout
the paper, we employ the \textquotedblleft fixed-effect" framework to
condition on the realizations of $\alpha_{i}^{0}$ and $\gamma_{t}^{0}$
(denoted by the same symbols) in the density, and treat them as unknown
nuisance parameters to be estimated together with $\theta_{0}$. Conditioning
on $\alpha_{i}^{0}$ and $\gamma_{t}^{0}$, we assume $(Y_{it},X_{it}^{\prime
})^{\prime}$ independent across $i$ but serially dependent over $t$. In
particular, we allow $X_{it}$ to contain both strictly exogenous components
and predetermined components w.r.t. $Y_{it}$. For instance, $X_{it}$ may
contain lagged values of $Y_{it}$. In this paper, we focus on the case where
all parameters in (\ref{Eq.IPP.DGP}) are heterogeneous. This is without loss
of generality, because a \textquotedblleft homogeneous parameter" can be
accommodated by setting the relevant component of $\alpha_{i}^{0}+\gamma
_{t}^{0}$ to $0$ for all $\left(  i,t\right)  $.

We are interested in the estimation of and the inference about the structural
parameter $\theta_{0}$ in (\ref{Eq.IPP.DGP}), under the presence of the
nuisance parameter $\phi^{0}$. Our approach is likelihood-based. Denote
$\phi:=(\alpha_{1}^{\prime},\ldots,\alpha_{N}^{\prime},\gamma_{1}^{\prime
},\ldots,\gamma_{T}^{\prime})^{\prime}$ and define the log-likelihood function
(\textquotedblleft likelihood" hereafter) as%
\begin{equation}
\ell(\theta,\phi):=\frac{1}{NT}\sum_{i=1}^{N}\sum_{t=1}^{T}l_{i,t}(\theta
,\phi),\qquad l_{i,t}(\theta,\phi):=\log f(Y_{it}|X_{it},\theta+\alpha
_{i}+\gamma_{t}). \label{Eq.IPP.OriginalLike}%
\end{equation}
Due to the additive structure $\theta+\alpha_{i}+\gamma_{t}$, $\theta$ is not
identified because the likelihood function is invariant to transformations%
\begin{equation}
\alpha_{i}\mapsto\alpha_{i}+c_{1},\qquad\gamma_{t}\mapsto\gamma_{t}%
+c_{2},\qquad\theta\mapsto\theta-c_{1}-c_{2} \label{Eq.IPP.RotInd}%
\end{equation}
for all pairs of constant $(c_{1},c_{2})$ and all $(i,t)$. To resolve this, we
follow \cite{ls2023} to reparameterize the likelihood by setting $\alpha
_{N}=-\sum_{i=1}^{N-1}\alpha_{i}$ and $\gamma_{T}=-\sum_{t=1}^{T-1}\gamma_{t}%
$, thereby imposing the normalization $\sum_{i=1}^{N}\alpha_{i}=0$ and
$\sum_{t=1}^{T}\gamma_{t}=0$. Any \textquotedblleft average effect" is
absorbed by $\theta$. Let $\psi:=\left(  \alpha_{1}^{\prime},\ldots
,\alpha_{N-1}^{\prime},\gamma_{1}^{\prime},\ldots,\gamma_{T-1}^{\prime
}\right)  ^{\prime}$, the reparameterized likelihood can be constructed from
the original $\ell\left(  \theta,\phi\right)  $ as%
\[
\ell\left(  \theta,D^{\prime}\psi\right)  =:l(\theta,\psi),\qquad
D:=\operatorname*{diag}\{D_{1},D_{2}\},
\]
where $D_{1}:=(\mathbb{I}_{N-1},-\iota_{N-1})\otimes\mathbb{I}_{K}$ and
$D_{2}:=(\mathbb{I}_{T-1},-\iota_{T-1})\otimes\mathbb{I}_{K}$. This rules out
the indeterminacy in (\ref{Eq.IPP.RotInd}) without explicitly using the Lagrangian.

\subsection{Incidental Parameter Problem}

\label{Section.IPP.IPPBias}

The dynamic nonlinear heterogeneous parameter, DN-HP, models generally suffer
from the incidental parameter problem (IPP). We briefly explain this problem
here, relegating to Section \ref{Section.TechDetails.BiasExpansion} of the
supplementary appendix i) the exact characterization of the remainders, ii) an
illustrative example, etc. We use $\mathbb{E}$ to denote the expectation
w.r.t. $\prod_{i=1}^{N}\prod_{t=1}^{T}f(y_{it}|X_{it},\theta_{0}+\alpha
_{i}^{0}+\gamma_{t}^{0})$, which is the true density evaluated at $\theta_{0}%
$, given: i) initial values of the predetermined regressors, ii) all strictly
exogenous regressors, and iii) the unobserved effect $\phi^{0}$. Following the
profiled likelihood framework of \cite{ps2006}, we continue our discussion by
defining, for a given $\theta$ and each pair of $(N,T)$,%
\begin{equation}
\widehat{\psi}(\theta):=\arg\max_{\psi}l(\theta,\psi),\qquad\psi(\theta
):=\arg\max_{\psi}\mathbb{E}[l(\theta,\psi)]. \label{Eq.IPP.PsiEstimator}%
\end{equation}
Here $\psi(\theta)$ is referred to as the pseudo-true value and $l(\theta
,\psi(\theta))$ is not subject to the IPP. See, e.g., \cite{hn2004} and
\cite{ah2016}. In practice, however, $\psi(\theta)$ is generally infeasible
and the plug-in version $l(\theta,\widehat{\psi}(\theta))$ is used instead.
Specifically, the MLE of $\theta_{0}$, $\widehat{\theta}:=\arg\max_{\theta
}\ell(\theta,\widehat{\psi}(\theta))$. The LR and LM test statistics (for
hypothesis about $\theta_{0}$) are also constructed using $l(\theta
,\widehat{\psi}(\theta))$. A fundamental problem is that $l(\theta
,\widehat{\psi}(\theta))$ contains a non-negligible asymptotic bias of order
$O(1/\sqrt{NT})$, due to the estimation errors in $\widehat{\psi}(\theta)$.
This is referred to as the IPP in the context of the DN-HP model. The
consequence is that, $\widehat{\theta}$ and many likelihood-based test
statistics deviate from their standard asymptotic distribution, leading to
incorrect inferences.

To demonstrate the non-negligible bias in $l(\theta,\widehat{\psi}(\theta))$,
consider a Taylor expansion of $l(\theta,\widehat{\psi}(\theta))$ around
$\psi(\theta)$:%
\begin{equation}
l(\theta,\widehat{\psi}(\theta))=l(\theta,\psi(\theta))+s^{\prime}%
(\theta)[\widehat{\psi}(\theta)-\psi(\theta)]+\frac{1}{2}[\widehat{\psi
}(\theta)-\psi(\theta)]^{\prime}H(\theta)[\widehat{\psi}(\theta)-\psi
(\theta)]+r^{l}(\theta), \label{Eq.IPP.LikeExpansion}%
\end{equation}
where $r^{l}(\theta)$ is a remainder, $\widehat{\psi}(\theta)-\psi(\theta)$
represents the estimation errors of the fixed effects,%
\begin{align*}
s(\theta)  &  :=s(\theta,\psi(\theta)),\qquad s(\theta,\psi):=\partial_{\psi
}l(\theta,\psi),\\
H\left(  \theta\right)   &  :=H(\theta,\psi(\theta)),\qquad H(\theta
,\psi):=\partial_{\psi\psi^{\prime}}l(\theta,\psi)
\end{align*}
are the score and Hessian. Here the estimation errors $\widehat{\psi}%
(\theta)-\psi(\theta)$ can be derived by another Taylor expansion. Seeing
$s(\theta,\widehat{\psi}(\theta))=0$ as $\widehat{\psi}(\theta)$ is the
maximizer, a Taylor expansion of $s(\theta,\widehat{\psi}(\theta))$ around
$\psi(\theta)$ gives%
\begin{align}
0  &  =s\left(  \theta\right)  +H\left(  \theta\right)  [\widehat{\psi}%
(\theta)-\psi(\theta)]+r^{s}(\theta),\nonumber\\
\widehat{\psi}(\theta)-\psi(\theta)  &  =-[H\left(  \theta\right)
]^{-1}s\left(  \theta\right)  -[H\left(  \theta\right)  ]^{-1}r^{s}(\theta),
\label{Eq.IPP.PsiExpansion}%
\end{align}
where $r^{s}(\theta)$ is a remainder. Combining (\ref{Eq.IPP.PsiExpansion})
and (\ref{Eq.IPP.LikeExpansion}), and taking expectation, we obtain%
\begin{align}
\sqrt{NT}\mathbb{E}[l(\theta,\widehat{\psi}(\theta))-l(\theta,\psi(\theta))]
&  =-\sqrt{NT}B(\theta)+\sqrt{NT}\mathbb{E}r(\theta),\nonumber\\
B(\theta)  &  :=\frac{1}{2}\mathbb{E\{}s^{\prime}(\theta)[H(\theta
)]^{-1}s(\theta)\}, \label{Eq.IPP.BiasExpansion}%
\end{align}
where $r(\theta)$ is the final remainder depending on $r^{l}(\theta)$ and
$r^{s}(\theta)$ satisfying $\mathbb{E}r(\theta)=o(1/\sqrt{NT})$, uniformly in
$\theta$, as $N,T\rightarrow\infty$ with $N/T\rightarrow\kappa$ for some
$0<\kappa<\infty$. This implies that $\sqrt{NT}\mathbb{E}r(\theta)=o(1)$ and
is negligible. $B(\theta)$ is the IPP bias arising from the estimation errors
of $\widehat{\psi}(\theta)$. As opposite to $\mathbb{E}r(\theta)$, the bias
$B(\theta)$ is not negligible, in the sense that $\sqrt{NT}B(\theta
)\not \rightarrow 0$ as $N,T\rightarrow\infty$ with $N/T\rightarrow\kappa$.
Consequently, $\widehat{\theta}$ and many likelihood-based test statistics
inherit the bias from the likelihood, leading to invalid inferences.

\section{Bias Correction and Asymptotic Theory}

\label{Section.Asymptotics}

The bias expansion of $\widehat{l}(\theta):=l(\theta,\widehat{\psi}(\theta))$
in Equation (\ref{Eq.IPP.BiasExpansion}) motivates our approach of correcting
the likelihood $l(\theta,\widehat{\psi}(\theta))$ by adding $B(\theta)$. Our
corrected likelihood can be viewed as an approximation to the IPP-free
infeasible likelihood $l(\theta):=l(\theta,\psi(\theta))$, with approximation
error $o_{\mathbb{P}}(1/\sqrt{NT})$. That is, it does not suffer from the IPP
to the first order. Intuitively, the estimator of $\theta_{0}$, and the LR,
LM, and Wald test statistics, will follow their standard asymptotic
distributions when they are derived from the corrected likelihood (instead of
$\widehat{l}(\theta)$). In what follows, we explain our correction procedure
and give the main result. We leave in Section
\ref{Section.TechDetails.BiasTerms} of the supplementary appendix i) more
technical details; ii) formal statements and discussions about our secondary
results (Equations \ref{Eq.Asymptotics.LikeExpansion},
\ref{Eq.Asymptotic.NormalTheta}, \ref{Eq.Asymptotic.Chi2Tests}, etc.); iii)
useful remarks; etc. In addition, we explain the relation of our method to
\cite{bh2009} and \cite{ah2016} in Section \ref{Section.AH2016} of the
supplementary appendix.

In this paper, we decompose $B(\theta)=B_{\alpha}(\theta)+B_{\gamma}%
(\theta)+o(1/\sqrt{NT})$ to obtain the bias expansion%
\begin{equation}
\mathbb{E}l(\theta)=\mathbb{E}\widehat{l}(\theta)+B_{\alpha}(\theta
)+B_{\gamma}(\theta)+o(1/\sqrt{NT}), \label{Eq.Asymptotics.LikeExpansion}%
\end{equation}
where%
\[
B_{\alpha}(\theta):=\frac{1}{2}\operatorname*{trace}[S_{\alpha\alpha}%
(\theta)H_{\alpha\alpha}^{\mathcal{\ast}}(\theta)],\qquad B_{\gamma}%
(\theta):=\frac{1}{2}\operatorname*{trace}\mathbb{[}S_{\gamma\gamma}%
(\theta)H_{\gamma\gamma}^{\ast}(\theta)]
\]
appear because of the estimation errors of, respectively, the
individual-specific heterogeneities $\alpha_{i}^{0}$ and the time-specific
heterogeneities $\gamma_{t}^{0}$.\footnote{\label{Footnote.AsymBiasInKappa}In
addition, we have $B_{\alpha}(\theta)=O\left(  T^{-1}\right)  $ and
$B_{\gamma}(\theta)=O\left(  N^{-1}\right)  $, indicating that $\sqrt
{NT}(\mathbb{E}l(\theta)-\mathbb{E}\widehat{l}(\theta))$ is proportional to
$\sqrt{\kappa}+\sqrt{\kappa^{-1}}$ as $N,T\rightarrow\infty$ provided that
$N/T\rightarrow\kappa$. This is similar to, e.g., \cite{fw2016}.} The terms
$H_{\alpha\alpha}^{\mathcal{\ast}}(\theta)$, $H_{\gamma\gamma}^{\ast}(\theta
)$, $S_{\alpha\alpha}(\theta)$, and $S_{\gamma\gamma}(\theta)$ are defined
from partitioning\footnote{Note that $\mathbb{E}s\left(  \theta\right)  =0$ by
definition. We keep $\mathbb{E}s\left(  \theta\right)  $ here for the
convenience of constructing the corrected likelihood.} $[\mathbb{E}%
H(\theta)]^{-1}$ and $s(\theta)-\mathbb{E}s\left(  \theta\right)  $ as%
\begin{equation}
\lbrack\mathbb{E}H(\theta)]^{-1}=:\left(
\begin{array}
[c]{cc}%
H_{\alpha\alpha}^{\mathcal{\ast}}(\theta) & H_{\alpha\gamma}^{\mathcal{\ast}%
}(\theta)\\
H_{\gamma\alpha}^{\mathcal{\ast}}(\theta) & H_{\gamma\gamma}^{\mathcal{\ast}%
}(\theta)
\end{array}
\right)  ,\qquad s(\theta)-\mathbb{E}s\left(  \theta\right)  =:\left(
\begin{array}
[c]{c}%
\widetilde{s}_{\alpha}(\theta)\\
\widetilde{s}_{\gamma}(\theta)
\end{array}
\right)  . \label{Eq.Asymptotics.Partitions}%
\end{equation}
Here $H_{\alpha\alpha}^{\mathcal{\ast}}(\theta)$ is $(N-1)K\times(N-1)K$,
which corresponds to the inverse expected Hessian w.r.t. $\left(  \alpha
_{1}^{\prime},\ldots,\alpha_{N-1}^{\prime}\right)  ^{\prime}$, and
$H_{\gamma\gamma}^{\ast}(\theta)$ is $(T-1)K\times(T-1)K$, corresponding to
the inverse expected Hessian w.r.t. $\left(  \gamma_{1}^{\prime},\ldots
,\gamma_{T-1}^{\prime}\right)  ^{\prime}$. The off-diagonal blocks
$H_{\alpha\gamma}^{\mathcal{\ast}}(\theta)$ and $H_{\gamma\alpha
}^{\mathcal{\ast}}(\theta)$ are defined accordingly (but are irrelevant to the
construction of the corrected likelihood). Similarly, $\widetilde{s}_{\alpha
}(\theta)$, the score w.r.t. $\left(  \alpha_{1}^{\prime},\ldots,\alpha
_{N-1}^{\prime}\right)  ^{\prime}$, is $(N-1)K\times1$ and $\widetilde{s}%
_{\gamma}(\theta)$, the score w.r.t. $\left(  \gamma_{1}^{\prime}%
,\ldots,\gamma_{T-1}^{\prime}\right)  ^{\prime}$ is $(T-1)K\times1$.
$\widetilde{s}_{\alpha}(\theta)$ and $\widetilde{s}_{\gamma}(\theta)$ are used
to construct%
\[
S_{\alpha\alpha}(\theta):=\mathbb{E\{}\widetilde{s}_{\alpha}(\theta
)\mathbb{[}\widetilde{s}_{\alpha}(\theta)]^{\prime}\},\qquad S_{\gamma\gamma
}(\theta):=\mathbb{E\{}\widetilde{s}_{\gamma}(\theta)\mathbb{[}\widetilde{s}%
_{\gamma}(\theta)]^{\prime}\},
\]
which are essentially the covariance matrices of the scores $\widetilde{s}%
_{\alpha}(\theta)$ and $\widetilde{s}_{\gamma}(\theta)$, respectively.

Estimating $B_{\alpha}(\theta)$ and $B_{\gamma}(\theta)$ by plug-in estimates,
the corrected likelihood is established as%

\begin{align}
L(\theta)  &  :=\widehat{l}(\theta)+\widehat{B}_{\alpha}(\theta)+\widehat{B}%
_{\gamma}(\theta),\label{Eq.Asymptotic.CorrectedLike}\\
\widehat{B}_{\alpha}(\theta)  &  :=\frac{1}{2}\operatorname*{trace}%
[\widehat{S}_{\alpha\alpha}(\theta)\widehat{H}_{\alpha\alpha}^{\mathcal{\ast}%
}(\theta)],\qquad\widehat{B}_{\gamma}(\theta):=\frac{1}{2}%
\operatorname*{trace}\mathbb{[}\widehat{S}_{\gamma\gamma}(\theta
)\widehat{H}_{\gamma\gamma}^{\ast}(\theta)].\nonumber
\end{align}
Here $\widehat{H}_{\alpha\alpha}^{\mathcal{\ast}}(\theta)$, $\widehat{H}%
_{\gamma\gamma}^{\ast}(\theta)$, $\widehat{S}_{\alpha\alpha}(\theta)$, and
$\widehat{S}_{\gamma\gamma}(\theta)$ are the plug-in versions of the
corresponding quantities in $B_{\alpha}(\theta)$ and $B_{\gamma}(\theta
)$.\ They are constructed as follows. First, denoting $\widehat{H}%
(\theta):=H(\theta,\widehat{\psi}(\theta))$, the two Hessian terms
$\widehat{H}_{\alpha\alpha}^{\mathcal{\ast}}(\theta)$ and $\widehat{H}%
_{\gamma\gamma}^{\ast}(\theta)$ are defined from the partition of
$[\widehat{H}(\theta)]^{-1}$ as%
\begin{equation}
\lbrack\widehat{H}(\theta)]^{-1}=:\left(
\begin{array}
[c]{cc}%
\widehat{H}_{\alpha\alpha}^{\mathcal{\ast}}(\theta) & \widehat{H}%
_{\alpha\gamma}^{\mathcal{\ast}}(\theta)\\
\widehat{H}_{\gamma\alpha}^{\mathcal{\ast}}(\theta) & \widehat{H}%
_{\gamma\gamma}^{\mathcal{\ast}}(\theta)
\end{array}
\right)  \label{Eq.Asymptotic.Hessian}%
\end{equation}
in the same manner as in (\ref{Eq.Asymptotics.Partitions}). Here
$\widehat{H}_{\alpha\alpha}^{\mathcal{\ast}}(\theta)$ and $\widehat{H}%
_{\gamma\gamma}^{\mathcal{\ast}}(\theta)$ estimate $H_{\alpha\alpha
}^{\mathcal{\ast}}(\theta)$ and $H_{\gamma\gamma}^{\mathcal{\ast}}(\theta)$,
respectively. Second, denote%
\[
s_{i,t}^{\gamma}(\theta,\phi):=\partial_{\gamma_{t}}l_{i,t}(\theta
,\phi),\qquad s_{i,t}^{\alpha}(\theta,\phi):=\partial_{\alpha_{i}}%
l_{i,t}(\theta,\phi),
\]
which are the derivatives of the original likelihood $l_{i,t}(\theta,\phi)$
w.r.t. $\alpha_{i}$ and $\gamma_{t}$, respectively, evaluated at $\phi
(\theta)$. $\widehat{S}_{\alpha\alpha}(\theta)$ and $\widehat{S}_{\gamma
\gamma}(\theta)$ are defined as:%
\begin{align}
\widehat{S}_{\alpha\alpha}(\theta)  &  :=D_{1}\widehat{\mathcal{S}}%
_{\alpha\alpha}(\theta)D_{1}^{\prime},\label{Eq.Asymptotics.SaaHat}\\
\widehat{\mathcal{S}}_{\alpha\alpha}(\theta)  &  :=\frac{1}{N^{2}T^{2}}%
\sum_{t=1}^{T}\sum_{s=1}^{T}%
\mathds{1}%
\{\left\vert t-s\right\vert \leq\tau\}\operatorname*{diag}\{\widehat{s}%
_{1,t}^{\alpha}(\theta)[\widehat{s}_{1,s}^{\alpha}(\theta)]^{\prime}%
,\ldots,\widehat{s}_{N,t}^{\alpha}(\theta)[\widehat{s}_{N,s}^{\alpha}%
(\theta)]^{\prime}\},\nonumber\\
\widehat{s}_{i,t}^{\alpha}(\theta)  &  :=s_{i,t}^{\alpha}(\theta
,\widehat{\phi}(\theta))-\frac{1}{T}\sum_{t=1}^{T}s_{i,t}^{\alpha}%
(\theta,\widehat{\phi}(\theta)),\nonumber
\end{align}
where $\widehat{\phi}(\theta):=D^{\prime}\widehat{\psi}\left(  \theta\right)
$ is the \textquotedblleft unparameterized" counterpart of $\widehat{\psi
}\left(  \theta\right)  $; and%
\begin{align}
\widehat{S}_{\gamma\gamma}(\theta)  &  :=D_{2}\widehat{\mathcal{S}}%
_{\gamma\gamma}(\theta)D_{2}^{\prime},\label{Eq.Asymptotics.SggHat}\\
\widehat{\mathcal{S}}_{\gamma\gamma}(\theta)  &  :=\frac{1}{N^{2}T^{2}}%
\sum_{i=1}^{N}\widehat{s}_{i}^{\gamma}(\theta)[\widehat{s}_{i}^{\gamma}%
(\theta)]^{\prime},\qquad\widehat{s}_{i}^{\gamma}(\theta):=\{[\widehat{s}%
_{i,1}^{\gamma}(\theta)]^{\prime},\ldots,[\widehat{s}_{i,T}^{\gamma}%
(\theta)]^{\prime}\}^{\prime},\nonumber\\
\widehat{s}_{i,t}^{\gamma}(\theta)  &  :=s_{i,t}^{\gamma}(\theta
,\widehat{\phi}(\theta))-\frac{1}{N}\sum_{i=1}^{N}s_{i,t}^{\gamma}%
(\theta,\widehat{\phi}(\theta)).\nonumber
\end{align}
The indicator $%
\mathds{1}%
\{\left\vert t-s\right\vert \leq\tau\}$ is a truncation mechanism, with
truncation parameter $\tau$, which is common in the relevant literature. In
our simulation, we use $\tau=1$ and $2$ and find the difference relatively insignificant.

\begin{remark}
[efficient computation of bias terms]\label{Remark.Computation1}The terms
$\widehat{S}_{\alpha\alpha}(\theta)$ and $\widehat{S}_{\gamma\gamma}(\theta)$
are constructed using the derivatives of the original (i.e., \textquotedblleft
unparameterized") likelihood. This is valid because the reparameterization
produces $\partial_{\psi}l(\theta,\psi)=D\partial_{\phi}\ell(\theta,\phi)$ and
$\partial_{\psi\psi^{\prime}}l(\theta,\psi)=D\partial_{\phi\phi^{\prime}}%
\ell(\theta,\phi)D^{\prime}$ holding true for every $\phi=D^{\prime}\psi$ and
every $\theta$. Generally, it may be easier to construct $\widehat{S}%
_{\alpha\alpha}(\theta)$ and $\widehat{S}_{\gamma\gamma}(\theta)$ from the
original likelihood, because analytical expressions for $\partial_{\phi}%
\ell(\theta,\phi)$ and $\partial_{\phi\phi^{\prime}}\ell(\theta,\phi)$ are
well-known for many frequently used models (e.g., the probit, the logit, and
the Poisson).

In addition, if the model is equipped with linear indices, these derivatives
may be calculated efficiently using the chain rule. We have $\partial_{\phi
}l_{i,t}(\theta,\phi)=\partial_{\pi}l_{i,t}(\pi_{i,t}\left(  \theta
,\phi\right)  )\partial_{\phi}\pi_{i,t}\left(  \theta,\phi\right)  $ by
viewing $l_{i,t}(\theta,\phi):=l_{i,t}(\pi_{i,t}\left(  \theta,\phi\right)
)$, where $\pi_{i,t}\left(  \theta,\phi\right)  :=X_{it}^{\prime}%
(\theta+\alpha_{i}+\gamma_{t})$ is the linear index. Here $\partial_{\phi}%
\pi_{i,t}\left(  \theta,\phi\right)  $ is high-dimensional but can be
calculated easily, because $\pi_{i,t}\left(  \theta,\phi\right)  $ is only
linear. The more complex component $\partial_{\pi}l_{i,t}(\cdot)$ is only
scalars and, therefore, can be calculated afforably, even with numerical
differentiation. The complexity of this only grows with the number of linear
indices, but not the dimension of $\phi$.
\end{remark}

Intuitively, by adding back $\widehat{B}_{\alpha}(\theta)$ and $\widehat{B}%
_{\gamma}(\theta)$, $L(\theta)$ serves as an approximation to the infeasible
IPP-free likelihood $l(\theta)$ with rate $o_{\mathbb{P}}(1/\sqrt{NT})$.
Denote $\phi(\theta):=D^{\prime}\psi(\theta)$, which is the \textquotedblleft
unparameterized" counterpart of $\psi(\theta)$. We impose the following
assumption and state this result formally in Theorem
\ref{Theorem.CorrectedLike} below.%

\begin{asu}
\
\begin{itemize}[leftmargin=*,itemindent=-2em]%
\label{Assumption.Expansion}%

\item[]\refstepcounter{subassumption}\thesubassumption\
\label{Assumption.Expansion.AsySeq}Suppose $N/T\rightarrow\kappa$ for some
$0<\kappa<\infty$ as $N,T\rightarrow\infty$.%

\item[]\refstepcounter{subassumption}\thesubassumption\
\label{Assumption.Expansion.Smooth}Let $\Theta$ be a compact subset of
$\mathbb{R}^{K}$ with $\theta_{0}$ in its interior and$\ \Phi$ be a compact
subset of $\mathbb{R}^{(N+T)K}$. For each $\theta\in\Theta$, $\Phi$ contains
both $\widehat{\phi}(\theta)$ and $\phi(\theta)$ in its interior.
$l_{i,t}(\theta,\phi)$ is three-time continuously differentiable w.r.t. to
$\phi\in\Phi$. There exists a function $g(w_{it})$, $w_{it}=(Y_{it}%
,X_{it}^{\prime})^{\prime}$, independent of $\theta$ and $\phi$, such that%
\[
\sup_{\theta\in\Theta}\sup_{\phi\in\Phi}\left\vert \partial_{\phi_{1}%
\cdots\phi_{S}}l_{i,t}(\theta,\phi)\right\vert <g(w_{it}),
\]
where $\phi_{s}$ represents any element of $\phi$ and $S\in\{0,1,2,3\}$ ($S=0$
is understood as not taking derivatives), and%
\[
\sup_{N,T}\max_{1\leq i\leq N,1\leq t\leq T}\mathbb{E}_{\phi}\{[g(w_{it}%
)]^{\eta}\}<\infty
\]
for some $\eta>2,$ where $\mathbb{E}_{\phi}$ denotes the conditional
expectation w.r.t. the joint distribution of $w_{it}$, given the heterogeneous
effects $\phi^{0}$.%

\item[]\refstepcounter{subassumption}\thesubassumption\
\label{Assumption.Expansion.Normalize}For each $k=1,\ldots,K$, $\sum_{i=1}%
^{N}\alpha_{k,i}^{0}=\sum_{t=1}^{T}\gamma_{k,t}^{0}=0$, where $\alpha
_{k,i}^{0}$ and $\gamma_{k,t}^{0}$ are the $k$-th component of $\alpha_{i}%
^{0}$ and $\gamma_{t}^{0}$, respectively.%

\item[]\refstepcounter{subassumption}\thesubassumption\
\label{Assumption.Expansion.Ident} For each $\theta\in\Theta$, $\mathbb{P}%
[l(\theta,\psi)\neq l(\theta,\psi(\theta))]>0$ for every $\psi$ such that
$D^{\prime}\psi\in\Phi$ and $\psi\neq\psi(\theta)$.%

\item[]\refstepcounter{subassumption}\thesubassumption\
\label{Assumption.Expansion.Data}Conditional on $\phi^{0}$, $\{w_{it}\}$ are
independent across $i$ and conditionally strong mixing with mixing coefficient%
\[
a_{i}(m):=\sup_{t\geq1}\sup_{A\in\mathcal{A}_{it},B\in\mathcal{B}_{it+m}%
}\left\vert \mathbb{P}(A\cap B|\phi^{0})-\mathbb{P}(A|\phi^{0})\mathbb{P}%
(B|\phi^{0})\right\vert
\]
such that, for $\eta>2$ as above,%
\[
\sup_{N}\max_{1\leq i\leq N}\sum_{m=0}^{\infty}[a_{i}(m)]^{1-2/\eta}<\infty,
\]
where $\mathcal{A}_{it}$ and $\mathcal{B}_{it}$ are the $\sigma$-algebras
generated by $\{w_{is}:1\leq s\leq t\}$ and $\{w_{is}:t\leq s\leq T\}$, respectively.%

\item[]\refstepcounter{subassumption}\thesubassumption\
\label{Assumption.Expansion.Hessian}The matrix $\sqrt{NT}\mathbb{E}%
[H(\theta)]$ has eigenvalues $\lambda_{p}(\theta)$ for $p=1,\ldots,(N+T-2)K$
which satisfy%
\[
\sup_{N>N_{0},T>T_{0}}\max_{1\leq p\leq(N+T-2)K}\sup_{\theta\in\Theta}%
\lambda_{p}(\theta)<0
\]
a.s. for some integers $N_{0}$ and $T_{0}$.%

\end{itemize}
\end{asu}%

\begin{theorem}
\label{Theorem.CorrectedLike}Under Assumption \ref{Assumption.Expansion} and
for some $\tau\rightarrow\infty$ such that $\tau/T\rightarrow0$, we have%
\[
L(\theta)-l(\theta)=o_{\mathbb{P}}(1/\sqrt{NT}),
\]
uniformly over $\theta\in\Theta$ as $N,T\rightarrow\infty$, where $L(\theta)$
is the corrected likelihood defined in Equation
(\ref{Eq.Asymptotic.CorrectedLike}).
\end{theorem}

Using the corrected likelihood $L(\theta)$, we may obtain the corrected
estimator of $\theta_{0}$ as%
\[
\widehat{\theta}_{L}:=\arg\max_{\theta}L(\theta).
\]
Intuitively, since $L(\theta)$ approximates $l(\theta)$ with an error
negligible relative to $1/\sqrt{NT}$, the estimator $\widehat{\theta}_{L}$
inherits this rate and is consistent and asymptotically normal with mean zero
under $N/T\rightarrow\kappa$ as $N,T\rightarrow\infty$. In particular,%
\begin{equation}
\sqrt{NT}[-\mathbb{E}\triangledown_{\theta\theta^{\prime}}l(\theta_{0}%
)]^{1/2}(\widehat{\theta}_{L}-\theta_{0})\overset{\mathbb{D}}{\longrightarrow
}\mathcal{N}(0,\mathbb{I}_{K}), \label{Eq.Asymptotic.NormalTheta}%
\end{equation}
where $\mathcal{N}(\mu,\Sigma)$ stands for the normal distribution with mean
$\mu$ and covariance matrix $\Sigma$, $\nabla_{\theta\theta^{\prime}}$ denotes
the second total derivative w.r.t. to $\theta$. Note that the result in
(\ref{Eq.Asymptotic.NormalTheta}) makes use of the information matrix equality
to simplify the presentation. We refer the reader to, e.g., \cite{w1982} when
such an equality does not hold.

For hypothesis testing procedures, we consider a generic null hypothesis
$H_{0}:R(\theta_{0})=0$, where $R(\theta)$ is a known $r\times1$ ($r\leq K$)
vector-valued non-random function with Jacobian $J(\theta):=\partial
_{\theta^{\prime}}R(\theta)$ satisfying $\operatorname*{rank}[J(\theta)]=r$.
The LR ($\widehat{\xi}_{LR}$), LM ($\widehat{\xi}_{LM}$), and Wald
($\widehat{\xi}_{LM}$) test statistics can be constructed as%
\begin{align*}
\widehat{\xi}_{LR}  &  :=-2NT\{L(\widehat{\theta}_{R})-L(\widehat{\theta}%
_{L})\},\\[4pt]
\widehat{\xi}_{LM}  &  :=-NT\{\triangledown_{\theta^{\prime}}L(\widehat{\theta
}_{R})[\triangledown_{\theta\theta^{\prime}}L(\widehat{\theta}_{R}%
)]^{-1}\triangledown_{\theta}L(\widehat{\theta}_{R})\},\\[4pt]
\widehat{\xi}_{Wald}  &  :=-NT\{R^{\prime}(\widehat{\theta}_{L}%
)[J(\widehat{\theta}_{L})[\triangledown_{\theta\theta^{\prime}}%
L(\widehat{\theta}_{L})]^{-1}J^{\prime}(\widehat{\theta}_{L})]^{-1}%
R(\widehat{\theta}_{L})\},
\end{align*}
where $\nabla_{\theta}$ denotes the first total derivative w.r.t. to $\theta$,
and $\widehat{\theta}_{R}:=\arg\max_{\theta}L(\theta)$ subject to
$R(\theta)=0$. The same argument above intuitively indicates that%
\begin{equation}
\widehat{\xi}_{LR},\widehat{\xi}_{LM},\widehat{\xi}_{Wald}\overset{\mathbb{D}%
}{\longrightarrow}\chi^{2}(r), \label{Eq.Asymptotic.Chi2Tests}%
\end{equation}
under $H_{0}$ as $N,T\rightarrow\infty$, where $\chi^{2}(r)$ is the $\chi^{2}%
$-distribution with degrees of freedom $r$.

\section{Simulation}

\label{Section.Simulation}

In this section, we present a simulation study. We consider dynamic binary
response models specified according to the data-generating process (Design 1):%
\begin{equation}
Y_{it}=%
\mathds{1}%
\{\rho_{0}Y_{it-1}+(\beta_{0}+\alpha_{1,i}^{0}+\gamma_{1,t}^{0})Z_{it}%
+\alpha_{2,i}^{0}+\gamma_{2,t}^{0}+\varepsilon_{it}>0\},
\label{Equation.Simulation.BinResp}%
\end{equation}
where the regressor vector is $X_{it}=\left(  Y_{it-1},Z_{it}\right)
^{\prime}$ ($Z_{it}$ is described below); the parameter of interests is
$\theta_{0}=(\rho_{0},\beta_{0})^{\prime}=(0.5,0.5)^{\prime}$; $\varepsilon
_{it}$ is standard normal\ (for the probit model) or standard logistic (for
the logit model), independent and identically distributed (i.i.d.) over
$\left(  i,t\right)  $ and independent from the regressor $Z_{it}$; for
$k=1,2$, $\{\alpha_{k,i}^{0},\gamma_{k,t}^{0}\}\sim\mathcal{N}(0,0.04)$ and
are demeaned\footnote{During the estimation, we do not normalize the
individual-effect parameters, because the model only contains fixed effects in
the intercept.} after being generated; the regressor $Z_{it}\sim
\mathcal{N}(\left(  \alpha_{1,i}^{0}+\alpha_{2,i}^{0}+\gamma_{1,t}^{0}%
+\gamma_{2,t}^{0}\right)  /2,1)$ i.i.d. over $(i,t)$; and the initial value
$Y_{i0}=%
\mathds{1}%
\{(\beta_{0}+\alpha_{1,i}^{0})Z_{i0}+\alpha_{2,i}^{0}+\varepsilon_{i0}>0\}$
with $Z_{i0}\sim\mathcal{N}(\left(  \alpha_{1,i}^{0}+\alpha_{2,i}^{0}\right)
/2,1)$ i.i.d. over $(i,t)$. We also simulate a static version of
(\ref{Equation.Simulation.BinResp}), where everything is the same, except that
we set $\rho_{0}=0$ (hence $\theta_{0}=\beta_{0}$) and remove $Y_{it-1}$ from
the regressors of estimated model. We consider $\left(  N,T\right)
\in\left\{  (30,30),(60,60),(90,90)\right\}  $ and run $1000$ replications,
comparing respectively the MLEs $\widehat{\rho}\,$and$\ \widehat{\beta}$ (of
$\rho_{0}\ $and $\beta_{0}$ respectively); the corrected estimators
$\widehat{\rho}_{L}^{\left(  \tau\right)  }\,$and$\ \widehat{\beta}%
_{L}^{\left(  \tau\right)  }$ (for dynamic models, setting $\tau\in\left\{
1,2\right\}  $), or $\widehat{\rho}_{L}\,$and$\ \widehat{\beta}_{L}$ (for
static models, setting $\tau=0$); the split-panel jackknife estimators
$\widehat{\rho}_{J}$ and $\widehat{\beta}_{J}$ of \cite{cfw2018}; and the
bootstrap-corrected estimators $\widehat{\rho}_{B}$ and $\widehat{\beta}_{B}$
of \cite{hj2024}, with $499$ repetitions. We also compare the empirical sizes
and powers, at the $5\%$ level, of the LR tests based on the uncorrected
likelihood $\widehat{l}\left(  \theta\right)  $ (reporting size only), on the
corrected likelihood $L\left(  \theta\right)  $ with $\tau=1$ (for dynamic
models) or $\tau=0$ (for static model), and on the bootstrap. For the powers,
the null hypotheses are $H_{0}:\theta_{0}=(0.5,0.5)^{\prime}+\delta$ for
$\delta\in\left\{  \pm0.2,\pm0.1\right\}  $.

Summarizing important insights in Tables \ref{Table.Simulation.Design1Est} and
\ref{Table.Simulation.Design1LR}, our findings are as follows.

\begin{enumerate}
\item The MLEs $\widehat{\beta}$ and $\widehat{\rho}$ may exhibit significant
bias, especially when the sample size is small. Our procedure reduces the bias
considerably without inflating the RMSE, even at $\left(  N,T\right)  =\left(
30,30\right)  $.

\item Our corrected estimators show smaller biases than the jackknife and the
bootstrap for $\beta$ when the sample size is small. The biases of all
candidate estimators, except the MLEs, become similar as the sample size
increases. For $\left(  N,T\right)  =\left(  30,30\right)  $, the jackknife
estimators may show larger biases than the MLEs for structural parameters
associated with two-way heterogeneities. We discuss the possible cause in
Section \ref{Section.Jackknife} of the supplementary appendix.

\item The LR test based on $\widehat{l}\left(  \theta\right)  $ has severe
size-distortions even with $\left(  N,T\right)  =\left(  90,90\right)  $. On
the contrary, the LR test based on our $L\left(  \theta\right)  $ is able to
deliver an empirical size close to the nominal level of $5\%$ as the sample
size increases, maintaining relatively strong powers. For small sample sizes,
$L\left(  \theta\right)  $ reduces the type-I error risk considerably. The
empirical sizes from the bootstrap are close to the nominal level with small
sample sizes. However, the powers are relatively low, especially near the true
null hypothesis.
\end{enumerate}

%

\begin{table}[H]
\begin{centering}%
\caption{Comparisons of Uncorrected and Corrected Logit Estimates, Design 1}%
\label{Table.Simulation.Design1Est}%
\setstretch{1.2}%
\footnotesize
%

\begin{tabular*}
{\linewidth}[c]{@{\extracolsep{\fill}}llrrcrrcrr}\hline\hline
$(N,T)$ &  & \multicolumn{2}{c}{$(30,30)$} &  & \multicolumn{2}{c}{$(60,60)$}
&  & \multicolumn{2}{c}{$(90,90)$}\\
&  & Bias & RMSE & \multicolumn{1}{r}{} & Bias & RMSE & \multicolumn{1}{r}{} &
Bias & RMSE\\\hline
&  & \multicolumn{8}{c}{Static Logit Model}\\
$\widehat{\beta}$ &  & $0.165$ & $0.197$ & \multicolumn{1}{r}{} & $0.073$ &
$0.086$ & \multicolumn{1}{r}{} & $0.045$ & $0.053$\\
$\widehat{\beta}_{L}$ &  & $0.055$ & $0.106$ & \multicolumn{1}{r}{} & $0.015$
& $0.044$ & \multicolumn{1}{r}{} & $0.006$ & $0.027$\\
$\widehat{\beta}_{J}$ &  & $-0.776$ & $1.716$ & \multicolumn{1}{r}{} &
$-0.022$ & $0.049$ & \multicolumn{1}{r}{} & $-0.008$ & $0.028$\\
$\widehat{\beta}_{B}$ &  & $-0.069$ & $0.098$ & \multicolumn{1}{r}{} &
$-0.008$ & $0.040$ & \multicolumn{1}{r}{} & $-0.004$ & $0.026$\\\hline
& \multicolumn{1}{c}{} & \multicolumn{8}{c}{Dynamic Logit Model}\\
$\widehat{\beta}$ &  & $0.170$ & $0.210$ & \multicolumn{1}{r}{} & $0.072$ &
$0.087$ & \multicolumn{1}{r}{} & $0.046$ & $0.054$\\
$\widehat{\beta}_{L}^{(1)}$ &  & $0.061$ & $0.120$ & \multicolumn{1}{r}{} &
$0.016$ & $0.046$ & \multicolumn{1}{r}{} & $0.007$ & $0.027$\\
$\widehat{\beta}_{L}^{(2)}$ &  & $0.063$ & $0.122$ & \multicolumn{1}{r}{} &
$0.017$ & $0.046$ & \multicolumn{1}{r}{} & $0.007$ & $0.027$\\
$\widehat{\beta}_{J}$ &  & $-1.014$ & $2.120$ & \multicolumn{1}{r}{} &
$-0.029$ & $0.111$ & \multicolumn{1}{r}{} & $-0.009$ & $0.029$\\
$\widehat{\beta}_{B}$ &  & $-0.079$ & $0.110$ & \multicolumn{1}{r}{} &
$-0.010$ & $0.042$ & \multicolumn{1}{r}{} & $-0.004$ & $0.026$\\
$\widehat{\rho}$ &  & $-0.109$ & $0.209$ & \multicolumn{1}{r}{} & $-0.051$ &
$0.097$ & \multicolumn{1}{r}{} & $-0.030$ & $0.059$\\
$\widehat{\rho}_{L}^{(1)}$ &  & $-0.047$ & $0.165$ & \multicolumn{1}{r}{} &
$-0.018$ & $0.079$ & \multicolumn{1}{r}{} & $-0.007$ & $0.049$\\
$\widehat{\rho}_{L}^{(2)}$ &  & $-0.042$ & $0.168$ & \multicolumn{1}{r}{} &
$-0.013$ & $0.079$ & \multicolumn{1}{r}{} & $-0.003$ & $0.049$\\
$\widehat{\rho}_{J}$ &  & $0.010$ & $0.177$ & \multicolumn{1}{r}{} & $0.002$ &
$0.081$ & \multicolumn{1}{r}{} & $0.003$ & $0.051$\\
$\widehat{\rho}_{B}$ &  & $0.009$ & $0.153$ & \multicolumn{1}{r}{} & $0.000$ &
$0.077$ & \multicolumn{1}{r}{} & $0.003$ & $0.049$\\\hline\hline
\end{tabular*}
%

\begin{footnotesize}
\justify
\end{footnotesize}%
%

\end{centering}
\end{table}%
%

\begin{table}[H]
\begin{centering}%
\caption{Empirical Size and Power from LR Test for Logit Model, Design 1}%
\label{Table.Simulation.Design1LR}%
\setstretch{1.2}%
\scriptsize
%

\begin{tabular*}
{\linewidth}[c]{@{\extracolsep{\fill}}cccccccccccc}\hline\hline
& \multicolumn{3}{c}{Size} & \multicolumn{4}{c}{Power at $\delta$
(Analytical)} & \multicolumn{4}{c}{Power at $\delta$ (Bootstrap)}\\
$(N,T)$ & Uncorrected & Analytical & Bootstrap & $-0.2$ & $-0.1$ & $0.1$ &
$0.2$ & $-0.2$ & $-0.1$ & $0.1$ & $0.2$\\\hline
& \multicolumn{11}{c}{Static Logit Model}\\
$(30,30)$ & $38$ & $13$ & $0$ & $88$ & $48$ & $12$ & $45$ & $53$ & $12$ & $0$
& $0$\\
$(60,60)$ & $40$ & $8$ & $2$ & $100$ & $85$ & $59$ & $99$ & $99$ & $74$ & $0$
& $32$\\
$(90,90)$ & $38$ & $5$ & $3$ & $100$ & $98$ & $95$ & $100$ & $100$ & $98$ &
$9$ & $97$\\\hline
& \multicolumn{11}{c}{Dynamic Logit Model}\\
$(30,30)$ & $37$ & $13$ & $1$ & $84$ & $42$ & $19$ & $55$ & $42$ & $8$ & $1$ &
$4$\\
$(60,60)$ & $36$ & $10$ & $3$ & $100$ & $79$ & $64$ & $99$ & $99$ & $65$ & $7$
& $68$\\
$(90,90)$ & $33$ & $6$ & $3$ & $100$ & $98$ & $95$ & $100$ & $100$ & $96$ &
$34$ & $99$\\\hline\hline
\end{tabular*}
%

\begin{footnotesize}
\justify
\end{footnotesize}%
%

\end{centering}
\end{table}%

In Section \ref{Section.AdditionalSim.Design1} of the supplementary appendix,
we report the empirical sizes and powers from the LM and the Wald test, which
are similar to the LR here. We also report results from the probit model in
Section \ref{Section.AdditionalSim.Design1}. In Table
\ref{Table.Simulation.Design1UnequalNT}, we present simulation results for
$\left(  N,T\right)  =\left(  90,10\right)  $ showing the performance of our
approach for very small $T$. Second, we present additional simulation results
in Section \ref{Section.AdditionalSim.Poisson} of the supplementary appendix
for models with heterogeneous autoregressive coefficient. Next, in Section
\ref{Section.AdditionalSim.Design2} of the supplementary appendix, we present
results from an alternative design (Design 2), where the biases of the MLEs
are larger. Our bias correction procedure is effective under both Designs 1
and 2. In Section \ref{Section.AdditionalSim.FVW2016} of the supplementary
appendix, we simulate a dynamic panel probit model with additive two-way fixed
effects in the intercept, under Design 1 of \cite{fw2016}. Finally, in Section
\ref{Section.APE} of the supplementary appendix, we present some additional
discussion and simulation results regarding the average partial effects.

\section{Empirical Analysis}

\label{Section.Empirical}

In this section we apply our likelihood-based bias correction method to
examining the determinants of the labor force participation (LFP) decision of
single mothers. In particular, we look at the impact of the number of children
on the decision of the mother to engage in paid employment.

An extensive literature in labor economics has studied the labor supply
decisions of married women
\citep{killingsworth_chapter_1986,ae1998,blau_changes_2007}. These studies
have uncovered the impacts of a variety of economic variables, including
female market wage and husband income \citep{mincer_labor_1962}, education
\citep{heath_causes_2016}, childcare costs \citep{connelly_effect_1992}, the
cost of home technology \citep{greenwood_technology_2016}, and culture norms
\citep{fernandez_cultural_2013}. Among these variables, the number of children
consistently emerges as one of the most important determinants of female labor
supply \citep{nakamura_econometrics_1992}. This is unsurprising since women
continue to bear a disproportionate share of childrearing responsibilities \citep{aguero_motherhood_2008}.

In contrast to the substantial body of research on the labor supply behavior
of married women, the labor supply decisions of single mothers have received
relatively limited attention, with only a few exceptions
\citep{kimmel_child_1998,blundell_female_2016}. Single mothers, however, face
distinctive challenges when it comes to balancing work and childrearing
responsibilities due to the absence of a second earner in the household.
Moreover, their employment decisions may have a more pronounced impact on
their children's well-being compared to the decisions of married women and
should thus be of great importance to economists and policymakers.

To study the labor supply decisions of single mothers, we compile a dataset
from waves 20--30 of the Panel Study of Income Dynamics (PSID), which span the
period of 1987 to 1997. Our sample includes only single mothers, defined as
unmarried female household heads with at least one child. The dependent
variable is labor force participation status, defined as whether the
individual worked any hours during the interview year. Our main explanatory
variable is the number of children under 18 in the household. Following
\citet{dj2015}, we focus on the \textit{informative sample} of individuals
aged 18--60 whose current and lagged LFP status each changed at least once
during the period. This restriction ensures within-individual variation in
both variables, which is necessary for identification of the dynamic model.
After applying this criterion, the estimation sample comprises $N=86$ single
mothers observed for $T=10$ consecutive years (1987--1996), with lagged
participation status measured from 1986 to 1995. Further details on data
construction are provided in Section \ref{sec:APP} of the supplementary
appendix. On average, 28 percent of these individuals switched into or out of
the labor force in a given year (Figure \ref{fig:lfpct}--\ref{fig:lfpflow}).
Table \ref{tab:DS} reports the summary statistics.

Let $Y_{it}\in\{0,1\}$ denote the labor-force-participation status of single
mother $i$ in year $t$, and let $X_{it}$ denote her number of children. To
assess the impact of fertility on labor supply, we estimate the following
dynamic probit models:
\begin{align}
Y_{it}  &  =\mathds{1}\!\left\{  \rho Y_{i,t-1}+\beta X_{it}+c_{i}+\tau
_{t}+\varepsilon_{it}>0\right\}  ,\label{eq:em01}\\
Y_{it}  &  =\mathds{1}\!\left\{  (\rho+\zeta_{i}+\eta_{t})Y_{i,t-1}%
+(\beta+\alpha_{i}+\gamma_{t})X_{it}+c_{i}+\tau_{t}+\varepsilon_{it}%
>0\right\}  , \label{eq:em02}%
\end{align}
where $\varepsilon_{it}$ follows the standard normal distribution, and
$\sum_{i}c_{i}=\sum_{i}\zeta_{i}=\sum_{i}\alpha_{i}=\sum_{t}\eta_{t}=\sum
_{t}\gamma_{t}=0$. Model \eqref{eq:em01} is the dynamic homogeneous-slope
model, which includes two-way fixed effects in the intercept. Model
\eqref{eq:em02} is the dynamic heterogeneous-slope model, which extends the
specification by allowing the slope coefficients---that is, the coefficients
on both the lagged dependent variable and the number of children---to vary
across individuals and over time: $\rho_{it}=\rho+\zeta_{i}+\eta_{t}$ and
$\beta_{it}=\beta+\alpha_{i}+\gamma_{t}$. The parameter $\beta_{it}$ measures
the heterogeneous impact of the number of children on a mother's latent
propensity to work, while $\rho_{it}$ captures heterogeneous state dependence
in labor force participation. Model \eqref{eq:em02} nests Model
\eqref{eq:em01} as a special case when $\rho_{it}$ and $\beta_{it}$ are
constant across $i$ and $t$.

As discussed in Section \ref{Section.Introduction}, when the true effect of
$X_{it}$ on the outcome, $\beta_{it}$, varies across the population, it is
important to account for its potential correlation with $X_{it}$. In practice,
such correlation often arises from self-selection. For instance, single
mothers who choose to have more children might be those with greater financial
resources or better access to childcare, such that their labor supply is less
affected by additional children. Alternatively, correlation between
$\beta_{it}$ and $X_{it}$ may stem from common time trends: over time,
fertility rates declined while concurrent factors such as rising childcare and
schooling costs, advances in home technology, and evolving welfare policies
(e.g., child tax credits) altered the effect of each additional child on their
mother's LFP. In both scenarios, Model \eqref{eq:em02} is the appropriate
specification. It controls for unobserved heterogeneity in $\beta_{it}$ by
incorporating individual and time fixed effects into the slope and allowing
them to correlate with $X_{it}$. Controlling for heterogeneous state
dependence further improves robustness by allowing the persistence of
labor-force participation to differ across individuals and periods, capturing
additional sources of dynamic heterogeneity.

Given our small sample size, direct estimation of both models could suffer
from severe incidental parameter problems, which we address using our bias
correction procedure. Table \ref{tab:M1} presents the estimation results.
Panel A reports results for Model \eqref{eq:em01} under the homogeneous slope
assumption. For each parameter, we provide the MLE and bias-corrected
estimates, along with their standard errors. Examining the bias-corrected
estimates, we observe that an increase in the number of children appears to
reduce a single mother's likelihood of labor force participation, while past
LFP status strongly predicts current LFP. However, this relationship shifts
dramatically once we account for slope heterogeneity. Panel B presents the
results for Model \eqref{eq:em02}, allowing for heterogeneous slopes. Here,
the bias-corrected estimates reveal a positive average impact of additional
children on a mother's labor supply: each additional child increases the
latent propensity to participate in the labor force by 0.436, which
corresponds to an average increase of about 7 percent in the probability of
employment. This indicates that, on average, single mothers with more children
are more likely to work, likely driven by the increased financial demands of
supporting multiple children independently. Additionally, unlike the
homogeneous-slope model, the serial correlation in labor-force participation
remains statistically significant but diminishes in magnitude, suggesting that
what appears to be strong state dependence in the homogeneous model partly
reflects unobserved heterogeneity in individuals' persistence of employment
rather than uniform structural dynamics. Together, these findings underscore
the importance of controlling for unobserved slope heterogeneity, as failing
to do so can lead to substantially different conclusions.

In addition to highlighting the difference between homogeneous and
heterogeneous slope models, Table \ref{tab:M1} demonstrates the importance of
bias-correction in estimating dynamic nonlinear panel data models. Examining
the results in Panel B, we observe that the MLE overestimates the impact of
the number of children while underestimating the degree of serial correlation
in LFP. Bias correction results in a 38 percent lower estimate of the former
and a 32 percent higher estimate of the latter.

\begin{center}
\begin{figure}[H]
\caption
{Estimated Heterogeneous Effects of Children on Single Mother Labor Force Participation}
\label{figure2}
\centering\begin{subfigure}[b]{0.48\textwidth}
\centering\includegraphics[width=\textwidth]{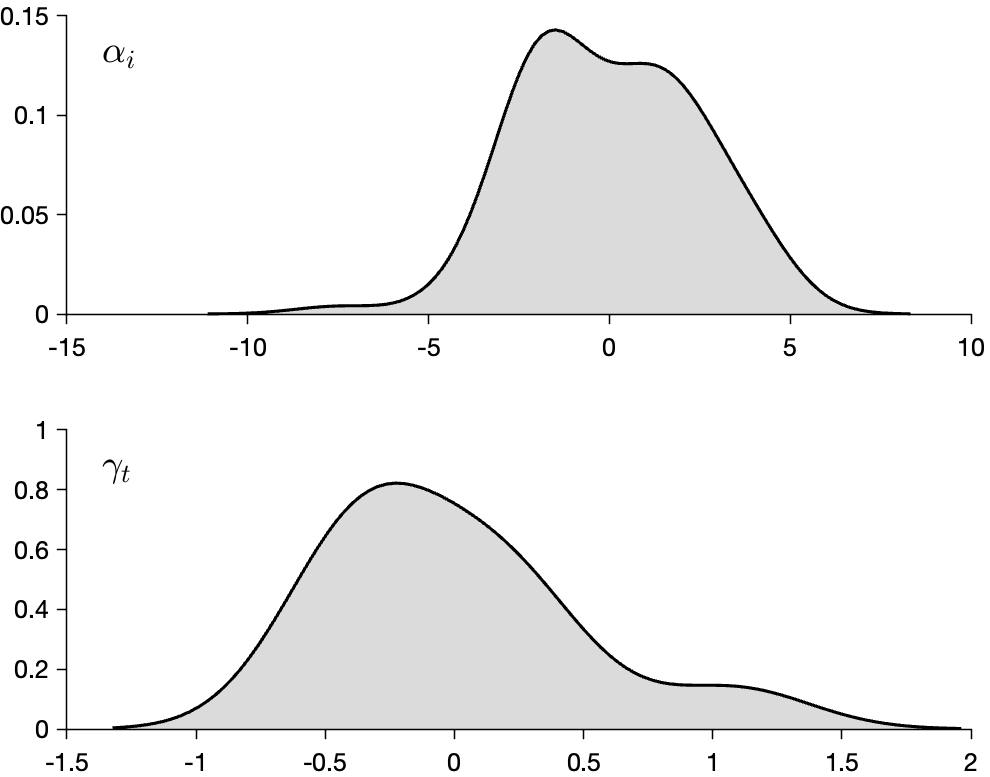}
\caption{\small Distribution of Estimated Effects}
\label{fig:tauA}
\end{subfigure}
\hfill\begin{subfigure}[b]{0.48\textwidth}
\centering\includegraphics[width=\textwidth]{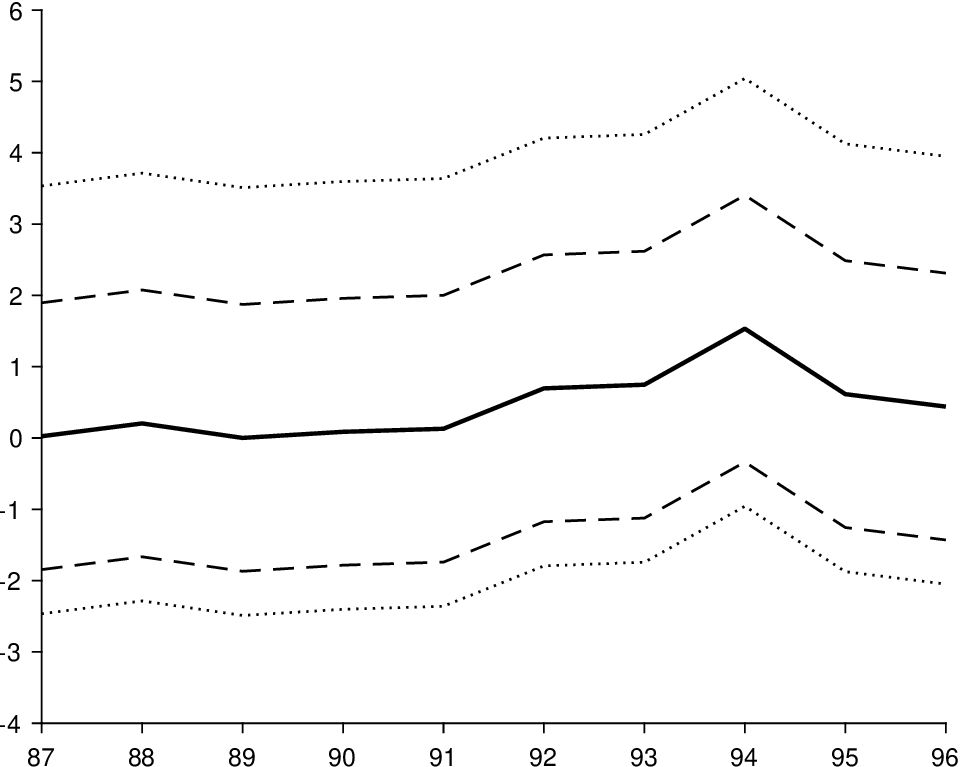}
\caption{\small Temporal Evolution of Estimated Effects}
\label{fig:tauB}
\end{subfigure}
\vspace{1mm}
\begin{flushleft}
\begin{justify}
\begin{footnotesize}
\noindent\textit{Notes}%
: Panel (a) shows the distributions of the estimated individual effects $\alpha
_i$ (top) and time effects $\gamma
_t$ (bottom) from the dynamic heterogeneous-slope model. Panel (b) plots the median (solid), interquartile range (25th\textendash
{}75th, dashed), and decile range (10th\textendash{}%
90th, dotted) of the slope coefficients $\beta_{it}$ over time. Each $\beta
_{it}$ is constructed as the sum of the bias-corrected average effect $\beta
$ and the estimated individual and time effects $\alpha_i$ and $\gamma_t$.
\end{footnotesize}
\end{justify}
\end{flushleft}
\end{figure}
\end{center}

Finally, in Figure \ref{fig:tauA}, we plot the distributions of the estimated
individual effects $\alpha_{i}$ and time effects $\gamma_{t}$ from Model
\eqref{eq:em02}, with corresponding summary statistics reported in Table
\ref{tab:M2d} (supplementary appendix).\footnote{\label{footnote.oy2020}We
acknowledge that the estimated densities of $\alpha_{i}$ and $\gamma_{t}$ may
themselves be affected by the incidental-parameter problem; see
\citet{oy2020}.} These estimates reveal substantial heterogeneity in the
effect of children on single mothers' labor force participation both across
individuals and over time. Figure \ref{fig:tauB} illustrates the temporal
evolution of the slope coefficients $\beta_{it}$, constructed as the sum of
the bias-corrected average effect $\beta$ and the estimated individual and
time effects $\alpha_{i}$ and $\gamma_{t}$. The median of $\beta_{it}$ rises
steadily from 1987 to 1994---implying an increasingly positive labor-supply
response to additional children---before declining modestly thereafter. The
interquartile and decile bands remain wide throughout, highlighting persistent
dispersion across individuals. Together, these results reinforce the
importance of accounting for both individual and temporal variations in slope
coefficients, as the labor supply effects of children can vary significantly
based on unobserved factors unique to each mother and time period.

In conclusion, our analysis demonstrates the efficacy of our likelihood-based
bias correction method for estimating dynamic nonlinear models of labor supply
with multi-dimensional heterogeneities. We show the importance of controlling
for unobserved slope heterogeneity when it may correlate with regressors and
the value of panel data models that accommodate two-way fixed effects in both
the intercept and the slope. Our findings reveal that the number of children
has, on average, a positive influence on a single mother's labor force
participation, but such effect is highly heterogeneous across individuals and
time periods. This heterogeneity underscores the complex nature of labor
supply decisions and the unique challenges each single mother faces in
balancing work and childrearing responsibilities.%

\setlength{\tabcolsep}{10pt}
\begin{table}[H]
\centering\begin{threeparttable}
\caption{Single Mother Labor Force Participation: Estimation Results \protect
\label{tab:M1}}
\begin{tabular}{V{\linewidth}V{\linewidth}V{\linewidth}V{\linewidth
}V{\linewidth}}
\hline\hline& \multicolumn{2}{c}{Number of Children} & \multicolumn{2}%
{c}{Lagged Participation}\tabularnewline\hline
& Uncorrected & Corrected & Uncorrected & Corrected\tabularnewline
\hline\multicolumn{5}{V{\linewidth}}{\vspace{1bp}
\emph{Panel A. Homogeneous Slope}
\ }\tabularnewline
Model parameter & -0.923 & -0.851 & 0.589 & 0.540\tabularnewline
& (0.232) & (0.228) & (0.108) & (0.107)\tabularnewline\hline\multicolumn
{5}{V{\linewidth}}{\vspace{1bp}
\emph{Panel B. Heterogeneous Slope}
\ }\tabularnewline
Model parameter & 0.704 & 0.436 & 0.191 & 0.253\tabularnewline
& (0.141) & (0.144) & (0.106) & (0.105)\tabularnewline\hline\hline
\end{tabular}
\vspace{1mm}
\begin{tablenotes}
\footnotesize\item\emph{Notes:}
Standard errors in parentheses. Panel A reports the estimation results of the dynamic homogeneous-slope model. Panel B reports the estimation results of the dynamic heterogeneous-slope model. The ``uncorrected'' and the ``corrected'' columns show respectively the MLE and the bias-corrected parameter estimates. Data source: PSID 1987--1997.
\end{tablenotes}
\end{threeparttable}
\end{table}%
%

\section{Final Remarks}

\label{Section.Conclusion}

In this paper, we propose a likelihood-based analytical bias correction
procedure for a general class of two-way dynamic nonlinear heterogeneous
parameter models subject to the incidental parameter problem. We give the
analytical form of a corrected likelihood and show that it delivers point
estimators that are asymptotically unbiased and test statistics that are
asymptotically $\chi^{2}$-distributed. Simulation studies and empirical
analyses support our claims.

Several issues deserve further studies. First, we have not rigorously
investigated the asymptotic properties of the average partial effects.
Admittedly, average partial effect is very important, especially for nonlinear
models. It is therefore worthwhile to formally establish its asymptotic
theory. Second, there has been various fixed-$T$ consistent solutions for the
logit model with fixed effects. However, to the best of our knowledge, we have
not seen any such developments for the two-way heterogeneous parameter logit
model. It could be useful to push such a development, because the logit model
is widely used in many studies. Finally, certain micro panels tend to have a
small $T$, in which case the higher-order bias terms may be non-negligible.
For this reason, it may be potentially important to develop high-order bias
correction approaches for two-way dynamic nonlinear heterogeneous parameter models.%

\section*{Acknowledgement}%

We would like to thank the editor Iv%
\'a%
n Fern%
\'a%
ndez-Val and two anonymous referees for their valuable comments and
suggestions. We would like to express our gratitude to comments and
suggestions from Geert Dhaene, Yanqin Fan, Jinyong Hahn, Chen Hsiao, Whitney
Newey, and Jun Yu. Our research assistants Yuchi Liu and Xinrui Zhang helped
us with various tasks. Yutao Sun is the sole corresponding author of this
paper. All correspondence shall be sent to Yutao Sun. Leng's research is
partially supported by the National Natural Science Foundation of China
(72573136), the Fujian Natural Science Foundation in China (2024J08013), and
the Basic Scientific Center Project of National Natural Science Foundation of
China (71988101). Mao acknowledges support from the NSFC Basic Science Center
Project for Econometric Modeling and Economic Policy Studies (Grant No.
71988101). Sun acknowledges the support from the National Natural Science
Foundation of China under Grant Number 72203032.%

\section*{\label{Section.Appendix}Appendix}
\sectionbookmark{Appendix}
\renewcommand{\thesubsection}{\Alph{subsection}}
\numberwithin{equation}{subsection}
\numberwithin{remark}{subsection}
\numberwithin{figure}{subsection}
\numberwithin{table}{subsection}
\numberwithin{lemma}{subsection}
\numberwithin{proposition}{subsection}
\numberwithin{example}{subsection}%

\subsection{Technical Details}

\label{Section.TechDetails}

\subsubsection{Model and Estimation}

\label{Section.TechDetails.ModelAndEstimation}

Our specification accommodates a general class of models that is not yet
covered in the current literature. We use the next example to illustrate the
generality of our settings.

\begin{example}
[heterogeneous parameter model]\label{Example.Motivate}Consider a probit model
with the data-generating process
\begin{equation}
Y_{it}=
\mathds{1}
\{\pi(X_{it},\theta_{0},\alpha_{i}^{0},\gamma_{t}^{0})+\varepsilon_{it}>0\},
\label{Eq.Example.Motivate1}%
\end{equation}
where $\varepsilon_{it}$ is a standard normal random variable independent of
$X_{it}$, $\theta_{0}$, $\alpha_{i}^{0}$, and $\gamma_{t}^{0}$. For arguments
$\theta$, $\alpha_{i}$, and $\gamma_{t}$ of dimension $K\times1$, we allow the
linear index\footnote{We use a model with a linear index solely for the
purpose of illustration. Our method applies to general models. One example is
the modified Neyman-Scott model in Example \ref{Example.HeteroNormalMean2}.}
$\pi(X_{it},\theta,\alpha_{i},\gamma_{t})$ to take the following general
form:
\[
\pi(X_{it},\theta,\alpha_{i},\gamma_{t})=\sum_{k=1}^{K}(\theta_{k}
+\alpha_{k,i}+\gamma_{k,t})X_{k,it},
\]
where $X_{k,it}$, $\theta_{k}$, $\alpha_{k,i}$, and $\gamma_{k,t}$ are the
$k$-th components of $X_{it}$, $\theta$, $\alpha_{i}$, and $\gamma_{t}$,
respectively, $k=1,\ldots,K$. In addition, when $X_{1,it}=1$,\ the
data-generating process in (\ref{Eq.Example.Motivate1}) consists of the
traditional individual- and time-effects in the intercept. When $X_{k,it}
=Y_{it-1}$ for some $k$, the data-generating process is dynamic because each
$Y_{it}$ depends on its past value.

Moreover, we also allow a subset of the coefficients to be homogeneous,
thereby accommodating both heterogeneous and homogeneous parameters. For
instance, when it is known that $\alpha_{1,i}^{0}=\gamma_{1,t}^{0}=0$, the
model can be estimated under constraints $\alpha_{1,i}=\gamma_{1,t}=0$ for all
$(i,t)$. In this case, the linear index is effectively
\[
\pi(X_{it},\theta,\alpha_{i},\gamma_{t})=\theta_{1}X_{1,it}+\sum_{k=2}%
^{K}(\theta_{k}+\alpha_{k,i}+\gamma_{k,t})X_{k,it},
\]
where $X_{1,it}$ is associated with a \textquotedblleft homogeneous parameter"
$\theta_{1}$ while all other $X_{k,it}$ ($k=2,\ldots,K$) are associated with
heterogeneous parameters $(\theta_{k}+\alpha_{k,i}+\gamma_{k,t})$. In a
similar fashion, our approach also applies to, e.g.,
\[
\pi(X_{it},\theta,\alpha_{i},\gamma_{t})=(\alpha_{1,i}+\gamma_{1,t}%
)X_{1,it}+\sum_{k=2}^{K}(\theta_{k}+\alpha_{k,i}+\gamma_{k,t})X_{k,it},
\]
with the constraint $\theta_{1}=0$. Other variants of $\pi(X_{it}%
,\theta,\alpha_{i},\gamma_{t})$ include:
\begin{align}
\pi(X_{it},\theta,\alpha_{i},\gamma_{t})  &  =(\alpha_{1,i}+\gamma
_{1,t})+\theta_{2}Y_{it-1}+\sum_{k=3}^{K}\theta_{k}X_{k,it}%
,\label{Eq.Example.Motivate5}\\
\pi(X_{it},\theta,\alpha_{i},\gamma_{t})  &  =(\theta_{1}+\alpha_{1,i}%
+\gamma_{1,t})+\sum_{k=2}^{K}(\theta_{k}+\alpha_{k,i}+\gamma_{k,t}%
)X_{k,it},\label{Eq.Example.Motivate6}\\
\pi(X_{it},\theta,\alpha_{i},\gamma_{t})  &  =(\theta_{1}+\alpha_{1,i}%
+\gamma_{1,t})+(\theta_{2}+\alpha_{2,i}+\gamma_{2,t})Y_{it-1}\nonumber\\
&  \quad+\sum_{k=3}^{K_{1}}(\theta_{k}+\alpha_{k,i}+\gamma_{k,t})X_{k,it}%
+\sum_{k=K_{1}+1}^{K}\theta_{k}X_{k,it} \label{Eq.Example.Motivate7}%
\end{align}
for some $3\leq K_{1}<K$. To our knowledge, except (\ref{Eq.Example.Motivate5}
) considered by \cite{fw2016}, none of the above specifications are fully
addressed under nonlinear models with predetermined regressors. The
specification in (\ref{Eq.Example.Motivate6}) is the same to \cite{kn2020} and
\cite{ls2023}, except that they consider only the linear regression model.
\cite{ko2018} provide a bias correction technique for general nonlinear models
with (\ref{Eq.Example.Motivate6}) only for strictly exogenous $X_{k,it}$.
Currently, no methods are available for nonlinear models with
(\ref{Eq.Example.Motivate7}).
\end{example}

The next example illustrates our reparameterization strategy.

\begin{example}
[reparameterization]\label{Example.HeteroNormalMean1}For some scalar
structural parameter $\mu_{0}$ of interests and $K=1$, consider $Y_{it}%
=\mu_{0}+\alpha_{i}^{0}+\gamma_{t}^{0}+\varepsilon_{it}$ where $\varepsilon
_{it}$ is standard normal independent across both $i$ and $t$, and independent
of $\alpha_{i}^{0}$ and $\gamma_{t}^{0}$. We restrict the variance of
$\varepsilon_{it}$ to be $1$ because we use this example only to demonstrate
the identification of $\mu_{0}$. In Example \ref{Example.HeteroNormalMean2}
below, we indeed allow $\varepsilon_{it}$ to have an unknown variance.

Conditioning on $\phi^{0}$, the original likelihood function is
\begin{align*}
\ell(\mu,\phi)  &  =-\frac{1}{2}\log(2\pi)-\frac{1}{NT}\sum_{i=1}^{N-1}%
\sum_{t=1}^{T-1}\frac{1}{2}(Y_{it}-\mu-\alpha_{i}-\gamma_{t})^{2}-\frac{1}%
{NT}\sum_{t=1}^{T-1}\frac{1}{2}(Y_{Nt}-\mu-\alpha_{N}-\gamma_{t})^{2}\\
&  \quad-\frac{1}{NT}\sum_{i=1}^{N-1}\frac{1}{2}(Y_{iT}-\mu-\alpha_{i}%
-\gamma_{T})^{2}-\frac{1}{NT}\frac{1}{2}(Y_{NT}-\mu-\alpha_{N}-\gamma_{T}%
)^{2}.
\end{align*}
Setting $\alpha_{N}=-\sum_{i=1}^{N-1}\alpha_{i}$ and $\gamma_{T}=-\sum
_{t=1}^{T-1}\gamma_{t}$, the reparameterized likelihood is
\begin{align*}
l(\mu,\psi)  &  =-\frac{1}{2}\log(2\pi)-\frac{1}{NT}\sum_{i=1}^{N-1}\sum
_{t=1}^{T-1}\frac{1}{2}(Y_{it}-\mu-\alpha_{i}-\gamma_{t})^{2}-\frac{1}{NT}%
\sum_{t=1}^{T-1}\frac{1}{2}\left(  Y_{Nt}-\mu+\sum_{i=1}^{N-1}\alpha
_{i}-\gamma_{t}\right)  ^{2}\\
&  \quad-\frac{1}{NT}\sum_{i=1}^{N-1}\frac{1}{2}\left(  Y_{iT}-\mu-\alpha
_{i}+\sum_{t=1}^{T-1}\gamma_{t}\right)  ^{2}-\frac{1}{NT}\frac{1}{2}\left(
Y_{NT}-\mu+\sum_{i=1}^{N-1}\alpha_{i}+\sum_{t=1}^{T-1}\gamma_{t}\right)  ^{2}.
\end{align*}
Calculating the first derivative of $l(\mu,\psi)$ w.r.t. $\alpha_{i}$, we
have
\[
\partial_{\alpha_{i}}l(\mu,\psi)=\frac{1}{NT}\sum_{t=1}^{T}(Y_{it}-\mu
-\alpha_{i})-\frac{1}{NT}\sum_{t=1}^{T}\left(  Y_{Nt}-\mu+\sum_{i=1}%
^{N-1}\alpha_{i}\right)
\]
for every $i=1,\ldots,N-1$. Equating $\partial_{\alpha_{i}}l(\mu,\psi)=0$, we
obtain
\begin{equation}
\alpha_{i}=\frac{1}{T}\sum_{t=1}^{T}Y_{it}-\frac{1}{T}\sum_{t=1}^{T}%
Y_{Nt}-\sum_{i=1}^{N-1}\alpha_{i}. \label{Eq.Example.HeteroNormalMean1.1}%
\end{equation}
Notice that (\ref{Eq.Example.HeteroNormalMean1.1}) is independent of $\mu$ and
all $\gamma_{t}$. Now, sum (\ref{Eq.Example.HeteroNormalMean1.1}) over
$i=1,\ldots,N-1$, we obtain
\begin{equation}
\sum_{i=1}^{N-1}\alpha_{i}=\frac{1}{NT}\sum_{i=1}^{N}\sum_{t=1}^{T}%
Y_{it}-\frac{1}{T}\sum_{t=1}^{T}Y_{Nt}. \label{Eq.Example.HeteroNormalMean1.2}%
\end{equation}
Combining (\ref{Eq.Example.HeteroNormalMean1.1}) and
(\ref{Eq.Example.HeteroNormalMean1.2}), it follows that
\[
\widehat{\alpha}_{i}=\frac{1}{T}\sum_{t=1}^{T}Y_{it}-\frac{1}{NT}\sum
_{i=1}^{N}\sum_{t=1}^{T}Y_{it},\qquad i=1,\ldots,N-1.
\]
It can be calculated in the same way that
\[
\widehat{\gamma}_{t}=\frac{1}{N}\sum_{i=1}^{N}Y_{it}-\frac{1}{NT}\sum
_{i=1}^{N}\sum_{t=1}^{T}Y_{it},\qquad t=1,\ldots,T-1.
\]
Subsequently,
\begin{equation}
\widehat{\mu}=\frac{1}{NT}\sum_{i=1}^{N}\sum_{t=1}^{T}Y_{it}.
\label{Eq.Example.HeteroNormalMean1.3}%
\end{equation}
In this simple example, $\widehat{\alpha}_{i}$ and $\widehat{\gamma}_{t}$ do
not depend on $\mu$. This is generally not the case for complex models.

In addition, maximizing $l(\mu,\psi)$ is identical to maximizing\ the original
likelihood $\ell\left(  \mu,\phi\right)  $ under constraints $\sum_{i=1}%
^{N}\alpha_{i}=0$ and $\sum_{t=1}^{T}\gamma_{t}=0$. Specifically, it can be
verified that the Lagrangian $\ell\left(  \mu,\phi\right)  +\lambda^{\alpha
}\sum_{i=1}^{N}\alpha_{i}+\lambda^{\gamma}\sum_{t=1}^{T}\gamma_{t}$, where
$\lambda^{\alpha}$ and $\lambda^{\gamma}$ are the two Lagrange-multipliers,
delivers upon maximization the same $\widehat{\alpha}_{i}$ and
$\widehat{\gamma}_{t}$ for $i=1,\ldots,N-1$ and $t=1,\ldots,T$, respectively.
For $\widehat{\alpha}_{N}$ and $\widehat{\gamma}_{T}$, the Lagrangian
produces
\[
\widehat{\alpha}_{N}=\frac{1}{T}\sum_{t=1}^{T}Y_{Nt}-\frac{1}{NT}\sum
_{i=1}^{N}\sum_{t=1}^{T}Y_{it},\qquad\widehat{\gamma}_{T}=\frac{1}{N}%
\sum_{i=1}^{N}Y_{iT}-\frac{1}{NT}\sum_{i=1}^{N}\sum_{t=1}^{T}Y_{it}.
\]
Indeed, $\widehat{\alpha}_{N}=-\sum_{i=1}^{N-1}\widehat{\alpha}_{i}$ as in
Equation (\ref{Eq.Example.HeteroNormalMean1.2}), and it can be verified that
$\widehat{\gamma}_{T}=-\sum_{t=1}^{T-1}\widehat{\gamma}_{t}$ as well.
\end{example}

\subsubsection{Incidental Parameter Problem}

\label{Section.TechDetails.BiasExpansion}

In this section, we give additional technical details about our discussion in
Section \ref{Section.IPP.IPPBias}. We start by giving the following remark on
our definition of $\mathbb{E}$.

\begin{remark}
[expectation]\label{Remark.Expectation}Our definition of $\mathbb{E}$ depends
on the model specification and is the same to \cite{ab2009} and \cite{fw2016}
.\ Consider the frequently used autoregressive model with $X_{it}%
=(Y_{it-1},Z_{it}^{\prime})^{\prime}$ for some strictly exogenous regressor
$Z_{it}$. It is common that, for each $i=1,\ldots,N$, the density of $Y_{i1}$
is missing and (\ref{Eq.IPP.OriginalLike}) becomes a conditional likelihood as
in Chapter 5 of \cite{h1994}. In this case, the expectation is taken w.r.t.
$f_{Y_{i2},\ldots,Y_{iT}}\left(  \cdot|Y_{i1},Z_{i1},\ldots,Z_{iT},\phi
^{0};\theta_{0}\right)  $, which is the density of only $Y_{i2},\ldots,Y_{iT}%
$, conditional on the initial value $Y_{i1}$, strictly exogenous regressors
$Z_{i1},\ldots,Z_{iT}$, and the unobserved effects $\phi^{0}$. However, when
the density of $Y_{i1}$ is indeed specified, so that
(\ref{Eq.IPP.OriginalLike}) is an exact likelihood, the expectation is taken
w.r.t. $f_{Y_{i1},\ldots,Y_{iT}}\left(  \cdot|Z_{i1},\ldots,Z_{iT},\phi
^{0};\theta_{0}\right)  $, which is the joint density of all $Y_{i1}%
,\ldots,Y_{iT}$, conditioning on the strictly exogenous regressors
$Z_{i1},\ldots,Z_{iT}$ and the unobserved effects $\phi^{0}$ only.
\end{remark}

Next, the exact expressions of the remainders are as follows:%
\begin{align}
r(\theta)  &  :=\frac{1}{2}(r^{s}(\theta,a))^{\prime}[H\left(  \theta\right)
]^{-1}r^{s}(\theta,a)+r^{l}(\theta,a_{0}),\label{Eq.Appendix.RemainderR}\\
r^{s}\left(  \theta\right)   &  :=r^{s}\left(  \theta,a\right)  :=\left(
r_{\psi_{1}}^{s}\left(  \theta,a_{1}\right)  ,\ldots,r_{\psi_{(N+T-2)K}}%
^{s}(\theta,a_{(N+T-2)K})\right)  ^{\prime},\label{Eq.Appendix.RemainderRS}\\
r^{l}\left(  \theta\right)   &  :=r^{l}\left(  \theta,a_{0}\right)  :=\frac
{1}{3}\left(  \widehat{\psi}(\theta)-\psi(\theta)\right)  ^{\prime}%
r^{s}(\theta,a_{0}\times\iota_{(N+T-2)K}), \label{Eq.Appendix.RemainderRL}%
\end{align}
for some $a_{0}\in(0,1)$, where\footnote{\cite{fwct2014} point out that the
expansion in (\ref{Eq.IPP.PsiExpansion}) needs to be done
component-by-component. This leads to a different $a_{j}$ for each component
of $r^{s}\left(  \theta,a\right)  $.}%
\begin{align*}
r_{\psi_{j}}^{s}\left(  \theta,a_{j}\right)   &  :=\frac{1}{2}\left(
\widehat{\psi}\left(  \theta\right)  -\psi(\theta)\right)  ^{\prime}%
G_{\psi_{j}}(\theta,a_{j})\left(  \widehat{\psi}(\theta)-\psi(\theta)\right)
,\\
G_{\psi_{j}}(\theta,a_{j})  &  :=\partial_{\psi_{j}}H(\theta,a_{j}%
\widehat{\psi}(\theta)+(1-a_{j})\psi\left(  \theta\right)  )
\end{align*}
with $a:=(a_{1},\ldots,a_{(N+T-2)K})^{\prime}$ for each $a_{j}\in(0,1)$;
$\psi_{j}$ denotes the $j$-th component of $\psi$; and $G_{\psi_{j}}%
(\theta,a_{j})$ is the derivative of each element of $H(\theta,\psi)$ with
respect to $\psi_{j}$, evaluated at the point $a_{j}\widehat{\psi}\left(
\theta\right)  +\left(  1-a_{j}\right)  \psi\left(  \theta\right)  $. In Lemma
\ref{Lemma.RemainderR}, we state that $\mathbb{E}r(\theta)$ is $O\left(
1/\left(  NT\right)  \right)  $ uniformly in $\theta$ and give the proof thereof.

Using Example \ref{Example.HeteroNormalMean2} below, we work out the
explicitly bias of the likelihood and the estimator for a modified
Neyman-Scott model, where $Y_{it}|\phi^{0}$ follows a normal distribution with
a homogeneous variance but heterogeneous means.

\begin{example}
[heterogeneous mean model]\label{Example.HeteroNormalMean2}Following the same
setting in Example \ref{Example.HeteroNormalMean1} but let $\varepsilon_{it}$
be normal with mean $0$ and variance $\sigma_{0}>0$. In the current example,
$\theta_{0}=(\mu_{0},\sigma_{0})^{\prime}$ is the structural parameter of
interests. This example is a modified version of \cite{ns1948} and is similar
to \cite{fw2016}, except that the mean of $Y_{it}$ is $\theta_{0}+\alpha
_{i}^{0}+\gamma_{t}^{0}$, conditional on $\phi^{0}$.

Define\footnote{In this simple example, $\varepsilon_{it}$ is independent
across both $i$ and $t$. Therefore, $\mathbb{E}$ is simply w.r.t. the normal
density having mean $\mu^{0}+\alpha_{i}^{0}+\gamma_{t}^{0}$ and variance
$\sigma^{0}$.} $\widetilde{Y}_{it}:=Y_{it}-\mathbb{E}Y_{it}$. Following the
same steps in Example \ref{Example.HeteroNormalMean1}, we can derive
\begin{align}
\widehat{\alpha}_{i}-\alpha_{i}^{0}  &  =\frac{1}{T}\sum_{t=1}^{T}
\widetilde{Y}_{it}-\frac{1}{NT}\sum_{i=1}^{N}\sum_{t=1}^{T}\widetilde{Y}
_{it},\qquad i=1,\ldots,N-1,\nonumber\\
\widehat{\gamma}_{t}-\gamma_{t}^{0}  &  =\frac{1}{N}\sum_{i=1}^{N}
\widetilde{Y}_{it}-\frac{1}{NT}\sum_{i=1}^{N}\sum_{t=1}^{T}\widetilde{Y}
_{it},\qquad t=1,\ldots,T-1, \label{Eq.Example.HeteroNormalMean2.1}%
\end{align}
where $\alpha_{i}^{0}$ and $\gamma_{t}^{0}$ can be obtained by maximizing
$\mathbb{E}l(\mu,\psi)$ in the same way. Denote additionally
\[
\widehat{\alpha}_{N}:=-\sum_{i=1}^{N-1}\widehat{\alpha}_{i},\qquad\alpha
_{N}^{0}:=-\sum_{i=1}^{N-1}\alpha_{i}^{0},\qquad\widehat{\gamma}_{T}
:=-\sum_{t=1}^{T-1}\widehat{\gamma}_{t},\qquad\gamma_{T}^{0}:=-\sum
_{t=1}^{T-1}\gamma_{t}^{0}.
\]
It follows that
\begin{equation}
\widehat{\alpha}_{N}-\alpha_{N}^{0}=\frac{1}{T}\sum_{t=1}^{T}\widetilde{Y}
_{Nt}-\frac{1}{NT}\sum_{i=1}^{N}\sum_{t=1}^{T}\widetilde{Y}_{it}
,\qquad\widehat{\gamma}_{T}-\gamma_{T}^{0}=\frac{1}{N}\sum_{i=1}
^{N}\widetilde{Y}_{iT}-\frac{1}{NT}\sum_{i=1}^{N}\sum_{t=1}^{T}\widetilde{Y}
_{it}. \label{Eq.Example.HeteroNormalMean2.2}%
\end{equation}
Now, for $\psi^{0}:=({\alpha}_{1}^{0},\ldots,{\alpha}_{N-1}^{0},{\gamma}
_{1}^{0},\ldots,{\gamma}_{T-1}^{0})^{\prime}$, $\widehat{\psi}
:=(\widehat{\alpha}_{1},\ldots,\widehat{\alpha}_{N-1},\widehat{\gamma}
_{1},\ldots,\widehat{\gamma}_{T-1})^{\prime}$, and $Y_{it}(\mu):=Y_{it}
-\mu-\alpha_{i}^{0}-\gamma_{t}^{0}$, we can calculate
\begin{align}
l(\theta,\widehat{\psi})-l(\theta,\psi^{0})  &  =\frac{1}{\sigma}\frac{1}
{NT}\sum_{i=1}^{N}\sum_{t=1}^{T}\left(  Y_{it}(\mu)-\frac{1}{2}
(\widehat{\alpha}_{i}-\alpha_{i}^{0})\right)  (\widehat{\alpha}_{i}-\alpha
_{i}^{0})\nonumber\\
&  \quad\quad+\frac{1}{\sigma}\frac{1}{NT}\sum_{i=1}^{N}\sum_{t=1}^{T}\left(
Y_{it}(\mu)-\frac{1}{2}(\widehat{\gamma}_{t}-\gamma_{t}^{0})\right)
(\widehat{\gamma}_{t}-\gamma_{t}^{0}), \label{Eq.Example.HeteroNormalMean2.3}%
\end{align}
where $\theta=(\mu,\sigma)^{\prime}$ with $\sigma>0$. Substituting
(\ref{Eq.Example.HeteroNormalMean2.1}) and
(\ref{Eq.Example.HeteroNormalMean2.2}) into
(\ref{Eq.Example.HeteroNormalMean2.3}), we obtain, after some algebra,
\begin{align*}
l(\theta,\widehat{\psi})-l(\theta,\psi^{0})  &  =\frac{1}{2\sigma}\frac
{1}{NT^{2}}\sum_{i=1}^{N}\sum_{t=1}^{T}\sum_{s=1}^{T}\widetilde{Y}
_{it}\widetilde{Y}_{is}+\frac{1}{2\sigma}\frac{1}{N^{2}T}\sum_{i=1}^{N}
\sum_{j=1}^{N}\sum_{t=1}^{T}\widetilde{Y}_{it}\widetilde{Y}_{jt}\\
&  \quad\quad-\frac{2}{\sigma}\frac{1}{NT}\sum_{i=1}^{N}\sum_{t=1}^{T}\left(
\mu-\mu^{0}\right)  \widetilde{Y}_{it}+O_{p}(1/(NT)).
\end{align*}
Here, $\mathbb{E}\widetilde{Y}_{it}=0$ by definition and $\mathbb{E}
\widetilde{Y}_{it}\widetilde{Y}_{js}=0$, for $i\neq j$ or $t\neq s$, by
independence. Therefore, calculating the expectation, we arrive at
\[
\mathbb{E[}l(\theta,\widehat{\psi})-l(\theta,\psi^{0})]=-B(\theta
)+O(1/(NT)),\qquad B(\theta)=-\left(  \frac{1}{T}+\frac{1}{N}\right)
\frac{\sigma^{0}}{2\sigma}.
\]
The term $B(\theta)$ is the IPP bias. Apparently,
\[
\sqrt{NT}B(\theta)=-\left(  \sqrt{\frac{N}{T}}+\sqrt{\frac{T}{N}}\right)
\frac{\sigma^{0}}{2\sigma}\not \rightarrow 0
\]
as $N,T\rightarrow\infty$ and $N/T\rightarrow\kappa$. When $N/T\rightarrow0$
(or $N/T\rightarrow\infty$), $\sqrt{T/N}\rightarrow\infty$ (or $\sqrt
{N/T}\rightarrow\infty$) and $\sqrt{NT}B(\theta)$ explodes.

The bias term $B(\theta)$ causes the variance estimator to be biased. We
illustrate this consequence here. Observe that
\begin{align*}
l(\theta,\widehat{\psi})  &  =-\frac{1}{2}\log(2\pi)-\frac{1}{2}\log
\sigma-\frac{1}{NT}\sum_{i=1}^{N}\sum_{t=1}^{T}\frac{1}{2\sigma}(Y_{it}%
-\mu-\widehat{\alpha}_{i}-\widehat{\gamma}_{t})^{2},\\
\partial_{\sigma}l(\theta,\widehat{\psi})  &  =-\frac{1}{2\sigma}+\frac{1}%
{NT}\sum_{i=1}^{N}\sum_{t=1}^{T}\frac{1}{2\sigma^{2}}(Y_{it}-\mu
-\widehat{\alpha}_{i}-\widehat{\gamma}_{t})^{2}.
\end{align*}
Following the same steps in Example \ref{Example.HeteroNormalMean1}, we obtain
exactly the same estimator of $\mu_{0}$, $\widehat{\mu}$, as in Equation
(\ref{Eq.Example.HeteroNormalMean1.3}); and it is independent of $\sigma$.
Therefore, equating $\partial_{\sigma}l(\theta,\widehat{\psi})=0$,
substituting $\mu$ with $\widehat{\mu}$, and solving for $\sigma$, the
estimator of $\sigma^{0}$ is
\begin{align*}
\widehat{\sigma}  &  =\frac{1}{NT}\sum_{i=1}^{N}\sum_{t=1}^{T}(Y_{it}%
-\widehat{\mu}-\widehat{\alpha}_{i}-\widehat{\gamma}_{t})^{2}\\
&  =\frac{1}{NT}\sum_{i=1}^{N}\sum_{t=1}^{T}[Y_{it}-\mu_{0}-\alpha_{i}%
^{0}-\gamma_{t}^{0}-(\widehat{\mu}-\mu_{0})-(\widehat{\alpha}_{i}-\alpha
_{i}^{0})-(\widehat{\gamma}_{t}-\gamma_{t}^{0})]^{2},
\end{align*}
where it is not difficult to see
\[
\widehat{\mu}-\mu_{0}=\frac{1}{NT}\sum_{i=1}^{N}\sum_{t=1}^{T}\widetilde{Y}%
_{it}.
\]
Therefore, seeing $\widetilde{Y}_{it}=Y_{it}-\mu_{0}-\alpha_{i}^{0}-\gamma
_{t}^{0}$ by definition, we finally obtain
\begin{align*}
\widehat{\sigma}  &  =\frac{1}{NT}\sum_{i=1}^{N}\sum_{t=1}^{T}\widetilde{Y}%
_{it}^{2}-\frac{1}{NT^{2}}\sum_{i=1}^{N}\sum_{t=1}^{T}\sum_{s=1}%
^{T}\widetilde{Y}_{it}\widetilde{Y}_{is}-\frac{1}{N^{2}T}\sum_{i=1}^{N}%
\sum_{j=1}^{N}\sum_{t=1}^{T}\widetilde{Y}_{it}\widetilde{Y}_{jt}\\
&  \quad\quad+\frac{1}{N^{2}T^{2}}\sum_{i=1}^{N}\sum_{j=1}^{N}\sum_{t=1}%
^{T}\sum_{s=1}^{T},\widetilde{Y}_{it}\widetilde{Y}_{js}%
\end{align*}
from where we conclude that
\[
\mathbb{E}\widehat{\sigma}=\sigma_{0}-\frac{\sigma_{0}}{T}-\frac{\sigma_{0}%
}{N}+\frac{\sigma_{0}}{NT},
\]
leading to
\[
\sqrt{NT}(\mathbb{E}\widehat{\sigma}-\sigma_{0})\rightarrow-\kappa\sigma
_{0}-\frac{\sigma_{0}}{\kappa}%
\]
as $N,T\rightarrow\infty$ with $N/T\rightarrow\kappa$. While $\widehat{\mu}$
is not affected by the IPP, the variance estimator $\widehat{\sigma}$ contains
two non-negligible bias terms, $-\sigma_{0}/T$ and $-\sigma_{0}/N$. These bias
terms cause $\sqrt{NT}(\widehat{\sigma}-\sigma_{0})$ to be centered at
$-\kappa\sigma_{0}-\sigma_{0}/\kappa\neq0$.
\end{example}

\subsubsection{Bias Correction and Asymptotic Theory}

\label{Section.TechDetails.BiasTerms}

\paragraph{Motivating Assumption \ref{Assumption.Expansion}.}

First, Assumption \ref{Assumption.Expansion.AsySeq} is the same to
\cite{fw2016}. It is used to ensure that the bias $\sqrt{NT}B(\theta)$ does
not explode as $N,T\rightarrow\infty$, as discussed above. Assumption
\ref{Assumption.Expansion.Smooth} is standard in the literature. See, e.g.,
\cite{hn2004}, \cite{hk2011}, and \cite{ah2016}. It simply requires that the
likelihood function $l_{i,t}(\theta,\phi)$ is differentiable in $\phi$ at
least three times. The term $B(\theta)$ in (\ref{Eq.IPP.BiasExpansion})
depends on the first and second derivatives, and the remainder $r(\theta)$
depends on the third. We require $l_{i,t}(\theta,\phi)$ and its derivatives to
be uniformly bounded by a function $g(w_{it})$ whose moment exists to an order
greater than the second. This is used to ensure, amongst other things, i) the
interchangeability of $\mathbb{E}$ and $\partial$; and ii) the existence of
moments, such as $\mathbb{E\{}s^{\prime}(\theta)[H(\theta)]^{-1}s(\theta)\}$
in (\ref{Eq.IPP.BiasExpansion}), for every $\theta\in\Theta$. It is also used
to establish the uniform rate of convergence, over $\theta\in\Theta$, of the
corrected likelihood $L(\theta)$. The uniform convergence of $L(\theta)$
guarantees that the bias correction effect carries over to both the estimators
and likelihood-based test statistics. Assumption
\ref{Assumption.Expansion.Normalize} states that the true effects $\alpha
_{i}^{0}$ and $\gamma_{t}^{0}$ sums up to $0$ over $i=1,\ldots,N$ and
$t=1,\ldots,T$. Given the additive structure of $\theta_{0}+\alpha_{i}%
^{0}+\gamma_{t}^{0}$ in (\ref{Eq.IPP.DGP}), this condition is without loss of
generality because we merely let $\theta_{0}$ capture the average effects of
$\alpha_{i}^{0}$ and $\gamma_{t}^{0}$. Assumption
\ref{Assumption.Expansion.Ident} is an identification condition for
$\psi(\theta)$. It is standard for maximum likelihood models. See, e.g.,
\cite{nm1994}. Assumption \ref{Assumption.Expansion.Data} is common in the
field of dynamic panels. See, for instance, \cite{hk2007}, \cite{fw2016}, and
\cite{dj2015}. Notice that, the argument is conditional on the true
heterogeneous effects $\phi^{0}$. Without conditioning, $w_{it}$ may not be,
for instance, independent over $i$, because $w_{it}$ and $w_{jt}$ may depend
on the same $\gamma_{t}^{0}$. In addition, we only require the mixing
coefficient $a_{i}(m)$ to decay at a polynomial rate such that $[a_{i}%
(m)]^{1-2/\eta}$ is summable as $m\rightarrow\infty$. The mixing condition is
used to guarantee that certain autocovariances which appear in our proof are
summable by Davydov's inequality. Notice that we are not requiring $w_{it}$ to
be identically distributed over $i$ or stationary over $t$. These are
generally not possible because $w_{it}$ depends on $\alpha_{i}^{0}$ and
$\gamma_{t}^{0}$. Finally, Assumption \ref{Assumption.Expansion.Hessian}
simply requires that the normalized expected Hessian $\sqrt{NT}\mathbb{E}%
H(\theta)$ has eigenvalues $\lambda_{p}(\theta)$ that are strictly less than
$0$ for all $N>N_{0}$ and $T>T_{0}$. This ensures that $\mathbb{E\{}s^{\prime
}(\theta)[H(\theta)]^{-1}s(\theta)\}$ is well defined for large $N$ and $T$.
This is a mild condition in that we only impose it on the expected Hessian
$\mathbb{E}H(\theta)$ of the reparameterized likelihood, which does not suffer
from the indeterminacy in Equation (\ref{Eq.IPP.RotInd}). We state the
condition with \textquotedblleft less than $0$" because the Hessian in a
maximum likelihood model is usually negative definite.

The following Lemma provides low-level sufficient conditions for Assumptions
\ref{Assumption.Expansion.Smooth} and \ref{Assumption.Expansion.Hessian} under
a logit model
\begin{equation}
\label{1013-1}Y_{it}=
\mathds{1}
\{X_{it}^{\prime}(\theta+\alpha_{i}+\gamma_{t})+\varepsilon_{it}>0\},
\end{equation}
where $\varepsilon_{it}$ follows a standard logistic distribution independent
of $X_{it}$, $\alpha_{i}$, and $\gamma_{t}.$

\begin{lemma}
\label{Lemma.LowLevelCondition} Under model \eqref{1013-1}, a sufficient
condition of Assumption \ref{Assumption.Expansion.Smooth} is that $\sup
_{i,t}\mathbb{E}_{\phi}\Vert X_{it}\Vert_{\max}^{3\eta}$$<\infty$ for some
$\eta>2.$ Let $X_{i\cdot}:=\left(  X_{i1},\ldots,X_{iT}\right)  ^{\prime}$ and
$X_{\cdot t}:=\left(  X_{1t},\ldots,X_{Nt}\right)  ^{\prime}$. Assumption
\ref{Assumption.Expansion.Hessian} is satisfied if, for all $i$, the columns
of $X_{i\cdot}$ are linearly independent and, for all $t$, the columns of
$X_{\cdot t}$ are linearly independent.
\end{lemma}

%

\begin{proof}%
In Section \ref{Section.Proof}.%
\end{proof}%

\noindent The first condition requires that certain higher moment of
$\left\Vert X_{it}\right\Vert _{\max}$ are bounded away from the infinity
uniformly. Note that $X_{it}$ has a fixed dimension, so that the choice of
norm is irrelevant. This is relatively a mild condition. The second condition
states that the regressors are linearly independent in both the individual and
time dimension. This condition is standard for models with two-way
heterogeneities, and it is usually satisfied if the regressors have sufficient
variations across both the cross-sectional and time dimension. See Remark 1 of
\cite{fw2016} for a similar argument. Notably, it rules out the possibility of
$Y_{it-1}$ being constant across $i$ or $t$.

\paragraph{The bias expansion in (\ref{Eq.Asymptotics.LikeExpansion}).}

The next theorem formalizes the bias expansion in
(\ref{Eq.Asymptotics.LikeExpansion}).

\begin{theorem}
\label{Theorem.LikeExpansion}Under Assumption \ref{Assumption.Expansion} and
uniformly over $\theta\in\Theta$,
\[
\mathbb{E}l(\theta)=\mathbb{E}\widehat{l}(\theta)+B_{\alpha}(\theta
)+B_{\gamma}(\theta)+o(1/\sqrt{NT}),
\]
as $N,T\rightarrow\infty$, where
\[
B_{\alpha}(\theta)=\frac{1}{2}\operatorname*{trace}[S_{\alpha\alpha}
(\theta)H_{\alpha\alpha}^{\mathcal{\ast}}(\theta)],\qquad B_{\gamma}
(\theta)=\frac{1}{2}\operatorname*{trace}\mathbb{[}S_{\gamma\gamma}
(\theta)H_{\gamma\gamma}^{\ast}(\theta)],
\]
and $B_{\alpha}(\theta)+B_{\gamma}(\theta)=O(1/\sqrt{NT})$.
\end{theorem}

%

\begin{proof}%
In Section \ref{Section.Proof}.%
\end{proof}%

\noindent In the view of Equation (\ref{Eq.IPP.BiasExpansion}), Theorem
\ref{Assumption.Expansion} essentially states $\mathbb{E}r(\theta
)=o(1/\sqrt{NT})$ and $B(\theta)-B_{\alpha}(\theta)-B_{\gamma}(\theta
)=o(1/\sqrt{NT})$ uniformly over $\theta\in\Theta$. The terms $B_{\alpha
}(\theta)$ and $B_{\gamma}(\theta)$ are, as discussed above, non-negligible.
Theorem \ref{Theorem.LikeExpansion} relies especially on Assumption
\ref{Assumption.Expansion.Data}. We first prove that the expected quadratic
form$\ \mathbb{E\{}s^{\prime}(\theta)[H(\theta)]^{-1}s(\theta)\}$ can be
decomposed into the sum of $B_{\alpha}(\theta)+B_{\gamma}(\theta)$ and a third
term. We show that the rates of the elements in $S_{\alpha\alpha}(\theta)$ and
$S_{\gamma\gamma}(\theta)$, which are double sums, can be controlled, because,
e.g., $\left.  \partial_{\alpha_{i}}l_{i,t}(\theta,\phi)-\mathbb{E}%
\partial_{\alpha_{i}}l_{i,t}(\theta,\phi)\right\vert _{\phi=\phi\left(
\theta\right)  }$ is summable over $t$ and independent over $i$, conditional
on $\phi^{0}$. Combined with properties of $[\mathbb{E}H(\theta)]^{-1}$, the
rates of the traces above are also controlled. Similar techniques are used to
show $\mathbb{E}r(\theta)=o(1/\sqrt{NT})$.

\paragraph{The relation of $\protect\widehat{S}_{\alpha\alpha}(\theta)$ and
$\protect\widehat{S}_{\gamma\gamma}(\theta)$ to $S_{\alpha\alpha}(\theta)$ and
$S_{\gamma\gamma}(\theta)$.}

The reparameterization ensures the following identity:%
\[
S_{\alpha\alpha}(\theta)=D_{1}\mathcal{S}_{\alpha\alpha}(\theta)D_{1}^{\prime
},
\]
where $\mathcal{S}_{\alpha\alpha}(\theta)$ is the outer product of the
$NK\times1$ vector of $\partial_{\alpha}\ell(\theta,\phi)-\mathbb{E}%
\partial_{\alpha}\ell(\theta,\phi)$ evaluated at $\phi=\phi(\theta)$, where
$\partial_{\alpha}\ell(\theta,\phi)$ is the derivative of the original
likelihood $\ell(\theta,\phi)$ w.r.t. $\alpha:=(\alpha_{1}^{\prime}%
,\ldots,\alpha_{N}^{\prime})^{\prime}$. Specifically,%
\begin{align*}
\mathcal{S}_{\alpha\alpha}(\theta)  &  :=\frac{1}{N^{2}T^{2}}\sum_{t=1}%
^{T}\sum_{s=1}^{T}\mathbb{E\{}\widetilde{s}_{t}^{\alpha}(\theta)[\widetilde{s}%
_{s}^{\alpha}(\theta)]^{\prime}\},\qquad\widetilde{s}_{t}^{\alpha}%
(\theta):=\{[\widetilde{s}_{1,t}^{\alpha}(\theta)]^{\prime},\ldots
,[\widetilde{s}_{N,t}^{\alpha}(\theta)]^{\prime}\}^{\prime},\\
\widetilde{s}_{i,t}^{\alpha}(\theta)  &  :=s_{i,t}^{\alpha}(\theta,\phi
(\theta))-\mathbb{E}s_{i,t}^{\alpha}(\theta,\phi(\theta)).
\end{align*}
Under cross-sectional independence, $\mathcal{S}_{\alpha\alpha}(\theta)$
simplifies to%
\[
\mathcal{S}_{\alpha\alpha}(\theta)=\frac{1}{N^{2}T^{2}}\sum_{t=1}^{T}%
\sum_{s=1}^{T}\mathbb{E}\operatorname*{diag}\{\widetilde{s}_{1,t}^{\alpha
}(\theta)[\widetilde{s}_{1,s}^{\alpha}(\theta)]^{\prime},\ldots,\widetilde{s}%
_{N,t}^{\alpha}(\theta)[\widetilde{s}_{N,s}^{\alpha}(\theta)]^{\prime}\}.
\]
Intuitively, $\widehat{S}_{\alpha\alpha}(\theta)$ in
(\ref{Eq.Asymptotics.SaaHat}) is a natural plug-in estimator of $S_{\alpha
\alpha}(\theta)$.

Similarly for $S_{\gamma\gamma}(\theta)$, the reparameterization produces the
identity:%
\[
S_{\gamma\gamma}(\theta)=D_{2}\mathcal{S}_{\gamma\gamma}(\theta)D_{2}^{\prime
},
\]
where $\mathcal{S}_{\gamma\gamma}(\theta)$ is the outer product of the
$TK\times1$ vector of $\partial_{\gamma}\ell(\theta,\phi)-\mathbb{E}%
\partial_{\gamma}\ell(\theta,\phi)$, for $\gamma:=(\gamma_{1}^{\prime}%
,\ldots,\gamma_{T}^{\prime})^{\prime}$, evaluated at $\phi=\phi(\theta)$:%
\begin{align*}
\mathcal{S}_{\gamma\gamma}(\theta)  &  =\frac{1}{N^{2}T^{2}}\sum_{i=1}^{N}%
\sum_{j=1}^{N}\mathbb{E\{}\widetilde{s}_{i}^{\gamma}(\theta)[\widetilde{s}%
_{j}^{\gamma}(\theta)]^{\prime}\},\qquad\widetilde{s}_{i}^{\gamma,T}%
(\theta):=\{[\widetilde{s}_{i,1}^{\gamma}(\theta)]^{\prime},\ldots
,[\widetilde{s}_{i,T}^{\gamma}(\theta)]^{\prime}\}^{\prime},\\
\widetilde{s}_{i,t}^{\gamma}(\theta)  &  :=s_{i,t}^{\gamma}(\theta,\phi
(\theta))-\mathbb{E}s_{i,t}^{\gamma}(\theta,\phi(\theta)).
\end{align*}
By cross-sectional independence, $\mathcal{S}_{\gamma\gamma}(\theta)$
similarly reduces to%
\begin{equation}
\mathcal{S}_{\gamma\gamma}(\theta)=\frac{1}{N^{2}T^{2}}\sum_{i=1}%
^{N}\mathbb{E\{}s_{i}^{\gamma}(\theta)[s_{i}^{\gamma}(\theta)]^{\prime}\}.
\label{Eq.Asymptotics.Sgg}%
\end{equation}
Naturally, $\widehat{S}_{\gamma\gamma}(\theta)$ in
(\ref{Eq.Asymptotics.SggHat}) is also an intuitive plug-in estimator of
$S_{\gamma\gamma}(\theta)$.

\paragraph{Useful remarks.}

We provide some elaborated discussions about various issues in the series of
remarks below.

\begin{remark}
[truncation]\label{Remark.Truncation}In the estimator $\widehat{S}
_{\alpha\alpha}(\theta)$, we use the truncation technique commonly found in
e.g., \cite{hk2011} (see also \cite{hk2007}), \cite{fw2016}, and
\cite{gk2016}. This is related to the literature on consistent estimations of
autocovariances, $\mathbb{E\{}\widetilde{s}_{i,t}^{\alpha}(\theta
)[\widetilde{s}_{i,s}^{\alpha}(\theta)]\}$ in our case. Detailed discussions
in this regard can be found in \cite{a1991}. Without truncation, we would have
$\widehat{S}_{\alpha\alpha}(\theta)=\widehat{s}_{\alpha}(\theta)[\widehat{s}
_{\alpha}(\theta)]^{\prime}=0$ for every $\theta$ because $\widehat{s}
_{\alpha}(\theta)=0$ by definition, where $\widehat{s}_{\alpha}(\theta)$ is
defined from $\widehat{s}(\theta):=s(\theta,\widehat{\psi}(\theta))$ under the
same partition in (\ref{Eq.Asymptotics.Partitions}). Here, the truncation
parameter $\tau\rightarrow\infty$ but sufficiently slower relative to $T$ (see
Theorem \ref{Theorem.CorrectedLike}). \cite{hk2011} suggests the default value
of $\tau=1$ for small $T$. For intermediate $T$, we recommend following the
advice from \cite{fw2016} to conduct a sensitivity test for different values
of $\tau$ and to avoid using very large $\tau$ values. In our simulation, we
use $\tau=1$ and $2$ and find the difference relatively insignificant.

On the other hand, the term $\widehat{S}_{\gamma\gamma}(\theta)$ does not
require any truncation, because we have already eliminated $\mathbb{E\{}%
\widetilde{s}_{i}^{\gamma}(\theta)[\widetilde{s}_{j}^{\gamma}(\theta
)]^{\prime}\}$ for $i\neq j$ from Equation (\ref{Eq.Asymptotics.Sgg}) using
cross-sectional independence. One may still opt to truncate
$\widehat{\mathcal{S}}_{\gamma\gamma}(\theta)$, such that all elements beyond
the $\tau$-th sub- and super-diagonal are symmetrically set to $0$. We have
experimented with such truncations but found little impact on $\widehat{\theta
}_{L}$ and test statistics. The comparison is, hence, not reported in the paper.
\end{remark}

\begin{remark}
[relation to existing literature]\label{Remark.BiasCorrection}The proposed
$L(\theta)$ can be viewed as an extension of the corrected likelihood of
\cite{bh2009} and \cite{ah2016} to two-way DN-HP models. This extension is,
however, non-trivial. We provide an elaborated discussion in Section
\ref{Section.AH2016} of the supplementary appendix for this matter. When the
model only contains individual effects in the intercept, our method reduces
to, e.g., \cite{ah2016}, applied on the original likelihood $\ell\left(
\theta,\phi\right)  $, except that we use a truncation-based (instead of a
bandwidth-based) approach to guarantee the positive definiteness of
$\widehat{\mathcal{S}}_{\alpha\alpha}(\theta)$. We illustrate this in Example
\ref{Example.DLogitFE} of Section \ref{Section.AH2016}, using a dynamic logit model.

Note that, for the logit model, there are fixed-$T$ consistent estimators
including \cite{c1980}. See also Remark \ref{Remark.LogitModel} for this
specific matter. As opposite to fixed-$T$ methods, our approach is a
large-$T$, \textquotedblleft first-order" bias correction procedure. Similar
to \cite{ah2016}, \cite{hn2004}, \cite{bh2009}, \cite{ah2016}, \cite{fw2016},
our approach only removes the bias up to $O(1/\sqrt{NT})$, making $L(\theta)$
(hence the estimators and test statistics) asymptotically unbiased under
$N/T\rightarrow\kappa$ as $N,T\rightarrow\infty$. For fixed $T$, our approach
generally does not eliminate the IPP. However, one advantage of our method is
that it mitigates the IPP for a wide range of models, especially for those
without fixed-$T$ consistent estimation techniques. For instance, there has
been no fixed-$T$ consistent procedure for the two-way heterogeneous parameter
logit model, to the best of our knowledge. Indeed, since our approach is only
first-order, caution shall be observed when the sample size is small, in which
case higher-order bias terms could be non-negligible, requiring higher-order
corrections. For models with individual effects, available methods include
\cite{dj2015}, \cite{ds2021}, and \cite{s2022}. However, we are not aware of
any higher-order methods for two-way DN-HP models.
\end{remark}

\paragraph{More asymptotic results about the corrected estimator and test
statistics.}

To establish the asymptotic properties of $\widehat{\theta}_{L}$ and
$L(\theta)$-based LR, LM, Wald test statistics, we enforce the following assumption.%

\begin{asu}
\
\begin{itemize}[leftmargin=*,itemindent=-2em]%
\label{Assumption.Statistic}%

\item[]\refstepcounter{subassumption}\thesubassumption\
\label{Assumption.Statistic.Smooth}The likelihood $l_{i,t}(\theta,\phi)$ is
four-time continuously differentiable w.r.t. to both $\theta\in\Theta$ and
$\phi\in\Phi$ almost surely, such that
\[
\sup_{\theta\in\Theta}\sup_{\phi\in\Phi}\left\vert \partial_{\nu_{1}\cdots
\nu_{S}}l_{i,t}(\theta,\phi)\right\vert <g(w_{it}),
\]
where $g(w_{it})$ is defined in Assumption \ref{Assumption.Expansion},
$\nu_{s}$ represents any element of $(\theta^{\prime},\phi^{\prime})^{\prime}%
$, and $S\in\{1,2,3,4\}$.%

\item[]\refstepcounter{subassumption}\thesubassumption\
\label{Assumption.Statistic.Ident}For every $\theta\in\Theta$ and $\theta
\neq\theta_{0}$, $\mathbb{P}[l(\theta)\neq l(\theta_{0})]>0$.%

\item[]\refstepcounter{subassumption}\thesubassumption\
\label{Assumption.Statistic.Dominance1}For large $N$ and $T$,
\[
\mathbb{E}\sup_{\theta\in\Theta}\left\vert l(\theta)\right\vert <\infty.
\]
%

\item[]\refstepcounter{subassumption}\thesubassumption\
\label{Assumption.Statistic.Dominance2}The sequence of functions $\{\left\Vert
\nabla_{\theta}l(\theta)\right\Vert :\theta\in\Theta\}$ is uniformly
integrable, and
\[
\mathbb{E}\sup_{\theta\in\Theta}\left\Vert \nabla_{\theta\theta^{\prime}%
}l(\theta)\right\Vert <\infty
\]
where $\left\Vert \cdot\right\Vert $ is the matrix Frobenius norm.%

\item[]\refstepcounter{subassumption}\thesubassumption\
\label{Assumption.Statistic.Hessian}The profiled Hessian $\mathbb{E}%
\nabla_{\theta\theta^{\prime}}l(\theta)$ is strictly negative definite for
large $N$ and $T$.%

\end{itemize}
\end{asu}%

\noindent Here Assumption \ref{Assumption.Statistic.Smooth} is similar to
Assumption \ref{Assumption.Expansion.Smooth}, except that we impose that
$l_{i,t}(\theta,\phi)$ is four-time differentiable, and require that the
derivatives of $l_{i,t}(\theta,\phi)$ w.r.t. to both $\theta$ and $\phi$ to be
uniformly dominated by $g(w_{it})$. This assumption ensures that the corrected
likelihood $L(\theta)$, which depends on second derivatives of $l_{i,t}%
(\theta,\phi)$, is twice differentiable. This is required by standard
techniques used to show, e.g., the asymptotic normality of $\widehat{\theta
}_{L}$. See., e.g., \cite{hf2000}. Similar to Assumption
\ref{Assumption.Expansion.Ident}, Assumption \ref{Assumption.Statistic.Ident}
is an identification condition for $\theta_{0}$. It guarantees that
$\operatorname*{plim}_{N,T\rightarrow\infty}l(\theta)$ is uniquely maximized
at $\theta_{0}$. Assumption \ref{Assumption.Statistic.Dominance1} is a
dominance condition implying that $l(\theta)-\mathbb{E}l(\theta)$ converges to
$0$ uniformly as $N,T\rightarrow\infty$. Together, Assumptions
\ref{Assumption.Statistic.Smooth} to \ref{Assumption.Statistic.Dominance1} are
used to establish the consistency of $\widehat{\theta}_{L}$. In addition to
these, Assumption \ref{Assumption.Statistic.Dominance2} establishes a
dominance condition for the profiled Hessian, so that $\triangledown
_{\theta\theta^{\prime}}l(\overline{\theta})\overset{\mathbb{P}%
}{\longrightarrow}\mathbb{E}[\triangledown_{\theta\theta^{\prime}}l(\theta
_{0})]$ for any consistent estimator $\overline{\theta}$ of $\theta_{0}$.
Assumption \ref{Assumption.Statistic.Hessian} requires that the expected
profiled Hessian is strictly negative definite for sufficiently large $N$ and
$T$. Assumptions \ref{Assumption.Statistic.Dominance2} and
\ref{Assumption.Statistic.Hessian} are additionally imposed to establish the
asymptotic distributions for the quantities of interests.

The corollary below states the asymptotic properties of $\widehat{\theta}_{L}$.

\begin{corollary}
\label{Theorem.Theta}Under Assumptions \ref{Assumption.Expansion} and
\ref{Assumption.Statistic}, $\widehat{\theta}_{L}\overset{\mathbb{P}
}{\rightarrow}\theta_{0}$ and
\[
\sqrt{NT}[-\mathbb{E}\triangledown_{\theta\theta^{\prime}}l(\theta_{0}
)]^{1/2}(\widehat{\theta}_{L}-\theta_{0})\overset{\mathbb{D}}{\longrightarrow
}\mathcal{N}(0,\mathbb{I}_{K})
\]
as $N,T\rightarrow\infty$.
\end{corollary}

%

\begin{proof}%
In Section \ref{Section.Proof}.%
\end{proof}%

\noindent Here, $[-\mathbb{E}\triangledown_{\theta\theta^{\prime}}l(\theta
_{0})]^{1/2}$ is the square root of $-\mathbb{E}\triangledown_{\theta
\theta^{\prime}}l(\theta_{0})$ which can be obtained by square-rooting the
eigenvalues in the eigendecomposition of $-\mathbb{E}\triangledown
_{\theta\theta^{\prime}}l(\theta_{0})$. By Assumption
\ref{Assumption.Statistic.Hessian}, $\mathbb{E}\triangledown_{\theta
\theta^{\prime}}l(\theta_{0})$ is strictly negative definite so that
$-\mathbb{E}\triangledown_{\theta\theta^{\prime}}l(\theta_{0})$ only has
strictly positive eigenvalues. In addition, since we focus on maximum
likelihood models, we use the information matrix equality to simplify the
presentation of Corollary \ref{Theorem.Theta}. In case the information matrix
equality does not hold\footnote{The information matrix equality generally does
not hold for, e.g., non-likelihood M-estimation models, misspecified models,
and quasi-likelihood models.}, $-\mathbb{E}\triangledown_{\theta\theta
^{\prime}}l(\theta_{0})$ shall be replaced by its sandwich form and
$[-\mathbb{E}\triangledown_{\theta\theta^{\prime}}l(\theta_{0})]^{1/2}$ shall
be constructed thereof. See, e.g., \cite{w1982} for detailed discussions.

We additionally provide some computational insights in the remark below.

\begin{remark}
[two-layer maximization]\label{Remark.Computation2}The computation of
$\widehat{\theta}_{L}$ involves a two-layer maximization: In the inner layer,
$\widehat{\psi}(\theta)$ is obtained by maximizing $l(\theta,\psi)$, keeping
$\theta$ fixed; while in the outer layer, $L(\theta)$ is maximized w.r.t.
$\theta$. This maximization problem can be done efficiently using the
Newton-Raphson algorithm in both layers. In the inner layer, we recommend
following Remark \ref{Remark.Computation1} to calculate the first and second
derivatives efficiently. In the outer layer, it is generally affordable to use
numerical derivatives, because the dimension of $\theta$ is usually small,
provided that the inner-layer maximization is done efficiently. As a
reference, with an AMD Ryzen 9 5950X CPU, the calculation of $\widehat{\theta
}_{L}$ for the dynamic logit model considered in our simulation (see Section
\ref{Section.Simulation} below) with $\left(  N,T\right)  =\left(
90,90\right)  $ (note, $NT=8100$) takes less than $50$ seconds.
\end{remark}

To continue, the next corollary states the asymptotic distribution of the LR
($\widehat{\xi}_{LR}$), the LM ($\widehat{\xi}_{LM}$), and the Wald
($\widehat{\xi}_{Wald}$) statistics.

\begin{corollary}
\label{Theorem.Inference}Under Assumptions \ref{Assumption.Expansion} to
\ref{Assumption.Statistic} and $H_{0}$,
\[
\widehat{\xi}_{LR},\widehat{\xi}_{LM},\widehat{\xi}_{Wald}\overset{\mathbb{D}
}{\longrightarrow}\chi^{2}(r)
\]
as $N,T\rightarrow\infty$.
\end{corollary}

%

\begin{proof}%
In Section \ref{Section.Proof}.%
\end{proof}%

\noindent Similarly, the test statistics $\widehat{\xi}_{LM}$ and
$\widehat{\xi}_{Wald}$ are constructed under the information matrix equality.
Adjustments are generally required if it does not hold. For instance,
\cite{w1982} shows that $-[\triangledown_{\theta\theta^{\prime}}%
L(\widehat{\theta}_{L})]^{-1}$ involved in $\widehat{\xi}_{Wald}$ shall be
replaced by its sandwich form for quasi-likelihood models.

Given Theorem \ref{Theorem.CorrectedLike}, Corollaries \ref{Theorem.Theta} and
\ref{Theorem.Inference} are established using standard techniques.
Particularly, Theorem \ref{Theorem.CorrectedLike} guarantees $L(\theta
)\rightarrow_{p}\mathbb{E}l(\theta)$, so that the consistency of
$\widehat{\theta}_{L}$ follows from \cite{nm1994}. Theorem
\ref{Theorem.CorrectedLike} also implies that $\nabla_{\theta}L(\theta
)-\nabla_{\theta}l(\theta)$ is $o_{\mathbb{P}}(1/\sqrt{NT})$, from which the
asymptotic normality is established using techniques in, e.g., \cite{hf2000}.
The proof of Corollary \ref{Theorem.Inference} also closely resembles the
standard techniques in, e.g., \cite{hf2000}, where it is shown that all
$\widehat{\xi}_{LR}$, $\widehat{\xi}_{LM}$, and $\widehat{\xi}_{Wald}$ are
equivalent, converging in probability to some $\widehat{\xi}$ which is
asymptotically $\chi^{2}(r)$.

In the remark below, we provide some discussion about the uniform inference issue.

\begin{remark}
[uniform inference]\label{Remark.UInference}The inference based on our
corrected likelihood is uniformly valid regardless of whether the underlying
data-generating process has a homogeneous or heterogeneous slope coefficient.
This is similar to \cite{ls2023}, except that they consider a linear
regression. For instance, consider the data-generating process%
\[
Y_{it}=\mathds{1}\{\left(  \beta_{0}+\alpha_{1,i}^{0}+\gamma_{1,t}^{0}\right)
X_{it}+\alpha_{2,i}^{0}+\gamma_{2,t}^{0}+\varepsilon_{it}>0\},
\]
where the slope coefficient of $X_{it}$ can be homogeneous, if $\alpha
_{1,i}^{0}=\gamma_{1,t}^{0}=0$ for all $i$ and $t$, or heterogeneous, if
otherwise. In either case, inference based on the corrected likelihood under
the model%
\[
Y_{it}=\mathds{1}\{(\beta+\alpha_{1,i}+\gamma_{1,t})X_{it}+\alpha_{2,i}%
+\gamma_{2,t}+\varepsilon_{it}>0\},\ \alpha_{1,i}\neq0\ \text{and }%
\gamma_{1,t}\neq0,
\]
remains valid. This uniform validity shares a similar spirit as the uniform
inference in \cite{ls2023}. In fact, the homogeneous data-generating process
is just a special case of the heterogeneous one. By employing the
bias-corrected likelihood for the heterogeneous model, inference on $\beta$
remains valid, albeit with a possible loss of efficiency when the true
data-generating process is homogeneous.

In addition, we treat the heterogeneous variables $\alpha_{k,i}^{0}$ and
$\gamma_{k,t}^{0}$ as parameters by conditioning on them, which is common for
panel maximum likelihood methods, we do not further distinguish between
locally heterogeneous and fully heterogeneous cases as in \cite{ls2023}.
\end{remark}

\subsection{Relation to Existing Approaches}

\label{Section.AH2016}

As discussed above, our method is a non-trivial extension of \cite{bh2009} and
\cite{ah2016} to two-way DN-HP models. In this section, we explain the
technical challenges in developing a likelihood-based bias correction approach
for the DN-HP models and provide Example \ref{Example.DLogitFE} to illustrate
how our method relates to the aforecited. We also leave Remark
\ref{Remark.LogitModel} to briefly discuss the difference of our procedure to
some existing methods providing fixed-$T$ consistency for the logit model

The primary technical challenge lies in the high dimensionality of the two-way
DN-HP models, due to incidental parameter vector $\phi$ (or $\psi$) having a
dimension increasing with $N$ and $T$. In what follows, we briefly re-visit
the procedure of \cite{ah2016}, pointing out the key difference, and explain
the challenges. For simplicity, we assume that the fixed effects are only
scalars\footnote{We will briefly discuss multi-dimensional fixed effects
later.}. Consider an arbitrary panel data model with individual effects only,
having likelihood function%
\[
\ell\left(  \theta,\alpha\right)  :=\frac{1}{N}\sum_{i=1}^{N}l_{i}\left(
\theta,\alpha_{i}\right)  ,\qquad l_{i}\left(  \theta,\alpha_{i}\right)
:=\frac{1}{T}\sum_{i=1}^{T}\log f\left(  Y_{it}|X_{it};\theta,\alpha
_{i}\right)  ,
\]
where $f\left(  Y_{it}|X_{it};\theta,\alpha_{i}\right)  $ is a known density
and $\alpha:=\left(  \alpha_{1},\ldots,\alpha_{N}\right)  ^{\prime}$. It is
important to note that $l_{i}\left(  \theta,\alpha_{i}\right)  $ are
independent across $i$, in the sense that $l_{i}\left(  \theta,\alpha
_{i}\right)  $ and $l_{j}\left(  \theta,\alpha_{j}\right)  $ share no common
arguments except $\theta$. Because of such independence, an expansion similar
to (\ref{Eq.IPP.LikeExpansion}) can be established individual by individual,
as%
\begin{align}
&  \quad\quad l_{i}\left(  \theta,\widehat{\alpha}_{i}\left(  \theta\right)
\right) \nonumber\\
&  =l_{i}\left(  \theta,\alpha_{i}\left(  \theta\right)  \right)
+\partial_{\alpha_{i}}l_{i}\left(  \theta,\alpha_{i}\left(  \theta\right)
\right)  \left(  \widehat{\alpha}_{i}\left(  \theta\right)  -\alpha_{i}\left(
\theta\right)  \right)  +\frac{1}{2}\partial_{\alpha_{i}^{2}}l_{i}\left(
\theta,\alpha_{i}\left(  \theta\right)  \right)  \left(  \widehat{\alpha}%
_{i}\left(  \theta\right)  -\alpha_{i}\left(  \theta\right)  \right)
^{2}+r_{i}^{l}\left(  \theta\right)  , \label{Eq.AH2016.LikeExpansion}%
\end{align}
where, for some $c_{i}\in\left(  0,1\right)  $,%
\begin{align}
r_{i}^{l}\left(  \theta\right)   &  :=\frac{1}{6}\partial_{\alpha_{i}^{3}%
}l_{i}\left(  \theta,c_{i}\alpha_{i}\left(  \theta\right)  +\left(
1-c_{i}\right)  \alpha_{i}\left(  \theta\right)  \right)  \left(
\widehat{\alpha}_{i}\left(  \theta\right)  -\alpha_{i}\left(  \theta\right)
\right)  ^{3},\nonumber\\
\widehat{\alpha}_{i}\left(  \theta\right)   &  :=\arg\max_{\alpha_{i}}%
l_{i}\left(  \theta,\alpha_{i}\right)  ,\qquad\alpha_{i}\left(  \theta\right)
:=\arg\max_{\alpha_{i}}\mathbb{E}l_{i}\left(  \theta,\alpha_{i}\right)  ,
\label{Eq.AH2016.AlphaHat}%
\end{align}
$\partial_{\alpha_{i}}l_{i}\left(  \theta,\alpha_{i}\left(  \theta\right)
\right)  $ denotes $\left[  \partial_{\alpha_{i}}l_{i}\left(  \theta
,\alpha_{i}\right)  \right]  |_{\alpha_{i}=\alpha_{i}\left(  \theta\right)  }%
$, and $\partial_{\alpha_{i}^{2}}l_{i}\left(  \cdot,\cdot\right)  $ and
$\partial_{\alpha_{i}^{3}}l_{i}\left(  \cdot,\cdot\right)  $ are defined
analogously. Here, notice importantly that $\widehat{\alpha}_{i}\left(
\theta\right)  $ and $\alpha_{i}\left(  \theta\right)  $ are obtained by
solving, respectively, $\partial_{\alpha_{i}}l_{i}\left(  \theta,\alpha
_{i}\right)  =0$ and $\mathbb{E}\partial_{\alpha_{i}}l_{i}\left(
\theta,\alpha_{i}\right)  =0$ individually. This is possible due to the
independence of $l_{i}\left(  \theta,\alpha_{i}\right)  $ across $i$.
Subsequently, a similar expansion to (\ref{Eq.IPP.PsiExpansion}) can be
constructed, again individually, as%
\begin{align}
0  &  =\partial_{\alpha_{i}}l_{i}\left(  \theta,\alpha_{i}\left(
\theta\right)  \right)  +\partial_{\alpha_{i}^{2}}l_{i}\left(  \theta
,\alpha_{i}\left(  \theta\right)  \right)  \left(  \widehat{\alpha}_{i}\left(
\theta\right)  -\alpha_{i}\left(  \theta\right)  \right)  +r_{i}^{s}\left(
\theta\right) \label{Eq.AH2016.PsiExpansion}\\
\widehat{\alpha}_{i}\left(  \theta\right)  -\alpha_{i}\left(  \theta\right)
&  =-\frac{1}{\partial_{\alpha_{i}^{2}}l_{i}\left(  \theta,\alpha_{i}\left(
\theta\right)  \right)  }\left(  \partial_{\alpha_{i}}l_{i}\left(
\theta,\alpha_{i}\left(  \theta\right)  \right)  +r_{i}^{s}\left(
\theta\right)  \right)  ,\nonumber
\end{align}
where, for some $d_{i}\in\left(  0,1\right)  $,%
\[
r_{i}^{s}\left(  \theta\right)  =\frac{1}{2}\partial_{\alpha_{i}^{3}}%
l_{i}\left(  \theta,d_{i}\alpha_{i}\left(  \theta\right)  +\left(
1-d_{i}\right)  \alpha_{i}\left(  \theta\right)  \right)  \left(
\widehat{\alpha}_{i}\left(  \theta\right)  -\alpha_{i}\left(  \theta\right)
\right)  ^{2}.
\]
Similarly combining (\ref{Eq.AH2016.PsiExpansion}) and
(\ref{Eq.AH2016.LikeExpansion}), we have%
\[
l_{i}\left(  \theta,\widehat{\alpha}_{i}\left(  \theta\right)  \right)
=l_{i}\left(  \theta,\alpha_{i}\left(  \theta\right)  \right)  -\frac{1}%
{2}\frac{\left[  \partial_{\alpha_{i}}l_{i}\left(  \theta,\alpha_{i}\left(
\theta\right)  \right)  \right]  ^{2}}{\partial_{\alpha_{i}^{2}}l_{i}\left(
\theta,\alpha_{i}\left(  \theta\right)  \right)  }-r_{i}\left(  \theta\right)
\]
for a remainder term $r_{i}\left(  \theta\right)  $ depending on $r_{i}%
^{s}\left(  \theta\right)  $ and $r_{i}^{l}\left(  \theta\right)  $, amongst
other things. The subsequent steps are omitted because they are similar to
their DN-HP counterparts.

The steps above are all followed only individually. This is possible due to
the independence of $l_{i}\left(  \theta,\alpha_{i}\right)  $; and the benefit
is that $r_{i}\left(  \theta\right)  $, $\partial_{\alpha_{i}^{2}}l_{i}\left(
\theta,\alpha_{i}\left(  \theta\right)  \right)  $, etc. are only scalars. For
two-way DN-HP models,\ the likelihood $l_{i,t}(\theta,\phi)=\log
f(Y_{it}|X_{it},\theta+\alpha_{i}+\gamma_{t})$ (defined in
\ref{Eq.IPP.OriginalLike}) depends on both $\alpha_{i}$ and $\gamma_{t}$. This
is unlike models with individual effects only, and it implies that
$\widehat{\psi}\left(  \theta\right)  $ must be obtained by solving the system
of equations $\partial_{\psi}l\left(  \theta,\psi\right)  =0$ jointly, unlike
$\widehat{\alpha}_{i}\left(  \theta\right)  $. It further implies that the
expansions in (\ref{Eq.IPP.LikeExpansion}) and (\ref{Eq.IPP.PsiExpansion})
must be carried out jointly w.r.t. the entire vector $\psi$. In other words,
these expansions are high-dimenional for DN-HP models. Consequently, it is
much more involved to develop a likelihood-based bias correction for two-way
DN-HP models, in the follows notable ways, all due to the high-dimensionality.

\begin{enumerate}
\item For models with individuals effects only, $r_{i}^{s}\left(
\theta\right)  $ is only a scalar and it is relatively straightforward to
establish its rate of convergence. On the other hand, its DN-HP counterpart,
given in\ (\ref{Eq.Appendix.RemainderRS}), is a high-dimensional vector, with
each component itself being a quadratic form of $\widehat{\psi}\left(
\theta\right)  -\psi(\theta)$ and $G_{\psi_{j}}(\theta,a_{j})$ that are
already high-dimensional objects. It is technically more challenging to
control the rate of convergence of (\ref{Eq.Appendix.RemainderRS}). For
instance, one obvious issue is that the choice of norm becomes relevant for
(\ref{Eq.Appendix.RemainderRS}) but not $r_{i}^{s}\left(  \theta\right)  $.
The same arguments apply to other remainder terms.

\item When the model only contains (scalar) individual effects, the incidental
parameter Hessian $\partial_{\alpha_{i}^{2}}l_{i}\left(  \theta,\alpha
_{i}\left(  \theta\right)  \right)  $ is only a scalar and the rate of
convergence of its inverse, $1/\partial_{\alpha_{i}^{2}}l_{i}\left(
\theta,\alpha_{i}\left(  \theta\right)  \right)  $, is relatively
straightforward. For the DN-HP models, however, the incidental parameter
Hessian $H\left(  \theta\right)  $ is a high-dimensional matrix. It is more
complex to control the rate of its inverse, $\left[  H\left(  \theta\right)
\right]  ^{-1}$.

\item When only individual effects are present, there is only one scalar bias
term. It is relatively straightforward to derive its rate of convergence and
propose an estimator for it. On the contrary, for the DN-HP models, the bias
correction term $B(\theta)$, defined in (\ref{Eq.IPP.BiasExpansion}), is a
quadratic form of $s\left(  \theta\right)  $ and $H\left(  \theta\right)  $
which are high-dimensional objects. In addition to this, we show that
$B(\theta)\ $can be approximated by $B_{\alpha}(\theta)+B_{\gamma}(\theta)$,
which is not needed for models with individual effects only.
\end{enumerate}

\noindent Other challenges include, e.g., the lack of stationarity across $t$
(due to the existence of $\gamma_{t}^{0}$), the need for the normalization and
reparameterization, etc. Note that, for models with individual effects only,
even if $\alpha_{i}$ is multi-dimensional, the steps above can still be
carried out individually, implying that the above-discussed
high-dimensionality problems still would not happen, as long as the dimension
of $\alpha_{i}$ is fixed.

Next, in Example \ref{Example.DLogitFE} below, we illustrate how our approach
reduces to \cite{ah2016} and apply it to a dynamic logit model with only
scalar individual effects in the intercept.

\begin{example}
[dynamic logit with individual effects only]\label{Example.DLogitFE}Consider a
model with only scalar individual effects in the intercept, whose likelihood
for individual $i$ at time $t$ is $l_{i,t}\left(  \theta,\alpha_{i}\right)  $,
and assume that the data is independent over $i$ while stationary across $t$.
Since, reparameterization is not needed, we work with the original likelihood
$\ell\left(  \theta,\phi\right)  \equiv N^{-1}T^{-1}\sum_{i=1}^{N}\sum
_{t=1}^{T}l_{i,t}\left(  \theta,\alpha_{i}\right)  $ for $\phi\equiv\left(
\alpha_{1},\ldots,\alpha_{N}\right)  ^{\prime}$. Given that $l_{i,t}\left(
\theta,\alpha_{i}\right)  $ are independent across $i$, $\widehat{\phi}%
(\theta)$ and $\phi(\theta)$ can be constructed individually, as%
\[
\widehat{\phi}(\theta)\equiv\left(  \widehat{\alpha}_{1}\left(  \theta\right)
,\ldots,\widehat{\alpha}_{N}\left(  \theta\right)  \right)  ^{\prime}%
,\qquad\phi(\theta)\equiv\left(  \alpha_{1}\left(  \theta\right)
,\ldots,\alpha_{N}\left(  \theta\right)  \right)  ^{\prime},
\]
where, $\widehat{\alpha}_{i}\left(  \theta\right)  $ and $\alpha_{i}\left(
\theta\right)  $ are defined in (\ref{Eq.AH2016.AlphaHat}). Next, notice that
$\mathbb{E}s_{i,t}^{\alpha}(\theta,\alpha_{i}(\theta))=\mathbb{E}T^{-1}%
\sum_{t=1}^{T}s_{i,t}^{\alpha}(\theta,\alpha_{i}(\theta))=0$, where the first
equality is because of stationarity and the second because $\alpha_{i}%
(\theta)$ maximizes $\mathbb{E}l_{i}\left(  \theta,\alpha_{i}\right)  $. This
implies that $S_{\alpha\alpha}(\theta)$ is a diagonal matrix defined as
\[
S_{\alpha\alpha}(\theta)=\frac{1}{N^{2}T}\operatorname*{diag}\left\{
S_{i}^{\alpha\alpha}\left(  \theta\right)  \right\}  ,\qquad S_{i}%
^{\alpha\alpha}\left(  \theta\right)  :=\frac{1}{T}\sum_{t=1}^{T}\sum
_{s=1}^{T}\mathbb{E}\left\{  s_{i,t}^{\alpha}(\theta,\alpha_{i}(\theta
))s_{i,s}^{\alpha}(\theta,\alpha_{i}(\theta))\right\}  .
\]
Similarly, the expected Hessian is also diagonal, as
\[
\mathbb{E}H(\theta)=\frac{1}{N}\operatorname*{diag}\left\{  \mathbb{E}%
h_{i}^{\alpha\alpha}(\theta)\right\}  ,\qquad h_{i}^{\alpha\alpha}%
(\theta):=\frac{1}{T}\sum_{t=1}^{T}\partial_{\alpha_{i}^{2}}l_{i,t}\left(
\theta,\alpha_{i}\left(  \theta\right)  \right)  .
\]
By stationarity, the inverse of $\mathbb{E}H(\theta)$ is simply%
\[
H_{\alpha\alpha}^{\mathcal{\ast}}(\theta)\equiv\lbrack\mathbb{E}%
H(\theta)]^{-1}=N\operatorname*{diag}\left\{  [ \mathbb{E}\partial_{\alpha
_{i}^{2}}l_{i,t}\left(  \theta,\alpha_{i}\left(  \theta\right)  \right)
]^{-1}\right\}  .
\]
Finally, seeing $B_{\gamma}(\theta)\equiv0$ without time effects and using the
properties of $\operatorname*{trace}$, we have
\[
B(\theta)\equiv B_{\alpha}(\theta)=\frac{1}{T}\left(  \frac{1}{N}\sum
_{i=1}^{N}B_{i}^{\alpha}(\theta)\right)  ,\qquad B_{i}^{\alpha}(\theta
):=\frac{1}{2}\frac{T^{-1}\sum_{t=1}^{T}\sum_{s=1}^{T}\mathbb{E}\left\{
s_{i,t}^{\alpha}(\theta,\alpha_{i}(\theta))s_{i,s}^{\alpha}(\theta,\alpha
_{i}(\theta))\right\}  }{\mathbb{E}\partial_{\alpha_{i}^{2}}l_{i,t}\left(
\theta,\alpha_{i}\left(  \theta\right)  \right)  }.
\]
This is identical to the bias term of \cite{ah2016}. The term $B_{i}^{\alpha
}(\theta)$ is estimated by
\begin{equation}
\widehat{B}_{i}^{\alpha}(\theta):=\frac{1}{2}\frac{\sum_{t=1}^{T}\sum
_{s=1}^{T}\left(
\mathds{1}%
\{\left\vert t-s\right\vert \leq\tau\} s_{i,t}^{\alpha}(\theta,\widehat{\alpha
}_{i}(\theta))s_{i,s}^{\alpha}(\theta,\widehat{\alpha}_{i}(\theta))\right)
}{\sum_{t=1}^{T}\partial_{\alpha_{i}^{2}}l_{i,t}\left(  \theta,\alpha
_{i}\left(  \theta\right)  \right)  }, \label{Eq.Example.DLogitFE.0}%
\end{equation}
which is the same to the trace-based approach of \cite{ah2016}, except that we
use a truncation here.

Next, consider a dynamic logit model $Y_{it}=%
\mathds{1}%
\left\{  \theta_{0}Y_{it-1}+\alpha_{i}^{0}+\varepsilon_{it}>0\right\}  $,
where $Y_{it}$ is independent over $i$ and stationary across $t$, and
$\varepsilon_{it}$ is standard logistic. The likelihood function for
individual $i$ at time $t$ is
\[
l_{i,t}\left(  \theta,\alpha_{i}\right)  =\left(  Y_{i,t}-1\right)  \pi
_{i,t}\left(  \theta,\alpha_{i}\right)  -\log\left(  1+\exp\left(  -\pi
_{i,t}\left(  \theta,\alpha_{i}\right)  \right)  \right)  ,\qquad\pi
_{i,t}\left(  \theta,\alpha_{i}\right)  :=\theta Y_{it-1}+\alpha_{i};
\]
while the first and second derivatives of $l_{i,t}\left(  \theta,\alpha
_{i}\right)  $ w.r.t. $\alpha_{i}$ are, respectively,
\begin{equation}
s_{i,t}^{\alpha}\left(  \theta,\alpha_{i}\right)  =Y_{it}-\frac{1}{\exp\left(
-\pi_{i,t}\left(  \theta,\alpha_{i}\right)  \right)  +1},\qquad\partial
_{\alpha_{i}^{2}}l_{i,t}\left(  \theta,\alpha_{i}\left(  \theta\right)
\right)  =\frac{-\exp\left(  -\pi_{i,t}\left(  \theta,\alpha_{i}\right)
\right)  }{\left[  \exp\left(  -\pi_{i,t}\left(  \theta,\alpha_{i}\right)
\right)  +1\right]  ^{2}}. \label{Eq.Example.DLogitFE.1}%
\end{equation}
Substituting (\ref{Eq.Example.DLogitFE.1}) into (\ref{Eq.Example.DLogitFE.0})
and letting $\widehat{l}(\theta)\equiv\ell(\theta,\widehat{\phi}(\theta))$, we
have
\[
L(\theta)\equiv\widehat{l}(\theta)+\frac{1}{T}\left(  \frac{1}{N}\sum
_{i=1}^{N}\widehat{B}_{i}^{\alpha}(\theta)\right)  ,
\]
where, denoting $\widehat{\pi}_{i,t}\left(  \theta\right)  :=\pi_{i,t}\left(
\theta,\widehat{\alpha}_{i}\left(  \theta\right)  \right)  $,
\[
\widehat{B}_{i}^{\alpha}(\theta) :=\frac{1}{2}\left(  \sum_{t=1}^{T}%
\frac{-\exp\left(  -\widehat{\pi}_{i,t}\left(  \theta\right)  \right)
}{\left[  \exp\left(  -\widehat{\pi}_{i,t}\left(  \theta\right)  \right)
+1\right]  ^{2}}\right)  ^{-1}\left(  \sum_{t=2}^{T}\sum_{s=2}^{T}%
\mathds{1}%
\{\left\vert t-s\right\vert \leq\tau\}[Y_{it}Y_{is}+\widehat{b}_{i,t,s}%
^{\alpha,1}(\theta)+\widehat{b}_{i,t,s}^{\alpha,2}(\theta)] \right)  ,
\]
\[
\widehat{b}_{i,t,s}^{\alpha,1}(\theta) :=-2Y_{it}\left[  \exp\left(
-\widehat{\pi}_{i,s}\left(  \theta\right)  \right)  +1\right]  ^{-1}%
,\qquad\widehat{b}_{i,t,s}^{\alpha,2}(\theta):=\left[  \exp\left(
-\widehat{\pi}_{i,t}\left(  \theta\right)  \right)  +1\right]  ^{-1}\left[
\exp\left(  -\widehat{\pi}_{i,s}\left(  \theta\right)  \right)  +1\right]
^{-1}.
\]

\end{example}

We use the following remark to briefly discuss the difference between our bias
correction method and several existing approaches that remove the IPP for
logit models.

\begin{remark}
[panel logit models]\label{Remark.LogitModel}For static logit models with
individual effects in the intercept, \cite{c1980} develops a conditional
maximum likelihood, on some sufficient statistics, that produces a fixed-$T$
consistent estimator for $\theta_{0}$. For dynamic models, \cite{c1985} shows
that such sufficient statistics can be found when the only regressor is the
lagged dependent variable. \cite{hk2000} and \cite{hw2024} provide methods for
dynamic logit models with both lagged and exogenous regressors. When an
additional time effect is included in the model, \cite{c2017} and
\cite{dhw2025} provide consistent estimation procedures. To our knowledge,
there has been no fixed-$T$ consistent estimator available for logit models
two-way heterogeneities. As discussed in Remark \ref{Remark.BiasCorrection},
our approach (and that of \cite{bh2009} and \cite{ah2016}) does not generally
eliminate the IPP bias. This is true even for the logit model: the simulation
results from \cite{ds2021} show that a \textquotedblleft second-order"
corrected logit estimator is less biased than the first-order corrected. This
implies that the first-order corrected estimator is likely not fixed-$T$ consistent.
\end{remark}

\subsection{Proofs}

\label{Section.Proof}

In this section, we provide all proofs. We first give the proof of Lemma
\ref{Lemma.LowLevelCondition}, which provide low-level conditions for
Assumptions \ref{Assumption.Expansion.Smooth} and
\ref{Assumption.Expansion.Hessian}.%

\begin{proof}[Proof of Lemma \ref{Lemma.LowLevelCondition}]%
Under the logit model \eqref{1013-1} the log-likelihood for individual $i$ at
time $t$ is
\[
l_{i,t}\left(  \theta,\alpha_{i},\gamma_{t}\right)  =\left(  Y_{it}-1\right)
X_{it}^{\prime}(\theta+\alpha_{i}+\gamma_{t})-\log\big(1+\exp\left(
-X_{it}^{\prime}(\theta+\alpha_{i}+\gamma_{t})\right)  \big).
\]
We assume that $X_{it}$ is scalar regressor for simplicity. The derivatives of
$l_{i,t}\left(  \theta,\alpha_{i},\gamma_{t}\right)  $ w.r.t. the
heterogeneous parameters are, without loss of generality,
\[
\partial_{\alpha_{i}}l_{i,t}\left(  \theta,\alpha_{i},\gamma_{t}\right)
=\left(  Y_{it}-\frac{1}{1+\exp(-X_{it}(\theta+\alpha_{i}+\gamma_{t}%
))}\right)  X_{it},
\]%
\[
\partial_{\alpha_{i}\gamma_{t}}l_{i,t}\left(  \theta,\alpha_{i},\gamma
_{t}\right)  =\frac{-\exp(-X_{it}(\theta+\alpha_{i}+\gamma_{t}))}%
{[1+\exp(-X_{it}(\theta+\alpha_{i}+\gamma_{t}))]^{2}}X_{it}^{2},
\]
and
\[
\partial_{\alpha_{i}\gamma_{t}\gamma_{t}}l_{i,t}\left(  \theta,\alpha
_{i},\gamma_{t}\right)  =\frac{\exp(-X_{it}(\theta+\alpha_{i}+\gamma
_{t}))\left[  1-\exp(-X_{it}(\theta+\alpha_{i}+\gamma_{t}))\right]  }%
{[1+\exp(-X_{it}(\theta+\alpha_{i}+\gamma_{t}))]^{3}}X_{it}^{3}.
\]
Note that the cross derivatives w.r.t. $\alpha_{i}$ and $\alpha_{j},$ or
$\gamma_{t}$ and $\gamma_{s},$ are all $0$ for $i\neq j$ or $t\neq s$.

Assumption \ref{Assumption.Expansion.Smooth}. We have
\begin{align}
\sup_{\theta,\alpha_{i},\gamma_{t}}\left\vert l_{i,t}\left(  \theta,\alpha
_{i},\gamma_{t}\right)  \right\vert  &  \leq\sup_{\theta,\alpha_{i},\gamma
_{t}}\left\vert X_{it}(\theta+\alpha_{i}+\gamma_{t})\right\vert +\sup
_{\theta,\alpha_{i},\gamma_{t}}\log\big(1+\exp\left(  -X_{it}(\theta
+\alpha_{i}+\gamma_{t})\right)  \big)\nonumber\\[3pt]
&  \leq2\sup_{\theta,\alpha_{i},\gamma_{t}}\left\vert X_{it}(\theta+\alpha
_{i}+\gamma_{t})\right\vert +\log2\nonumber\\[3pt]
&  \leq C|X_{it}|+\log2, \label{Eq.Lemma.LowLevelCondition.1}%
\end{align}
where $C<\infty$ is a positive constant depending on the compact parameter
spaces $\Theta$ and $\Phi.$ It is easy to see that
\begin{equation}
\sup_{\theta,\alpha_{i},\gamma_{t}}\left\vert \partial_{\alpha_{i}}%
l_{i,t}\left(  \theta,\alpha_{i},\gamma_{t}\right)  \right\vert \leq\left\vert
X_{it}\right\vert . \label{Eq.Lemma.LowLevelCondition.2}%
\end{equation}
For the second derivative, observe that
\[
\sup_{\theta,\alpha_{i},\gamma_{t}}\left\vert \partial_{\alpha_{i}\gamma_{t}%
}l_{i,t}\left(  \theta,\alpha_{i},\gamma_{t}\right)  \right\vert \leq\sup
_{\pi\in\mathbb{R}}\frac{\exp(-\pi)}{(\exp(-\pi)+1)^{2}}X_{it}^{2}\leq\frac
{1}{4}X_{it}^{2}.
\]
For the third derivative, we have
\begin{align}
\sup_{\theta,\alpha_{i},\gamma_{t}}\left\vert \partial_{\alpha_{i}\gamma
_{t}\gamma_{t}}l_{i,t}\left(  \theta,\alpha_{i},\gamma_{t}\right)
\right\vert  &  \leq\sup_{\pi\in\mathbb{R}}\left\vert \frac{\exp(-\pi)\left(
1-\exp(-\pi)\right)  }{(\exp(-\pi)+1)^{3}}\right\vert \left\vert
X_{it}\right\vert ^{3}\nonumber\\[3pt]
&  \leq\sup_{s\geq0}\frac{s+s^{2}}{(1+s)^{3}}\left\vert X_{it}\right\vert
^{3}\nonumber\\[3pt]
&  \leq\left(  \frac{1}{5}+\frac{1}{3}\right)  \left\vert X_{it}\right\vert
^{3}. \label{Eq.Lemma.LowLevelCondition.4}%
\end{align}
Combining (\ref{Eq.Lemma.LowLevelCondition.1}%
)--(\ref{Eq.Lemma.LowLevelCondition.4}), Assumption
\ref{Assumption.Expansion.Smooth} holds with
\[
g\left(  w_{it}\right)  =C(\left\Vert X_{it}\right\Vert _{\max}+1)^{3}+\log2,
\]
where $C>1.$ Therefore, the condition $\sup_{i,t}\mathbb{E}_{\phi}\left\Vert
X_{it}\right\Vert _{\max}^{3\eta}<\infty$ is sufficient for Assumption
\ref{Assumption.Expansion.Smooth}.

Assumption \ref{Assumption.Expansion.Hessian}. Under the logit model
\eqref{1013-1}, the Hessian can be expressed as
\begin{equation}
\sqrt{NT}H(\theta)=\frac{1}{\sqrt{NT}}DX^{\prime}H^{\pi}(\theta)XD^{\prime},
\label{1014-1}%
\end{equation}
where $D$ is the deterministic matrix defined in Section \ref{ME}, and
$X=(X^{\alpha},X^{\gamma})$ with
\[
X^{\gamma}=\left(
\begin{array}
[c]{ccc}%
X_{11}^{\prime} &  & \\
& \ddots & \\
&  & X_{1T}^{\prime}\\
X_{21}^{\prime} &  & \\
& \ddots & \\
&  & X_{2T}^{\prime}\\
\vdots & \ddots & \vdots\\
X_{N1}^{\prime} &  & \\
& \ddots & \\
&  & X_{NT}^{\prime}%
\end{array}
\right)  ,\qquad X^{\alpha}=\left(
\begin{array}
[c]{cccc}%
X_{11}^{\prime} &  &  & \\
\vdots &  &  & \\
X_{1T}^{\prime} &  &  & \\
& X_{21}^{\prime} &  & \\
& \vdots &  & \\
& X_{2T}^{\prime} &  & \\
&  & \ddots & \\
&  &  & X_{N1}^{\prime}\\
&  &  & \\
&  &  & X_{NT}^{\prime}%
\end{array}
\right)  .
\]
Moreover, $H^{\pi}(\theta)=\operatorname*{diag}\left\{  H_{1}^{\pi}%
(\theta),\ldots,H_{N}^{\pi}(\theta)\right\}  $ is a diagnal matrix, where
block $H_{i}^{\pi}(\theta)=\operatorname*{diag}\left\{  h_{i,1}^{\pi}%
(\theta),\ldots,h_{i,T}^{\pi}(\theta)\right\}  $ and each element is given by
\[
h_{i,t}^{\pi}(\theta)=-\frac{\exp(-\pi_{i,t}(\theta))}{[1+\exp(-\pi
_{i,t}(\theta))]^{2}}<0,\quad\pi_{i,t}(\theta)=X_{it}^{\prime}(\theta
+\alpha_{i}(\theta)+\gamma_{t}(\theta)).
\]
Here, $\alpha_{i}(\theta)$ and $\gamma_{t}(\theta)$ denote the pseudo-true
values defined by \eqref{Eq.IPP.PsiEstimator}. Let $\lambda_{\max}(A)$ denote
the maximum eigenvalue of a matrix $A.$ Using \eqref{1014-1}, we have
\[
\lambda_{\max}(\sqrt{NT}\mathbb{E}H(\theta))\leq\sup_{i,t}h_{i,t}^{\pi}%
(\theta)\lambda_{\max}(DX^{\prime}XD^{\prime}/\sqrt{NT}).
\]
Therefore, Assumption \ref{Assumption.Expansion.Hessian} is implied if
$XD^{\prime}$ has full column rank. Given the structure of $X$, this requires
that, for all $i$, the columns of $X_{i\cdot}$ are linearly independent and
that, for all $t$, the columns of $X_{\cdot t}$ are linearly independent.%
\end{proof}%

Next, prove the asymptotic properties of $L(\theta)$, the corrected
likelihood, $\widehat{\theta}_{L}$ the corrected estimator, and
$\{\widehat{\xi}_{LR},\widehat{\xi}_{LM},\widehat{\xi}_{Wald}\}$, the three
test statistics we consider in this paper. To continue, we first introduce
three Lemmas. These lemmas are useful in establishing Theorems
\ref{Theorem.LikeExpansion} (asymptotic bias expansion) and
\ref{Theorem.CorrectedLike} (corrected likelihood). The proof of Corollaries
\ref{Theorem.Theta} (estimators) and \ref{Theorem.Inference} (tests) then follow.

Define the partition%
\[
H(\theta)=:%
\begin{pmatrix}
H_{\alpha\alpha^{\prime}}\left(  \theta\right)  , & H_{\alpha\gamma^{\prime}%
}\left(  \theta\right) \\
H_{\gamma\alpha^{\prime}}\left(  \theta\right)  , & H_{\gamma\gamma^{\prime}%
}\left(  \theta\right)
\end{pmatrix}
,
\]
where $H_{\alpha\alpha^{\prime}}\left(  \theta\right)  $ is $\left(
N-1\right)  K\times\left(  N-1\right)  K$, $H_{\gamma\gamma^{\prime}}\left(
\theta\right)  $ is $\left(  T-1\right)  K\times\left(  T-1\right)  K$,
$H_{\alpha\gamma^{\prime}}\left(  \theta\right)  $ is $\left(  N-1\right)
K\times\left(  T-1\right)  K$, and $H_{\gamma\alpha^{\prime}}\left(
\theta\right)  =\left[  H_{\alpha\gamma^{\prime}}\left(  \theta\right)
\right]  ^{\prime}$. The next lemma states that the quadratic form can be
decomposed into two dominating terms of order $O(1/\sqrt{NT})$. They establish
the two additive bias terms $B_{\alpha}(\theta)$ and $B_{\gamma}(\theta)$,
arising from estimating individual effects and time effects, respectively.

\begin{lemma}
\label{Lemma.QuadForm}Under Assumption \ref{Assumption.Expansion}, we have
\[
\mathbb{E}\left\{  s^{\prime}(\theta)[{H}\left(  \theta\right)  ]^{-1}
s(\theta)\right\}  =\mathbb{E}\left\{  \widetilde{s}_{\alpha}^{\prime}
(\theta)H_{\alpha\alpha^{\prime}}^{\ast}(\theta)\widetilde{s}_{\alpha}
(\theta)\right\}  +\mathbb{E\{}\widetilde{s}_{\gamma}^{\prime}(\theta
)H_{\gamma\gamma^{\prime}}^{\ast}(\theta)\widetilde{s}_{\gamma}(\theta
)\}+o(1/\sqrt{NT}),
\]
where $o(\cdot)$ is uniform in $\theta$, and
\[
\sup_{\theta\in\Theta}\mathbb{E}\left\{  \widetilde{s}_{\alpha}^{\prime
}(\theta)H_{\alpha\alpha^{\prime}}^{\ast}(\theta)\widetilde{s}_{\alpha}
(\theta)\right\}  =O\left(  1/T\right)  ,\qquad\sup_{\theta\in\Theta
}\mathbb{E}\left[  \widetilde{s}_{\gamma}^{\prime}(\theta)H_{\gamma
\gamma^{\prime}}^{\ast}(\theta)\widetilde{s}_{\gamma}(\theta)\right]
=O(1/N).
\]

\end{lemma}

%

\begin{proof}%
Seeing $\mathbb{E}s(\theta)=0$, the claim without $\mathbb{E}$ is
\begin{align}
&  \quad\quad s^{\prime}(\theta)[-{H}(\theta)]^{-1}s\left(  \theta\right)
\nonumber\\
&  =[s(\theta)-\mathbb{E}s(\theta)]^{\prime}[-{H}\left(  \theta\right)
]^{-1}[s(\theta)-\mathbb{E}s(\theta)]\nonumber\\
&  =[s(\theta)-\mathbb{E}s(\theta)]^{\prime}[-\mathbb{E}{H}\left(
\theta\right)  ]^{-1}[s(\theta)-\mathbb{E}s(\theta)]^{\prime}(1+o_{\mathbb{P}%
}(1))\nonumber\\
&  =-\left[  \widetilde{s}_{\alpha}^{\prime}(\theta)H_{\alpha\alpha^{\prime}%
}^{\ast}(\theta)\widetilde{s}_{\alpha}(\theta)+\widetilde{s}_{\gamma}^{\prime
}(\theta)H_{\gamma\gamma^{\prime}}^{\ast}(\theta)\widetilde{s}_{\gamma}%
(\theta)+2\widetilde{s}_{\alpha}^{\prime}(\theta)H_{\alpha\gamma^{\prime}%
}^{\ast}(\theta)\widetilde{s}_{\gamma}\left(  \theta\right)  \right]
(1+o_{\mathbb{P}}(1)). \label{Eq.Lemma.Quad.0}%
\end{align}
Since $K$ is fixed, we assume without loss of generality that $K=1$. In what
follows, we show
\begin{equation}
\sup_{\theta\in\Theta}\mathbb{E}\left[  \widetilde{s}_{\alpha}^{\prime}%
(\theta)H_{\alpha\gamma^{\prime}}^{\ast}(\theta)\widetilde{s}_{\gamma}\left(
\theta\right)  \right]  =O\left(  1/(NT)\right)  , \label{Eq.Lemma.Quad.1}%
\end{equation}
and
\begin{equation}
\sup_{\theta\in\Theta}\mathbb{E}\left[  \widetilde{s}_{\alpha}^{\prime}%
(\theta)H_{\alpha\alpha^{\prime}}^{\ast}(\theta)\widetilde{s}_{\alpha}%
(\theta)\right]  =O(1/T),\qquad\sup_{\theta\in\Theta}\mathbb{E}\left[
\widetilde{s}_{\gamma}^{\prime}(\theta)H_{\gamma\gamma^{\prime}}^{\ast}%
(\theta)\widetilde{s}_{\gamma}(\theta)\right]  =O(1/N).
\label{Eq.Lemma.Quad.2}%
\end{equation}
Thus, $\mathbb{E}\left[  \widetilde{s}_{\alpha}^{\prime}(\theta)H_{\alpha
\gamma^{\prime}}^{\ast}(\theta)\widetilde{s}_{\gamma}\left(  \theta\right)
\right]  $ is negligible in the expansion relative to the other two terms from
the quadratic form.

Under Assumptions \ref{Assumption.Expansion.Smooth} and
\ref{Assumption.Expansion.Data}, using Davydov's inequality, we have
$\sup_{i,j}\sup_{\theta\in\Theta}\operatorname*{Var}[h_{i,j}^{\alpha}%
(\theta)]=o(1)$ as $N,T\rightarrow\infty$, where $\operatorname*{Var}$ denotes
the variance and $h_{i,j}^{\alpha}(\theta)$ is the $\left(  i,j\right)  $-th
element of $N\cdot H_{\alpha\alpha^{\prime}}\left(  \theta\right)  $ with
$i,j=1,\ldots,N-1$. Invoking the Markov's inequality,
\[
\sup_{i,j}\sup_{\theta\in\Theta}\left\{  h_{i,j}^{\alpha}(\theta
)-\mathbb{E}[h_{i,j}^{\alpha}(\theta)]\right\}  =o_{\mathbb{P}}(1).
\]
Similarly, we have
\begin{align*}
\sup_{t,s}\sup_{\theta\in\Theta}\left\{  h_{t,s}^{\gamma}(\theta
)-\mathbb{E}[h_{t,s}^{\gamma}(\theta)]\right\}   &  =o_{\mathbb{P}}(1),\\
\sup_{i,t}\sup_{\theta\in\Theta}\left\{  h_{i,t}^{\alpha\gamma}(\theta
)-\mathbb{E}[h_{i,t}^{\alpha\gamma}(\theta)]\right\}   &  =o_{\mathbb{P}}(1),
\end{align*}
where $h_{t,s}^{\gamma}(\theta)$ denote the $(t,s)$-th element of $T\cdot
H_{\gamma\gamma^{\prime}}(\theta)$, with $t,s=1,\ldots,T-1$, and
$h_{i,t}^{\alpha\gamma}(\theta)$ the $(i,t)$-th element of $NT\cdot
H_{\alpha\gamma^{\prime}}(\theta)$, for $i=1,\ldots,N-1$ and $t=1,\ldots,T-1$.
Therefore, we have
\begin{align*}
N\left\Vert H_{\alpha\alpha^{\prime}}(\theta)-\mathbb{E}[H_{\alpha
\alpha^{\prime}}(\theta)]\right\Vert _{\max}  &  =o_{\mathbb{P}}(1),\qquad
T\left\Vert H_{\gamma\gamma^{\prime}}(\theta)-\mathbb{E}[H_{\gamma
\gamma^{\prime}}(\theta)]\right\Vert _{\max}=o_{\mathbb{P}}(1),\\
NT\left\Vert H_{\alpha\gamma^{\prime}}(\theta)-\mathbb{E}[H_{\alpha
\gamma^{\prime}}(\theta)]\right\Vert _{\max}  &  =o_{\mathbb{P}}(1),
\end{align*}
uniformly for $\theta\in\Theta$, where $\Vert A\Vert_{\max}$ denotes the
$\max$-norm of a matrix $A$. In addition, by matrix inversion in block form,
it follows that
\begin{align*}
\sup_{\theta\in\Theta}\left\Vert H_{\alpha\alpha^{\prime}}^{\ast}%
(\theta)\right\Vert _{\max}  &  =O(T),\qquad\sup_{\theta\in\Theta}\left\Vert
H_{\gamma\gamma^{\prime}}^{\ast}\left(  \theta\right)  \right\Vert _{\max
}=O(N),\\
\sup_{\theta\in\Theta}\left\Vert H_{\alpha\gamma^{\prime}}^{\ast}%
(\theta)\right\Vert _{\max}  &  =O(1).
\end{align*}
For $i=1,\ldots,N-1$ and $t=1,\ldots,T-1$, define $h_{i,t}^{\ast\alpha\gamma
}(\theta)$ as the $(i,t)$-th element of $H_{\alpha\gamma^{\prime}}^{\ast
}(\theta)$. We first show (\ref{Eq.Lemma.Quad.1}). We have
\begin{align}
&  \quad\quad\mathbb{E}\left\{  (N\sqrt{T}\widetilde{s}_{\alpha}^{\prime
}\left(  \theta\right)  )H_{\alpha\gamma^{\prime}}^{\ast}\left(
\theta\right)  (\sqrt{N}T\widetilde{s}_{\gamma}(\theta))\right\} \nonumber\\
&  =\frac{1}{\sqrt{NT}}\mathbb{E}\left\{  \sum_{i=1}^{N-1}\sum_{t=1}%
^{T-1}h_{i,t}^{\ast\alpha\gamma}(\theta)\cdot\left[  \sum_{s=1}^{T}%
\widetilde{s}_{i,s}^{\alpha}(\theta)-\widetilde{s}_{N,s}^{\alpha}%
(\theta)\right]  \left[  \sum_{j=1}^{N}\widetilde{s}_{j,t}^{\gamma}%
(\theta)-\widetilde{s}_{j,T}^{\gamma}(\theta)\right]  \right\} \nonumber\\
&  =\frac{1}{\sqrt{NT}}\sum_{i=1}^{N-1}\mathbb{E}\left[  \left(  \sum
_{s=1}^{T}\widetilde{s}_{i,s}^{\alpha}(\theta)\right)  \left(  \sum
_{t=1}^{T-1}\widetilde{s}_{i,t}^{\gamma}(\theta)h_{i,t}^{\ast\alpha\gamma
}(\theta)\right)  \right] \nonumber\\
&  \quad\quad-\frac{1}{\sqrt{NT}}\sum_{i=1}^{N-1}\left[  \left(  \sum
_{t=1}^{T-1}h_{i,t}^{\ast\alpha\gamma}(\theta)\right)  \mathbb{E}\left(
\sum_{s=1}^{T}\widetilde{s}_{i,s}^{\alpha}(\theta)\widetilde{s}_{j,T}^{\gamma
}(\theta)\right)  \right] \nonumber\\
&  \quad\quad-\frac{1}{\sqrt{NT}}\mathbb{E}\left\{  \left(  \sum_{s=1}%
^{T}\widetilde{s}_{N,s}^{\alpha}(\theta)\right)  \left[  \sum_{t=1}%
^{T-1}\widetilde{s}_{N,t}^{\gamma}(\theta)\left(  \sum_{i=1}^{N-1}%
h_{i,t}^{\ast\alpha\gamma}(\theta)\right)  \right]  \right\} \nonumber\\
&  \quad\quad+\frac{1}{\sqrt{NT}}\left(  \sum_{i=1}^{N-1}\sum_{t=1}%
^{T-1}h_{i,t}^{\ast\alpha\gamma}(\theta)\right)  \mathbb{E}\left(  \sum
_{s=1}^{T}\widetilde{s}_{N,s}^{\alpha}(\theta)\widetilde{s}_{N,T}^{\gamma
}(\theta)\right) \nonumber\\
&  =:A_{NT,1}(\theta)-A_{NT,2}(\theta)-A_{NT,3}(\theta)+A_{NT,4}(\theta).
\label{Eq.Lemma.Quad.3}%
\end{align}
By Assumptions \ref{Assumption.Expansion.Smooth} and
\ref{Assumption.Expansion.Data}, and using Davydov's inequality, we obtain for
some $\eta>2$ that
\begin{align*}
\left\vert A_{NT,1}(\theta)\right\vert  &  =\frac{1}{\sqrt{NT}}\sum
_{i=1}^{N-1}\left\vert \mathbb{E}\left[  \left(  \sum_{s=1}^{T-1}%
\widetilde{s}_{i,s}^{\alpha}(\theta)\right)  \left(  \sum_{t=1}^{T-1}%
\widetilde{s}_{i,t}^{\gamma}(\theta)h_{i,t}^{\ast\alpha\gamma}(\theta)\right)
+\sum_{t=1}^{T-1}\widetilde{s}_{i,T}^{\alpha}(\theta)\widetilde{s}%
_{i,t}^{\gamma}(\theta)h_{i,t}^{\ast\alpha\gamma}(\theta)\right]  \right\vert
\\
&  \leq\sqrt{NT}\left(  \sup_{\theta\in\Theta}\left\Vert H_{\alpha
\gamma^{\prime}}^{\ast}(\theta)\right\Vert _{\max}\right)  \left(  \sup
_{i,t}\{\mathbb{E}[g^{\eta}(w_{it})]\}^{2/\eta}\right) \\
&  \quad\quad\times\sup\left(  [a_{i}(0)]^{1-2/\eta}+2\sum_{m=1}^{T-2}\left[
\left(  1-\frac{m}{T-1}\right)  [a_{i}(m)]^{1-2/\eta}\right]  +\frac{1}%
{T-1}\sum_{m=1}^{T-1}[a_{i}(m)]^{1-2/\eta}\right) \\
&  =O(\sqrt{NT}).
\end{align*}
Similarly,
\begin{align*}
\left\vert A_{NT,2}(\theta)\right\vert  &  =\frac{T}{\sqrt{NT}}\sum
_{i=1}^{N-1}\left(  \frac{1}{T}\sum_{t=1}^{T-1}\left\vert h_{i,t}^{\ast
\alpha\gamma}(\theta)\right\vert \right)  \left\vert \mathbb{E}\left(
\sum_{s=1}^{T}\widetilde{s}_{i,s}^{\alpha}(\theta)\widetilde{s}_{i,T}^{\gamma
}(\theta)\right)  \right\vert \\
&  \leq\sqrt{NT}\left(  \sup_{\theta\in\Theta}\left\Vert H_{\alpha
\gamma^{\prime}}^{\ast}(\theta)\right\Vert _{\max}\right)  \left(  \sup
_{i,t}\{\mathbb{E}[g^{\eta}(w_{it})]\}^{2/\eta}\right)  \left(  \sup_{i}%
\sum_{m=0}^{T-1}[a_{i}(m)]^{1-2/\eta}\right) \\
&  =O(\sqrt{NT});\\
\left\vert A_{NT,3}(\theta)\right\vert  &  =\frac{N}{\sqrt{NT}}\left\vert
\mathbb{E}\left[  \left(  \sum_{s=1}^{T}\widetilde{s}_{N,s}^{\alpha}%
(\theta)\right)  \left(  \sum_{t=1}^{T-1}\widetilde{s}_{N,t}^{\gamma}%
(\theta)\left(  \frac{1}{N}\sum_{i=1}^{N-1}h_{i,t}^{\ast\alpha\gamma}%
(\theta)\right)  \right)  \right]  \right\vert \\
&  =\sqrt{NT}\left(  \sup_{\theta\in\Theta}\left\Vert H_{\alpha\gamma^{\prime
}}^{\ast}(\theta)\right\Vert _{\max}\right)  \left(  \sup_{i,t}\{\mathbb{E}%
[g^{\eta}(w_{it})]\}^{2/\eta}\right) \\
&  \quad\quad\times\sup_{i}\left(  [a_{i}(0)]^{1-2/\eta}+2\sum_{m=1}%
^{T-2}\left[  \left(  1-\frac{m}{T-1}\right)  [a_{i}(m)]^{1-2/\eta}\right]
+\frac{1}{T-1}\sum_{m=1}^{T-1}[a_{i}(m)]^{1-2/\eta}\right) \\
&  =O(\sqrt{NT});\\
\left\vert A_{NT,4}(\theta)\right\vert  &  \leq\sqrt{NT}\left(  \sup
_{\theta\in\Theta}\left\Vert H_{\alpha\gamma^{\prime}}^{\ast}(\theta
)\right\Vert _{\max}\right)  \left(  \sup_{i,t}\{\mathbb{E}[g^{\eta}%
(w_{it})]\}^{2/\eta}\right)  \sup_{i}\left(  \sum_{m=0}^{T-1}[a_{i}%
(m)]^{1-2/\eta}\right) \\
&  =O(\sqrt{NT}).
\end{align*}
Therefore, (\ref{Eq.Lemma.Quad.1}) follows from (\ref{Eq.Lemma.Quad.3}).

Let $h_{i,j}^{\ast\alpha}(\theta)$ denote the $(i,j)$-th element of
$T^{-1}H_{\alpha\alpha^{\prime}}^{\ast}(\theta)$, $i,j=1,\ldots,N-1$. By
Assumption \ref{Assumption.Expansion.Smooth}, we know that $\sup_{i\geq
1}h_{i,i}^{\ast\alpha}(\theta)=O(1)$ uniformly in $\theta$. Recall,
$\alpha:=(\alpha_{1}^{\prime},\ldots,\alpha_{N}^{\prime})^{\prime}$ and
$\gamma:=(\gamma_{1}^{\prime},\ldots,\gamma_{T}^{\prime})^{\prime}$. Note
that
\[
N\cdot H_{\alpha\alpha^{\prime}}(\theta)=D_{1}[N\cdot\partial_{\alpha
\alpha^{\prime}}l(\theta,\phi)|_{\phi=\phi(\theta)}]D_{1}^{\prime},
\]
implying $N\cdot H_{\alpha\alpha^{\prime}}(\theta)$ can be written as the sum
of a diagonal matrix and a rank-one matrix. Thus, following from \cite{m1981},
we have $\sup_{i\neq j}h_{i,j}^{\ast\alpha}(\theta)=O(1/N)$ uniformly in
$\theta$. Now,
\begin{align}
&  \quad\quad T\mathbb{E}\left[  \widetilde{s}_{\alpha}^{\prime}%
(\theta)H_{\alpha\alpha^{\prime}}^{\ast}(\theta)\widetilde{s}_{\alpha}%
(\theta)\right] \nonumber\\
&  =\frac{1}{N}\mathbb{E}\left\{  (N\sqrt{T}\widetilde{s}_{\alpha}^{\prime
}(\theta))[N^{-1}H_{\alpha\alpha^{\prime}}^{\ast}(\theta)](N\sqrt
{T}\widetilde{s}_{\alpha}(\theta))\right\} \nonumber\\
&  =\frac{1}{N}\sum_{i=1}^{N-1}h_{i,i}^{\ast\alpha}(\theta)\mathbb{E}\left(
\frac{1}{\sqrt{T}}\sum_{t=1}^{T}\left[  \widetilde{s}_{i,t}^{\alpha}%
(\theta)-\widetilde{s}_{N,t}^{\alpha}(\theta)\right]  \right)  ^{2}\nonumber\\
&  \quad\quad+\frac{2}{N}\sum_{1\leq i<j\leq N-1}h_{i,j}^{\ast\alpha}%
(\theta)\mathbb{E}\left[  \left(  \frac{1}{\sqrt{T}}\sum_{t=1}^{T}\left[
\widetilde{s}_{i,t}^{\alpha}(\theta)-\widetilde{s}_{N,t}^{\alpha}%
(\theta)\right]  \right)  \left(  \frac{1}{\sqrt{T}}\sum_{t=1}^{T}\left[
\widetilde{s}_{j,t}^{\alpha}(\theta)-\widetilde{s}_{N,t}^{\alpha}%
(\theta)\right]  \right)  \right] \nonumber\\
&  =:I_{NT,1}(\theta)+I_{NT,2}(\theta). \label{Eq.Lemma.Quad.4}%
\end{align}
Here, by Assumptions \ref{Assumption.Expansion.Smooth} and
\ref{Assumption.Expansion.Data}, applying Davydov's inequality gives, for some
$\eta>2$,
\begin{align*}
&  \quad\quad\sup_{\theta\in\Theta}\left\vert I_{NT,1}(\theta)\right\vert \\
&  =\sup_{\theta\in\Theta}\frac{1}{NT}\sum_{i=1}^{N-1}\left\vert h_{i,i}%
^{\ast\alpha}(\theta)\right\vert \left\vert \sum_{t=1}^{T}\sum_{s=1}%
^{T}\mathbb{E}\left[  \widetilde{s}_{i,t}^{\alpha}(\theta)\widetilde{s}%
_{i,s}^{\alpha}(\theta)\right]  +\sum_{t=1}^{T}\sum_{s=1}^{T}\mathbb{E}\left[
\widetilde{s}_{N,t}^{\alpha}(\theta)\widetilde{s}_{N,s}^{\alpha}%
(\theta)\right]  \right\vert \\
&  \leq\left(  \frac{1}{N}\sum_{i=1}^{N-1}|h_{i,i}^{\ast\alpha}(\theta
)|\right)  \left(  \sup_{i,t}\{\mathbb{E}[g^{\eta}(w_{it})]\}^{2/\eta}\right)
\sup_{i}\left(  [a_{i}(0)]^{1-2/\eta}+2\sum_{m=1}^{T-1}\left(  1-\frac{m}%
{T}\right)  [a_{i}(m)]^{1-2/\eta}\right) \\
&  =O\left(  1\right)
\end{align*}
as $N,T\rightarrow\infty$, where the conditional cross-sectional independence
leads to the first equality. Similarly
\begin{align*}
&  \quad\quad\sup_{\theta\in\Theta}\left\vert I_{NT,2}(\theta)\right\vert \\
&  =\frac{2}{N}\sum_{1\leq i<j\leq N-1}\left\{  \left\vert h_{i,j}^{\ast
\alpha}(\theta)\right\vert \mathbb{E}\left[  \left(  \frac{1}{\sqrt{T}}%
\sum_{t=1}^{T}\widetilde{s}_{N,t}^{\alpha}(\theta)\right)  ^{2}\right]
\right\} \\
&  \leq\left(  \frac{2}{N}\sum_{1\leq i<j\leq N-1}|h_{i,j}^{\ast\alpha}%
(\theta)|\right)  \left(  \sup_{i,t}\{\mathbb{E}[g^{\eta}(w_{it})]\}^{2/\eta
}\right)  \sup_{i}\left(  [a_{i}(0)]^{1-2/\eta}+2\sum_{m=1}^{T-1}\left(
1-\frac{m}{T}\right)  [a_{i}(m)]^{1-2/\eta}\right) \\
&  =O\left(  1\right)  ,
\end{align*}
noting that $\sup_{i\neq j}h_{i,j}^{\ast\alpha}(\theta)=O(1/N)$ uniformly in
$\theta$. Thus, the first part of (\ref{Eq.Lemma.Quad.2}) follows from
(\ref{Eq.Lemma.Quad.4}). The second part of (\ref{Eq.Lemma.Quad.2}) follows
from a similar procedure.

Combining (\ref{Eq.Lemma.Quad.0}), (\ref{Eq.Lemma.Quad.1}), and
(\ref{Eq.Lemma.Quad.2}),
\begin{align*}
\mathbb{E}\left\{  s^{\prime}(\theta)[-{H}(\theta)]^{-1}s(\theta)\right\}   &
=\mathbb{E}\left\{  \widetilde{s}_{\alpha}^{\prime}(\theta)H_{\alpha
\alpha^{\prime}}^{\ast}(\theta)\widetilde{s}_{\alpha}(\theta)\right\}
+\mathbb{E}\left\{  \widetilde{s}_{\gamma}^{\prime}(\theta)H_{\gamma
\gamma^{\prime}}^{\ast}(\theta)\widetilde{s}_{\gamma}(\theta)\right\} \\
&  \quad\quad+\mathbb{E}\left\{  \left[  \widetilde{s}_{\alpha}^{\prime
}(\theta)H_{\alpha\alpha^{\prime}}^{\ast}(\theta)\widetilde{s}_{\alpha}%
(\theta)+\widetilde{s}_{\gamma}^{\prime}\left(  \theta\right)  H_{\gamma
\gamma^{\prime}}^{\ast}\left(  \theta\right)  \widetilde{s}_{\gamma}%
(\theta)\right]  \cdot o_{\mathbb{P}}(1)\right\}  .
\end{align*}
By Assumption \ref{Assumption.Expansion.Smooth}, for $S\in\{0,1,2\}$, we have
\[
\sup_{i,t}\mathbb{E}\left(  |\partial_{\phi_{1}\ldots\phi_{S}}l_{i,t}\left(
\theta,\phi\right)  |%
\mathds{1}%
\{|\partial_{\phi_{1}\ldots\phi_{S}}l_{i,t}\left(  \theta,\phi\right)
|>a\}\right)  \rightarrow0
\]
as $a\rightarrow\infty$, which shows that $\{\partial_{\phi_{1}\ldots\phi_{S}%
}l_{it}\left(  \theta,\phi\right)  \}_{i,t}$ are uniformly integrable. By the
$L_{1}$ convergence theorem, we hence have
\[
\sqrt{NT}\mathbb{E}\left\{  \left[  \widetilde{s}_{\alpha}^{\prime}%
(\theta)H_{\alpha\alpha^{\prime}}^{\ast}(\theta)\widetilde{s}_{\alpha}%
(\theta)+\widetilde{s}_{\gamma}^{\prime}\left(  \theta\right)  H_{\gamma
\gamma^{\prime}}^{\ast}\left(  \theta\right)  \widetilde{s}_{\gamma}%
(\theta)\right]  \cdot o_{\mathbb{P}}(1)\right\}  \rightarrow0,
\]
as $N,T\rightarrow\infty$. This completes the proof.%
\end{proof}%

The next lemma shows that the remainder term in Equation
(\ref{Eq.IPP.PsiExpansion}) $r^{s}\left(  \theta\right)  :=r^{s}\left(
\theta,a\right)  $ is $O_{\mathbb{P}}\left(  1/\left(  NT\right)  \right)  $ uniformly.

\begin{lemma}
\label{Lemma.RemainderRS} Under Assumption \ref{Assumption.Expansion},
\[
\sup_{\theta\in\Theta}\sup_{a}\left\Vert r^{s}(\theta,a)\right\Vert _{\max
}=O_{\mathbb{P}}\left(  1/\left(  NT\right)  \right)  ,
\]
as $N$ and $T\rightarrow\infty$, where $\sup_{a}$ is taken over $a_{j}%
\in(0,1)$ for all $j=1,\ldots,(N+T-2)K$.
\end{lemma}

%

\begin{proof}%
In the proof, we retain the notation and setup from Lemma \ref{Lemma.QuadForm}%
. From (\ref{Eq.IPP.PsiExpansion}), it is evident that the dominant term in
the expansion of $\widehat{\psi}(\theta)-\psi\left(  \theta\right)  $ is
$[-{H}(\theta)]^{-1}s(\theta)$. Specifically,
\begin{align}
\lbrack-{H}(\theta)]^{-1}s(\theta)  &  =[-\mathbb{E}{H}(\theta)]^{-1}%
[{s}(\theta)-\mathbb{E}{s}(\theta)](1+o_{\mathbb{P}}(1))\nonumber\\[0.04in]
&  =-%
\begin{pmatrix}
H_{\alpha\alpha^{\prime}}^{\ast}(\theta)\widetilde{s}_{\alpha}(\theta
)+H_{\alpha\gamma^{\prime}}^{\ast}(\theta)\widetilde{s}_{\gamma}(\theta)\\
H_{\gamma\alpha^{\prime}}^{\ast}(\theta)\widetilde{s}_{\alpha}(\theta
)+H_{\gamma\gamma^{\prime}}^{\ast}(\theta)\widetilde{s}_{\gamma}(\theta)
\end{pmatrix}
+[-\mathbb{E}{H}(\theta)]^{-1}[{s}(\theta)-\mathbb{E}{s}(\theta)]\cdot
o_{\mathbb{P}}(1) \label{Eq.Lemma.RemainderRS.1}%
\end{align}
where $o_{\mathbb{P}}(1)$ is uniform in both $i$ and $t$. Let $u_{i}^{\alpha
}(\theta)$ denote the $i$-th element of $H_{\alpha\alpha^{\prime}}^{\ast
}(\theta)\widetilde{s}_{\alpha}(\theta)$. It follows that
\begin{equation}
N\sqrt{T}u_{i}^{\alpha}(\theta)=N\sum_{j=1}^{N-1}\left[  h_{i,j}^{\ast\alpha
}(\theta)\left(  \frac{1}{\sqrt{T}}\sum_{t=1}^{T}\left[  \widetilde{s}%
_{j,t}^{\alpha}(\theta)-\widetilde{s}_{N,t}^{\alpha}(\theta)\right]  \right)
\right]  . \label{Eq.Lemma.RemainderRS.2}%
\end{equation}
Using the result from \cite{m1981}, we know that $\sup_{i\geq1}h_{i,i}%
^{\ast\alpha}(\theta)=O(1)$ and $\sup_{i\neq j}h_{i,j}^{\ast\alpha}%
(\theta)=O(1/N)$ uniformly in $\theta$, implying $\sum_{j=1}^{N-1}%
h_{i,j}^{\ast\alpha}(\theta)=O(1/N)$ uniformly in both $i$ and $\theta$. Since
$1/\sqrt{T}\sum_{t=1}^{T}\widetilde{s}_{i,t}^{\alpha}(\theta)=O_{\mathbb{P}%
}(1)$ uniformly over $i$ by Assumptions \ref{Assumption.Expansion.Smooth} and
\ref{Assumption.Expansion.Data}, we obtain
\[
N\sum_{j=1}^{N-1}\left[  h_{i,j}^{\ast\alpha}(\theta)\left(  \frac{1}{\sqrt
{T}}\sum_{t=1}^{T}\widetilde{s}_{j,t}^{\alpha}(\theta)\right)  \right]
=O_{\mathbb{P}}(1),\qquad N\left(  \sum_{j=1}^{N-1}h_{i,j}^{\ast\alpha}%
(\theta)\right)  \left(  \frac{1}{\sqrt{T}}\sum_{t=1}^{T}\widetilde{s}%
_{N,t}^{\alpha}(\theta)\right)  =O_{\mathbb{P}}(1).
\]
Therefore, $N\sqrt{T}u_{i}^{\alpha}(\theta)=O_{\mathbb{P}}(1)$ uniformly over
$i$. Let $u_{i}^{\alpha\gamma}(\theta)$ denote the $i$-th element of
$H_{\alpha\gamma^{\prime}}^{\ast}\left(  \theta\right)  \widetilde{s}_{\gamma
}(\theta)$. Given that $\sup_{\theta\in\Theta}\Vert H_{\alpha\gamma^{\prime}%
}^{\ast}(\theta)\Vert_{\max}=O(1)$, we have
\begin{equation}
T\sqrt{N}u_{i}^{\alpha\gamma}(\theta)=\sum_{t=1}^{T-1}\left[  h_{i,t}%
^{\ast\alpha\gamma}(\theta)\left(  \frac{1}{\sqrt{N}}\sum_{j=1}^{N}\left[
\widetilde{s}_{j,t}^{\gamma}(\theta)-\widetilde{s}_{j,T}^{\gamma}%
(\theta)\right]  \right)  \right]  . \label{Eq.Lemma.RemainderRS.3}%
\end{equation}
Since $\sum_{t=1}^{T-1}h_{i,t}^{\ast\alpha\gamma}(\theta)=O(1)$, which follows
from $\mathbb{E}[\partial_{\breve{\alpha}\breve{\gamma}^{\prime}}\ell
(\theta,\psi)|_{\psi=\psi(\theta)}]=\partial_{\breve{\gamma}^{\prime}%
}\mathbb{E}[\partial_{\breve{\alpha}}\ell(\theta,\psi)|_{\psi=\psi(\theta
)}]=0$, for $\breve{\alpha}:=\left(  \alpha_{1}^{\prime},\ldots,\alpha
_{N-1}^{\prime}\right)  ^{\prime}$ and $\breve{\gamma}:=\left(  \gamma
_{1}^{\prime},\ldots,\gamma_{T-1}^{\prime}\right)  ^{\prime}$, it holds that
\[
\sum_{t=1}^{T-1}\left[  h_{i,t}^{\ast\alpha\gamma}(\theta)\left(  \frac
{1}{\sqrt{N}}\sum_{j=1}^{N}\widetilde{s}_{j,t}^{\gamma}(\theta)\right)
\right]  =O_{\mathbb{P}}(1),\qquad\left(  \sum_{t=1}^{T-1}h_{i,t}^{\ast
\alpha\gamma}(\theta)\right)  \left(  \frac{1}{\sqrt{N}}\sum_{j=1}%
^{N}\widetilde{s}_{j,T}^{\gamma}(\theta)\right)  =O_{\mathbb{P}}(1)
\]
uniformly in $\theta$. From (\ref{Eq.Lemma.RemainderRS.2}) and
(\ref{Eq.Lemma.RemainderRS.3}), we conclude that
\begin{equation}
\sup_{\theta\in\Theta}\Vert H_{\alpha\alpha^{\prime}}^{\ast}(\theta
)\widetilde{s}_{\alpha}(\theta)\Vert_{\max}=O_{\mathbb{P}}\left(  1/(N\sqrt
{T})\right)  ,\qquad\sup_{\theta\in\Theta}\Vert H_{\alpha\gamma^{\prime}%
}^{\ast}(\theta)\widetilde{s}_{\gamma}(\theta)\Vert_{\max}=O_{\mathbb{P}%
}\left(  1/(T\sqrt{N})\right)  . \label{Eq.Lemma.RemainderRS.4}%
\end{equation}
Following a similar proof, we also have
\begin{equation}
\sup_{\theta\in\Theta}\Vert H_{\gamma\gamma^{\prime}}^{\ast}(\theta
)\widetilde{s}_{\gamma}(\theta)\Vert_{\max}=O_{\mathbb{P}}\left(  1/(T\sqrt
{N})\right)  ,\qquad\sup_{\theta\in\Theta}\Vert H_{\gamma\alpha^{\prime}%
}^{\ast}(\theta)\widetilde{s}_{\alpha}(\theta)\Vert_{\max}=O_{\mathbb{P}%
}\left(  1/(N\sqrt{T})\right)  . \label{Eq.Lemma.RemainderRS.5}%
\end{equation}
Thus, combining (\ref{Eq.Lemma.RemainderRS.1}), (\ref{Eq.Lemma.RemainderRS.4}
), and (\ref{Eq.Lemma.RemainderRS.5}), we have
\[
\left\Vert \lbrack-{H}(\theta)]^{-1}s(\theta)+[\mathbb{E}{H}(\theta)]^{-1}%
[{s}(\theta)-\mathbb{E}{s}(\theta)]\right\Vert _{\max}=o_{\mathbb{P}}\left(
1/(NT)^{3/4}\right)  ,
\]
where $o_{\mathbb{P}}(\cdot)$ is uniform in $\theta$. Noting that
\[
\widehat{\psi}(\theta)-\psi(\theta)=[-{H}\left(  \theta\right)  ]^{-1}%
s(\theta)(1+o_{\mathbb{P}}(1)),
\]
we further have
\begin{equation}
\left\Vert \widehat{\psi}(\theta)-\psi(\theta)+[\mathbb{E}{H}(\theta
)]^{-1}[{s}(\theta)-\mathbb{E}{s}(\theta)]\right\Vert _{\max}=o_{\mathbb{P}%
}\left(  1/(NT)^{3/4}\right)  . \label{Eq.Lemma.RemainderRS.6}%
\end{equation}

From (\ref{Eq.Lemma.RemainderRS.6}), we conclude that $\sup_{\theta\in\Theta
}\Vert G_{\psi_{j}}\left(  \theta,a_{j}\right)  -\partial_{\psi_{j}}%
H(\theta)\Vert_{\max}=o_{\mathbb{P}}(1)$. Without loss of generality, let
$\psi_{j}=\alpha_{i}$ for $i=1,\ldots,N-1$. We can express $\partial
_{\alpha_{i}}H(\theta)$ as
\[
\partial_{\alpha_{i}}H(\theta)=%
\begin{pmatrix}
\partial_{\alpha_{i}}{H}_{\alpha\alpha^{\prime}}\left(  \theta\right)  , &
\partial_{\alpha_{i}}{H}_{\alpha\gamma^{\prime}}(\theta)\\
\partial_{\alpha_{i}}{H}_{\gamma\alpha^{\prime}}\left(  \theta\right)  , &
\partial_{\alpha_{i}}{H}_{\gamma\gamma^{\prime}}(\theta)
\end{pmatrix}
.
\]
We then have
\begin{align*}
N\cdot\partial_{\alpha_{i}}{H}_{\alpha\alpha^{\prime}}\left(  \theta\right)
&  =\frac{1}{T}\sum_{t=1}^{T}\operatorname*{diag}\left(  0,\ldots
,0,\partial_{\alpha_{i}\alpha_{i}\alpha_{i}}l_{i,t}(\theta,\phi)|_{\phi
=\phi(\theta)},0,\ldots,0\right) \\[3pt]
&  \quad\quad-\left(  \frac{1}{T}\sum_{t=1}^{T}\partial_{\alpha_{N}\alpha
_{N}\alpha_{N}}l_{N,t}(\theta,\phi)|_{\phi=\phi(\theta)}\right)  \iota
_{N-1}\iota_{N-1}^{\prime},\\
NT\cdot\partial_{\alpha_{i}}{H}_{\gamma\gamma^{\prime}}\left(  \theta\right)
&  =\operatorname*{diag}\left\{  \partial_{\gamma_{1}\gamma_{1}\alpha_{i}%
}l_{i,1}(\theta,\phi)|_{\phi=\phi(\theta)},\ldots,\partial_{\gamma_{T-1}%
\gamma_{T-1}\alpha_{i}}l_{i,T-1}(\theta,\phi)|_{\phi=\phi(\theta)}\right\} \\
&  \quad\quad+\partial_{\gamma_{T}\gamma_{T}\alpha_{i}}l_{i,T}(\theta
,\phi)|_{\phi=\phi(\theta)}\,\iota_{N-1}\iota_{N-1}^{\prime},\\
NT\cdot\partial_{\alpha_{i}}{H}_{\alpha\gamma^{\prime}}\left(  \theta\right)
&  =\left(  \partial_{\alpha_{i}\alpha_{i}\gamma_{1}}l_{i,1}(\theta
,\phi)|_{\phi=\phi(\theta)},\ldots,\partial_{\alpha_{i}\alpha_{i}\gamma_{T-1}%
}l_{i,T-1}(\theta,\phi)|_{\phi=\phi(\theta)}\right)  \otimes e_{i}\\
&  \quad\quad+\left(  \partial_{\alpha_{N}\alpha_{N}\gamma_{1}}l_{N,1}%
(\theta,\phi)|_{\phi=\phi(\theta)},\ldots,\partial_{\alpha_{N}\alpha_{N}%
\gamma_{T-1}}l_{N,T-1}(\theta,\phi)|_{\phi=\phi(\theta)}\right)  \otimes
\iota_{N-1},
\end{align*}
where $e_{i}$ is an $(N-1)\times1$ vector with the $i$-th component equal to
$1$ and all other elements equal to $0$. Using (\ref{Eq.Lemma.RemainderRS.6}),
for each $j=1,\ldots,N-1$, we write
\begin{align}
&  \quad\quad\sup_{a_{j}\in(0,1)}r_{\psi_{j}}^{s}(\theta,a_{j}%
)\nonumber\\[0.04in]
&  =\sup_{a_{j}\in(0,1)}\frac{1}{2}\left(  \widehat{\psi}(\theta)-\psi
(\theta)\right)  ^{\prime}G_{\psi_{j}}(\theta,a_{j})\left(  \widehat{\psi
}(\theta)-\psi\left(  \theta\right)  \right) \nonumber\\[0.04in]
&  =\frac{1}{2}%
\begin{pmatrix}
H_{\alpha\alpha^{\prime}}^{\ast}(\theta)\widetilde{s}_{\alpha}(\theta
)+H_{\alpha\gamma^{\prime}}^{\ast}(\theta)\widetilde{s}_{\gamma}(\theta)\\
H_{\gamma\alpha^{\prime}}^{\ast}(\theta)\widetilde{s}_{\alpha}(\theta
)+H_{\gamma\gamma^{\prime}}^{\ast}(\theta)\widetilde{s}_{\gamma}(\theta)
\end{pmatrix}
^{\prime}\partial_{\alpha_{i}}H(\theta)%
\begin{pmatrix}
H_{\alpha\alpha^{\prime}}^{\ast}(\theta)\widetilde{s}_{\alpha}(\theta
)+H_{\alpha\gamma^{\prime}}^{\ast}(\theta)\widetilde{s}_{\gamma}(\theta)\\
H_{\gamma\alpha^{\prime}}^{\ast}(\theta)\widetilde{s}_{\alpha}(\theta
)+H_{\gamma\gamma^{\prime}}^{\ast}(\theta)\widetilde{s}_{\gamma}(\theta)
\end{pmatrix}
\left[  1+o_{\mathbb{P}}(1)\right] \nonumber\\[0.04in]
&  =\frac{1}{2}\left\{  \left(  H_{\alpha\alpha^{\prime}}^{\ast}%
(\theta)\widetilde{s}_{\alpha}(\theta)+H_{\alpha\gamma^{\prime}}^{\ast}%
(\theta)\widetilde{s}_{\gamma}(\theta)\right)  ^{\prime}\partial_{\alpha_{i}%
}{H}_{\alpha\alpha^{\prime}}(\theta)\left(  H_{\alpha\alpha^{\prime}}^{\ast
}(\theta)\widetilde{s}_{\alpha}(\theta)+H_{\alpha\gamma^{\prime}}^{\ast
}(\theta)\widetilde{s}_{\gamma}(\theta)\right)  \right. \nonumber\\[3pt]
&  \quad\quad+\left(  H_{\gamma\alpha^{\prime}}^{\ast}(\theta)\widetilde{s}%
_{\alpha}(\theta)+H_{\gamma\gamma^{\prime}}^{\ast}(\theta)\widetilde{s}%
_{\gamma}(\theta)\right)  ^{\prime}\partial_{\alpha_{i}}{H}_{\gamma
\gamma^{\prime}}(\theta)\left(  H_{\gamma\alpha^{\prime}}^{\ast}%
(\theta)\widetilde{s}_{\alpha}(\theta)+H_{\gamma\gamma^{\prime}}^{\ast}%
(\theta)\widetilde{s}_{\gamma}(\theta)\right) \nonumber\\[3pt]
&  \quad\quad\left.  +2\left(  H_{\alpha\alpha^{\prime}}^{\ast}(\theta
)\widetilde{s}_{\alpha}(\theta)+H_{\alpha\gamma^{\prime}}^{\ast}%
(\theta)\widetilde{s}_{\gamma}(\theta)\right)  ^{\prime}\partial_{\alpha_{i}%
}{H}_{\alpha\gamma^{\prime}}(\theta)\left(  H_{\gamma\alpha^{\prime}}^{\ast
}(\theta)\widetilde{s}_{\alpha}(\theta)+H_{\gamma\gamma^{\prime}}^{\ast
}(\theta)\widetilde{s}_{\gamma}(\theta)\right)  \right\}  \left[
1+o_{\mathbb{P}}(1)\right] \nonumber\\[0.04in]
&  =:\frac{1}{2}\left\{  C_{NT,1}(\theta)+C_{NT,2}(\theta)+C_{NT,3}%
(\theta)\right\}  \left[  1+o_{\mathbb{P}}(1)\right]  .
\label{expansion remainder}%
\end{align}
Let $v_{i}^{\alpha}(\theta)$ denote the $i$-th element of $H_{\alpha
\alpha^{\prime}}^{\ast}(\theta)\widetilde{s}_{\alpha}(\theta)+H_{\alpha
\gamma^{\prime}}^{\ast}(\theta)\widetilde{s}_{\gamma}(\theta)$, for
$i=1,\ldots,N-1$. Using (\ref{Eq.Lemma.RemainderRS.2}) and
(\ref{Eq.Lemma.RemainderRS.3}), it can be shown that
\begin{align*}
&  \quad\quad C_{NT,1}(\theta)\\[0.04in]
&  =\frac{1}{N}\left(  \sum_{i=1}^{N-1}v_{i}^{\alpha}(\theta)\right)
^{2}\left(  \frac{1}{T}\sum_{t=1}^{T}\partial_{\alpha_{i}\alpha_{i}\alpha_{i}%
}l_{i,t}(\theta,\phi)|_{\phi=\phi(\theta)}\right)  -\frac{1}{N}\left(
\sum_{i=1}^{N-1}v_{i}^{\alpha}(\theta)\right)  ^{2}\left(  \frac{1}{T}%
\sum_{t=1}^{T}\partial_{\alpha_{N}\alpha_{N}\alpha_{N}}l_{N,t}(\theta
,\phi)|_{\phi=\phi(\theta)}\right) \\[3pt]
&  =O_{\mathbb{P}}\left(  1/(NT)\right)  ,
\end{align*}
uniformly in $\theta$. By similar arguments, we also have
\[
\sup_{\theta\in\Theta}C_{NT,2}(\theta)=O_{\mathbb{P}}\left(  1/(NT)^{3/2}%
\right)  ,\qquad\sup_{\theta\in\Theta}C_{NT,3}(\theta)=O_{\mathbb{P}}\left(
1/(NT)^{3/2}\right)  .
\]
Therefore, when $\psi_{j}=\alpha_{i}$, we have
\begin{align}
&  \quad\quad\sup_{\theta\in\Theta}\sup_{a_{j}\in(0,1)}r_{\psi_{j}}^{s}%
(\theta,a_{j})\nonumber\\[0.04in]
&  =-\frac{1}{2N}\left(  \sum_{i=1}^{N-1}v_{i}^{\alpha}(\theta)\right)
^{2}\left(  \frac{1}{T}\sum_{t=1}^{T}\partial_{\alpha_{N}\alpha_{N}\alpha_{N}%
}l_{N,t}(\theta,\phi)|_{\phi=\phi(\theta)}\right)  +o_{\mathbb{P}}\left(
1/(NT)\right) \nonumber\\[0.04in]
&  =O_{\mathbb{P}}\left(  1/(NT)\right)  . \label{Eq.Lemma.RemainderRS.7}%
\end{align}
Similarly, when $\psi_{j}=\gamma_{t}$ ($t=1,\ldots,T-1$),
\begin{align}
&  \quad\quad\sup_{\theta\in\Theta}\sup_{a_{j}\in(0,1)}r_{\psi_{j}}^{s}%
(\theta,a_{j})\nonumber\\[0.04in]
&  =-\frac{1}{2T}\left(  \sum_{t=1}^{T-1}v_{t}^{\gamma}(\theta)\right)
^{2}\left(  \frac{1}{N}\sum_{i=1}^{N}\partial_{\gamma_{T}\gamma_{T}\gamma_{T}%
}l_{i,T}(\theta,\phi)|_{\phi=\phi(\theta)}\right)  +o_{\mathbb{P}}\left(
1/(NT)\right) \nonumber\\[0.04in]
&  =O_{\mathbb{P}}\left(  1/(NT)\right)  , \label{Eq.Lemma.RemainderRS.8}%
\end{align}
where $v_{t}^{\gamma}(\theta)$ denotes the $t$-th element of $H_{\gamma
\alpha^{\prime}}^{\ast}(\theta)\widetilde{s}_{\alpha}(\theta)+H_{\gamma
\gamma^{\prime}}^{\ast}(\theta)\widetilde{s}_{\gamma}(\theta)$ for
$t=1,\ldots,T-1$. This leads to the desired result.%
\end{proof}%

The next lemma shows that $\mathbb{E}r(\theta)$ is $O\left(  1/\left(
NT\right)  \right)  $ uniformly in $\theta$.

\begin{lemma}
\label{Lemma.RemainderR} Under Assumption \ref{Assumption.Expansion},
$r(\theta)$ defined by (\ref{Eq.Appendix.RemainderR}) satisfies
\[
\sup_{\theta\in\Theta}\mathbb{E}[r(\theta)]=O\left(  1/\left(  NT\right)
\right)  ,
\]
as $N$ and $T\rightarrow\infty$.
\end{lemma}

%

\begin{proof}%
In the proof, we retain the notation and setup from Lemmas
\ref{Lemma.QuadForm} and \ref{Lemma.RemainderRS}. From Lemma
\ref{Lemma.RemainderRS}, we know that $NT\cdot r^{s}\left(  \theta,a\right)  $
is comparable to $(NT)^{3/4}\cdot s(\theta)$. Following similar arguments as
in Lemma \ref{Lemma.QuadForm}, we can show that
\[
\mathbb{E}\left\{  (r^{s}(\theta,a))^{\prime}[{H}\left(  \theta\right)
]^{-1}r^{s}(\theta,a)\right\}  =O\left(  1/(NT)\right)  .
\]
Moreover, from (\ref{Eq.Lemma.RemainderRS.6}), (\ref{Eq.Lemma.RemainderRS.7}),
and (\ref{Eq.Lemma.RemainderRS.7}), we have
\begin{align}
&  \quad\quad\left(  \widehat{\psi}(\theta)-\psi(\theta)\right)  ^{\prime
}r^{s}(\theta,a_{0}\,\iota_{N+T-2})\nonumber\\[0.04in]
&  =\frac{1}{2N}\left(  \sum_{i=1}^{N-1}v_{i}^{\alpha}(\theta)\right)
^{2}\left(  \frac{1}{T}\sum_{t=1}^{T}\partial_{\alpha_{N}\alpha_{N}\alpha_{N}%
}l_{N,t}(\theta,\phi)|_{\phi=\phi(\theta)}\right) \nonumber\\
&  \quad\quad+\frac{1}{2T}\left(  \sum_{t=1}^{T-1}v_{t}^{\gamma}%
(\theta)\right)  ^{2}\left(  \frac{1}{N}\sum_{i=1}^{N}\partial_{\gamma
_{T}\gamma_{T}\gamma_{T}}l_{i,T}(\theta,\phi)|_{\phi=\phi(\theta)}\right)
\nonumber\\[0.04in]
&  \quad\quad+\left(  \sum_{i=1}^{N-1}v_{i}^{\alpha}(\theta)+\sum_{t=1}%
^{T-1}v_{t}^{\gamma}(\theta)\right)  \cdot o_{\mathbb{P}}\left(  1/(NT)\right)
\nonumber\\[0.04in]
&  =O_{\mathbb{P}}\left(  1/(NT)^{5/4}\right)  . \label{Eq.Lemma.RemainderR.1}%
\end{align}
By Assumption \ref{Assumption.Expansion.Smooth} and applying the $L_{1}$
convergence theorem to \ref{Eq.Lemma.RemainderR.1}, we find that
\[
\sup_{\theta\in\Theta}\mathbb{E}\left[  \left(  \widehat{\psi}(\theta
)-\psi(\theta)\right)  ^{\prime}r^{s}\left(  \theta,a_{0}\,\iota
_{N+T-2}\right)  \right]  =o\left(  1/(NT)\right)  .
\]
Thus,
\[
\mathbb{E}\left[  r(\theta)\right]  =\frac{1}{2}\mathbb{E}\left[
(r^{s}\left(  \theta,a\right)  )^{\prime}[{H}(\theta)]^{-1}r^{s}\left(
\theta,a\right)  \right]  +\frac{1}{3}\mathbb{E}\left[  \left(  \widehat{\psi
}(\theta)-\psi(\theta)\right)  ^{\prime}r^{s}\left(  \theta,a_{0}%
\,\iota_{N+T-2}\right)  \right]  =O\left(  1/(NT)\right)
\]
uniformly in $\theta$.%
\end{proof}%

Next, we prove Theorem \ref{Theorem.LikeExpansion} and
\ref{Theorem.CorrectedLike} below.%

\begin{proof}[Proof of Theorem \ref{Theorem.LikeExpansion}]%
Following the expansion in (\ref{Eq.IPP.LikeExpansion}) and
(\ref{Eq.IPP.PsiExpansion}), we have
\[
\mathbb{E[}\widehat{l}(\theta)-l(\theta)]=-\frac{1}{2}\mathbb{E\{}s^{\prime
}(\theta)[H(\theta)]^{-1}s(\theta)\}+\mathbb{E}r(\theta)
\]
for $r\left(  \theta\right)  $ defined in (\ref{Eq.Appendix.RemainderR}). The
claim in Theorem \ref{Theorem.LikeExpansion} follows directly from Lemmas
\ref{Lemma.QuadForm} and \ref{Lemma.RemainderR}.%
\end{proof}%

The proof of Theorem \ref{Theorem.CorrectedLike} is as follows.%

\begin{proof}[Proof of Theorem \ref{Theorem.CorrectedLike}]%
We retain the notation introduced in Lemmas \ref{Lemma.QuadForm}%
-\ref{Lemma.RemainderR}. We first show
\begin{equation}
\widehat{B}_{\alpha}(\theta)-B_{\alpha}(\theta)=o_{\mathbb{P}}(1/\sqrt
{NT}),\qquad\widehat{B}_{\gamma}(\theta)-B_{\gamma}(\theta)=o_{\mathbb{P}%
}(1/\sqrt{NT}). \label{Eq.Theorem.LikeExpansion.1}%
\end{equation}

To show the first part of (\ref{Eq.Theorem.LikeExpansion.1}), note that
\[
\widehat{B}_{\alpha}(\theta)-B_{\alpha}(\theta)=\frac{1}{2}%
\operatorname*{trace}\left\{  \widehat{H}_{\alpha\alpha^{\prime}}^{\ast
}\left(  \theta\right)  \left[  S_{\alpha\alpha}\left(  \theta\right)
-\widehat{S}_{\alpha\alpha}\left(  \theta\right)  \right]  \right\}  +\frac
{1}{2}\operatorname*{trace}\left\{  \left[  \widehat{H}_{\alpha\alpha^{\prime
}}^{\ast}(\theta)-H_{\alpha\alpha^{\prime}}^{\ast}(\theta)\right]
S_{\alpha\alpha}\left(  \theta\right)  \right\}  ,
\]
where%
\begin{align*}
S_{\alpha\alpha}\left(  \theta\right)   &  =S_{\alpha\alpha}^{1}\left(
\theta\right)  +S_{\alpha\alpha}^{2}\left(  \theta\right)  ,\\
S_{\alpha\alpha}^{1}\left(  \theta\right)   &  :=\operatorname*{diag}\left\{
\mathbb{E}\left[  \widetilde{S}_{1}^{\alpha}\left(  \theta\right)  \left(
\widetilde{S}_{1}^{\alpha}\left(  \theta\right)  \right)  ^{\prime}\right]
,\ldots,\mathbb{E}\left[  \widetilde{S}_{N-1}^{\alpha}\left(  \theta\right)
\left(  \widetilde{S}_{N-1}^{\alpha}\left(  \theta\right)  \right)  ^{\prime
}\right]  \right\}  ,\\
S_{\alpha\alpha}^{2}\left(  \theta\right)   &  :=(\iota_{N-1}\iota
_{N-1}^{\prime})\otimes\mathbb{E}\left[  \widetilde{S}_{N}^{\alpha}\left(
\theta\right)  \left(  \widetilde{S}_{N}^{\alpha}\left(  \theta\right)
\right)  ^{\prime}\right]  ;\\
\widehat{S}_{\alpha\alpha}\left(  \theta\right)   &  =\widehat{S}%
_{\alpha\alpha}^{1}\left(  \theta\right)  +\widehat{S}_{\alpha\alpha}%
^{2}\left(  \theta\right)  ,\\
\widehat{S}_{\alpha\alpha}^{1}\left(  \theta\right)   &
:=\operatorname*{diag}\left\{  \widehat{S}_{1}^{\alpha}\left(  \theta\right)
\left(  \widehat{S}_{1}^{\alpha}\left(  \theta\right)  \right)  ^{\prime
},\ldots,\widehat{S}_{N-1}^{\alpha}\left(  \theta\right)  \left(
\widehat{S}_{N-1}^{\alpha}\left(  \theta\right)  \right)  ^{\prime}\right\}
,\\
\widehat{S}_{\alpha\alpha}^{2}\left(  \theta\right)   &  :=(\iota_{N-1}%
\iota_{N-1}^{\prime})\otimes\left[  \widehat{S}_{N}^{\alpha}\left(
\theta\right)  \left(  \widehat{S}_{N}^{\alpha}\left(  \theta\right)  \right)
^{\prime}\right]  ,
\end{align*}
for $\widetilde{s}_{\alpha}(\theta)=:\left[  \left(  \widetilde{S}_{1}%
^{\alpha}\left(  \theta\right)  \right)  ^{\prime},\ldots,\left(
\widetilde{S}_{N}^{\alpha}\left(  \theta\right)  \right)  ^{\prime}\right]
^{\prime}$, $\widehat{s}_{\alpha}(\theta)=:\left[  \left(  \widehat{S}%
_{1}^{\alpha}\left(  \theta\right)  \right)  ^{\prime},\ldots,\left(
\widehat{S}_{N}^{\alpha}\left(  \theta\right)  \right)  ^{\prime}\right]
^{\prime}$, and $\widehat{s}_{\alpha}(\theta)$ being $s_{\alpha}(\theta)$ but
with $\phi\left(  \theta\right)  $ replaced by $\widehat{\phi}\left(
\theta\right)  $. Here we have
\begin{align}
&  \quad\quad N^{2}\left[  S_{\alpha\alpha}^{1}\left(  \theta\right)
-\widehat{S}_{\alpha\alpha}^{1}\left(  \theta\right)  \right]
\nonumber\\[0.04in]
&  =\frac{1}{T^{2}}\sum_{t=1}^{T}\sum_{s=1}^{T}%
\mathds{1}%
\left\{  \left\vert t-s\right\vert \leq\tau\right\} \nonumber\\
&  \quad\quad\times\operatorname*{diag}\left\{  \widetilde{s}_{1,t}^{\alpha
}(\theta)\left(  \widetilde{s}_{1,s}^{\alpha}(\theta)\right)  ^{\prime
}-\widehat{s}_{1,t}^{\alpha}(\theta)\left(  \widehat{s}_{1,s}^{\alpha}%
(\theta)\right)  ^{\prime},\ldots,\widetilde{s}_{N-1,t}^{\alpha}%
(\theta)\left(  \widetilde{s}_{N-1,s}^{\alpha}(\theta)\right)  ^{\prime
}-\widehat{s}_{N-1,t}^{\alpha}(\theta)\left(  \widehat{s}_{N-1,s}^{\alpha
}(\theta)\right)  ^{\prime}\right\} \nonumber\\[0.04in]
&  \quad\quad+\frac{1}{T^{2}}\sum_{t=1}^{T}\sum_{s=1}^{T}%
\mathds{1}%
\left\{  \left\vert t-s\right\vert >\tau\right\} \nonumber\\
&  \quad\quad\times\operatorname*{diag}\left\{  \mathbb{E}\left[
\widetilde{s}_{1,t}^{\alpha}(\theta)\left(  \widetilde{s}_{1,s}^{\alpha
}(\theta)\right)  ^{\prime}\right]  ,\ldots,\mathbb{E}\left[  \widetilde{s}%
_{N-1,t}^{\alpha}(\theta)\left(  \widetilde{s}_{N-1,s}^{\alpha}(\theta
)\right)  ^{\prime}\right]  \right\}  . \label{Eq.Theorem.LikeExpansion.2}%
\end{align}
Under Assumptions \ref{Assumption.Expansion.Smooth} and
\ref{Assumption.Expansion.Data}, by Davydov's inequality,
\[
\mathbb{E}\left[  \widetilde{s}_{i,t}^{\alpha}(\theta)\left(  \widetilde{s}%
_{i,s}^{\alpha}(\theta)\right)  ^{\prime}\right]  \leq8[a_{i}%
(|t-s|)]^{1-2/\eta}\sup_{i,t}\{\mathbb{E}[g^{\eta}(w_{it})]\}^{2/\eta},
\]
where we assume without loss of generality that $K=1$. Thus, as $T\rightarrow
\infty$,
\begin{align*}
&  \quad\quad\frac{1}{T}\sum_{t=1}^{T}\sum_{s=1}^{T}%
\mathds{1}%
\left\{  \left\vert t-s\right\vert >\tau\right\}  \mathbb{E}\left[
\widetilde{s}_{i,t}^{\alpha}(\theta)\left(  \widetilde{s}_{i,s}^{\alpha
}(\theta)\right)  ^{\prime}\right] \\[0.04in]
&  \leq8\sup_{i,t}\{\mathbb{E}[g^{\eta}(w_{it})]\}^{2/\eta}\left(  \frac{1}%
{T}\sum_{t=1}^{T}\sum_{s=1}^{T}%
\mathds{1}%
\left\{  \left\vert t-s\right\vert >\tau\right\}  [a_{i}(|t-s|)]^{1-2/\eta
}\right) \\[3pt]
&  \leq8\sup_{i,t}\{\mathbb{E}[g^{\eta}(w_{it})]\}^{2/\eta}\left(
\frac{T-\tau}{T}\sum_{m=\tau+1}^{\infty}[a_{i}(m)]^{1-2/\eta}\right) \\[3pt]
&  =o(1)
\end{align*}
given that $\tau/T\rightarrow0$. Combining this with
(\ref{Eq.Theorem.LikeExpansion.2}), we have
\begin{equation}
\sup_{\theta\in\Theta}N^{2}\left\Vert S_{\alpha\alpha}^{1}\left(
\theta\right)  -\widehat{S}_{\alpha\alpha}^{1}\left(  \theta\right)
\right\Vert _{\max}=o_{\mathbb{P}}(1/T). \label{Eq.Theorem.LikeExpansion.3}%
\end{equation}
From the proof of Lemma \ref{Lemma.RemainderRS}, we also have $\sup_{\theta
\in\Theta}\Vert\widehat{\psi}(\theta)-\psi(\theta)\Vert_{\max}=o_{\mathbb{P}%
}(1)$. Moreover, under the conditional strong mixing assumption, each element
of $T^{-1}\widehat{H}_{\alpha\alpha^{\prime}}^{\ast}(\theta)$ converges in
probability to the corresponding element in $T^{-1}H_{\alpha\alpha^{\prime}%
}^{\ast}(\theta)$ by the law of large numbers, where Therefore
(\ref{Eq.Theorem.LikeExpansion.3}) further leads to
\begin{align*}
\operatorname*{trace}\left\{  \widehat{H}_{\alpha\alpha^{\prime}}^{\ast
}\left(  \theta\right)  \left[  S_{\alpha\alpha}^{1}\left(  \theta\right)
-\widehat{S}_{\alpha\alpha}^{1}\left(  \theta\right)  \right]  \right\}   &
=\frac{T}{N^{2}}\operatorname*{trace}\left\{  T^{-1}\widehat{H}_{\alpha
\alpha^{\prime}}^{\ast}(\theta)\cdot N^{2}\left[  S_{\alpha\alpha}^{1}\left(
\theta\right)  -\widehat{S}_{\alpha\alpha}^{1}\left(  \theta\right)  \right]
\right\} \\
&  =\frac{T}{N}\frac{1}{N}\sum_{i=1}^{N-1}h_{i,i}^{\ast\alpha}(\theta)\cdot
o_{\mathbb{P}}(1/T)\\
&  =o_{\mathbb{P}}(1/T).
\end{align*}
A similar procedure can be followed to show
\[
\operatorname*{trace}\left\{  \widehat{H}_{\alpha\alpha^{\prime}}^{\ast
}\left(  \theta\right)  \left[  S_{\alpha\alpha}^{2}\left(  \theta\right)
-\widehat{S}_{\alpha\alpha}^{2}\left(  \theta\right)  \right]  \right\}
=o_{\mathbb{P}}(1/N^{2}),
\]
noting $\sum_{j=1}^{N-1}h_{i,j}^{\ast\alpha}(\theta)=O(1/N)$ uniformly in both
$i$ and $\theta$ from Lemma \ref{Lemma.RemainderRS}. The above leads to
\begin{equation}
\operatorname*{trace}\left\{  \widehat{H}_{\alpha\alpha^{\prime}}^{\ast
}(\theta)\left[  S_{\alpha\alpha}\left(  \theta\right)  -\widehat{S}%
_{\alpha\alpha}\left(  \theta\right)  \right]  \right\}  =o_{\mathbb{P}%
}(1/\sqrt{NT}). \label{Eq.Theorem.LikeExpansion.4}%
\end{equation}

Second, from Theorem \ref{Theorem.LikeExpansion}, we also have
\begin{equation}
\operatorname*{trace}\left\{  \left[  \widehat{H}_{\alpha\alpha^{\prime}%
}^{\ast}(\theta)-H_{\alpha\alpha^{\prime}}^{\ast}(\theta)\right]
S_{\alpha\alpha}\left(  \theta\right)  \right\}  =o_{\mathbb{P}}%
(1)\cdot\operatorname*{trace}\left\{  H_{\alpha\alpha^{\prime}}^{\ast}%
(\theta)S_{\alpha\alpha}\left(  \theta\right)  \right\}  =o_{\mathbb{P}%
}(1)\cdot B_{\alpha}(\theta)=o_{\mathbb{P}}(1/\sqrt{NT}).
\label{Eq.Theorem.LikeExpansion.5}%
\end{equation}
Combining \ref{Eq.Theorem.LikeExpansion.4} and
\ref{Eq.Theorem.LikeExpansion.5} leads to the first part of
\ref{Eq.Theorem.LikeExpansion.1}. The second part of
\ref{Eq.Theorem.LikeExpansion.1} can be shown similarly.

Noting that
\[
l(\theta)=\widehat{l}(\theta)+\frac{1}{2}s_{\alpha}^{\prime}(\theta
)H_{\alpha\alpha^{\prime}}^{\ast}(\theta)s_{\alpha}(\theta)+\frac{1}%
{2}s_{\gamma}^{\prime}(\theta)H_{\gamma\gamma^{\prime}}^{\ast}(\theta
)s_{\gamma}(\theta)+o_{\mathbb{P}}(1/\sqrt{NT})
\]
as implied by the proof of Theorem \ref{Theorem.LikeExpansion}, we have
\begin{align*}
L(\theta)-l(\theta)  &  =\left[  \widehat{B}_{\alpha}(\theta)-B_{\alpha
}(\theta)\right]  +\left[  \widehat{B}_{\gamma}(\theta)-B_{\gamma}%
(\theta)\right] \\[3pt]
&  \quad\quad-\left[  \frac{1}{2}\widetilde{s}_{\alpha}^{\prime}%
(\theta)H_{\alpha\alpha^{\prime}}^{\ast}(\theta)\widetilde{s}_{\alpha}%
(\theta)-B_{\alpha}(\theta)\right]  -\left[  \frac{1}{2}\widetilde{s}_{\gamma
}^{\prime}(\theta)H_{\gamma\gamma^{\prime}}^{\ast}(\theta)\widetilde{s}%
_{\gamma}(\theta)-B_{\gamma}(\theta)\right] \\[3pt]
&  \quad\quad+o_{\mathbb{P}}(1/\sqrt{NT}).
\end{align*}
As from Lemma \ref{Lemma.QuadForm}, we have
\[
\frac{1}{2}\widetilde{s}_{\alpha}^{\prime}(\theta)H_{\alpha\alpha^{\prime}%
}^{\ast}(\theta)\widetilde{s}_{\alpha}(\theta)-B_{\alpha}(\theta
)=o_{\mathbb{P}}(1/\sqrt{NT}),\qquad\frac{1}{2}\widetilde{s}_{\gamma}^{\prime
}(\theta)H_{\gamma\gamma^{\prime}}^{\ast}(\theta)\widetilde{s}_{\gamma}%
(\theta)-B_{\gamma}(\theta)=o_{\mathbb{P}}(1/\sqrt{NT}).
\]
The conclusion follows by combining this and (\ref{Eq.Theorem.LikeExpansion.1}%
).%
\end{proof}%

We continue to establishing the asymptotitc properties of $\{\widehat{\theta
}_{L},\widehat{\xi}_{LR},\widehat{\xi}_{LM},\widehat{\xi}_{Wald}\}$. We first
prove Corollary \ref{Theorem.Theta} below which states the consistency and
asymptotic normality of $\widehat{\theta}_{L}$.%

\begin{proof}[Proof of Corollary \ref{Theorem.Theta}]%
Consider a mean value expansion of $\triangledown_{\theta}L(\widehat{\theta
}_{L})$ around $\theta_{0}$:
\[
0=\triangledown_{\theta}L(\widehat{\theta}_{L})=\triangledown_{\theta}%
L(\theta_{0})+\triangledown_{\theta\theta^{\prime}}L(\overline{\theta
})(\widehat{\theta}_{L}-\theta_{0}),
\]
where each component of $\overline{\theta}$ lies between the corresponding
components of $\widehat{\theta}_{L}$ and $\theta_{0}$. From Theorem
\ref{Theorem.CorrectedLike}, we further have
\begin{equation}
0=\triangledown_{\theta}l(\theta_{0})+\triangledown_{\theta\theta^{\prime}%
}l(\overline{\theta})(\widehat{\theta}_{L}-\theta_{0})+o_{\mathbb{P}}%
(1/\sqrt{NT}) \label{Eq.Theorem.Theta.1}%
\end{equation}
Under Assumption \ref{Assumption.Statistic.Ident}, $\theta_{0}$ is the unique
maximizer of $\mathbb{E}[l(\theta)]$. Since $L(\theta)-\mathbb{E}%
[l(\theta)]=o_{\mathbb{P}}(1)$, under Assumption
\ref{Assumption.Statistic.Dominance1}, we conclude that
\[
\widehat{\theta}_{L}\overset{\mathbb{P}}{\longrightarrow}\theta_{0}.
\]
This also implies $\overline{\theta}\overset{\mathbb{P}}{\longrightarrow
}\theta_{0}$. By Assumption \ref{Assumption.Statistic.Dominance2}, it follows
that
\[
\triangledown_{\theta\theta^{\prime}}l(\overline{\theta})-\mathbb{E}%
[\triangledown_{\theta\theta^{\prime}}l(\theta_{0})]=o_{\mathbb{P}}(1).
\]
Thus, using Assumption \ref{Assumption.Statistic.Hessian} and Slutsky's
theorem, we obtain from (\ref{Eq.Theorem.Theta.1}) that
\begin{equation}
\sqrt{NT}(\widehat{\theta}_{L}-\theta_{0})=-\left\{  \mathbb{E}[\triangledown
_{\theta\theta^{\prime}}l(\theta_{0})]\right\}  ^{-1}\sqrt{NT}\triangledown
_{\theta}l(\theta_{0})+o_{\mathbb{P}}(1). \label{Eq.Theorem.Theta.2}%
\end{equation}
Given the conditional cross-sectional independence and the conditional strong
mixing condition (Assumption \ref{Assumption.Expansion.Data}),
(\ref{Eq.Theorem.Theta.2}) leads to
\begin{equation}
\left[  \Sigma^{-1}\Omega\Sigma^{-1}\right]  ^{-1/2}\sqrt{NT}(\widehat{\theta
}_{L}-\theta_{0})\overset{\mathbb{D}}{\longrightarrow}\mathcal{N}\left(
0,\mathbb{I}_{K}\right)  \label{Eq.Theorem.Theta.4}%
\end{equation}
as $N,T\rightarrow\infty$, where $\mathbb{E}[\triangledown_{\theta}%
l(\theta_{0})]=0$ and
\[
\Sigma:=-\mathbb{E}\left[  \triangledown_{\theta\theta^{\prime}}l(\theta
_{0})\right]  ,\qquad\Omega:=\left(  NT\right)  \mathbb{E}[\triangledown
_{\theta}l(\theta_{0})\triangledown_{\theta^{\prime}}l(\theta_{0})].
\]
Since we consider a likelihood model, $\Sigma^{-1}\Omega\Sigma^{-1}$ in
(\ref{Eq.Theorem.Theta.4}) simplifies to $\Sigma^{-1}$ under the information
matrix equality ($\Sigma=\Omega$).%
\end{proof}%

To show the asymptotic distribution of $\widehat{\xi}_{LR},\widehat{\xi}%
_{LM},\widehat{\xi}_{Wald}$, we find the following lemma useful.

\begin{lemma}
\label{Lemma.ConstrainedTheta} Under Assumptions \ref{Assumption.Expansion}
and \ref{Assumption.Statistic}, and under the null hypothesis $H_{0}
:R(\theta_{0})=0$, as $N$ and $T\rightarrow\infty,$
\[
\widehat{\theta}_{R}-\widehat{\theta}_{L}=\Xi(\theta_{0})\triangledown
_{\theta}l(\theta_{0})+o_{\mathbb{P}}(1/\sqrt{NT}),
\]
where
\[
\Xi(\theta_{0}):=\left\{  \mathbb{E[}\triangledown_{\theta\theta^{\prime}
}l(\theta_{0})]\right\}  ^{-1}[J(\theta_{0})]^{\prime}\left\{  J(\theta
_{0})\left(  \mathbb{E[}\triangledown_{\theta\theta^{\prime}}l(\theta
_{0})]\right)  ^{-1}[J(\theta_{0})]^{\prime}\right\}  ^{-1}J(\theta
_{0})\left\{  \mathbb{E[}\triangledown_{\theta\theta^{\prime}}l(\theta
_{0})]\right\}  ^{-1}.
\]

\end{lemma}

%

\begin{proof}%
Consider the constrained maximization problem:
\[
\left\{  \widehat{\theta}_{R},\widehat{\lambda}\right\}  =\arg\max_{\theta
\in\Theta,\lambda\in\mathbb{R}}\left[  L(\theta)+\lambda^{\prime}%
R(\theta)\right]
\]
where $\lambda$ is the Lagrangian multiplier. The first-order conditions for
this constrained maximization are
\begin{align}
\sqrt{NT}\triangledown_{\theta}L(\widehat{\theta}_{R})+[J(\widehat{\theta}%
_{R})]^{\prime}\sqrt{NT}\widehat{\lambda}  &
=0,\label{Eq.Lemma.ConstrainedTheta.1}\\[3pt]
\sqrt{NT}R(\widehat{\theta}_{R})  &  =0. \label{Eq.Lemma.ConstrainedTheta.2}%
\end{align}
Following a similar proof as in Corollary \ref{Theorem.Theta}, we have
$\sqrt{NT}(\widehat{\theta}_{R}-\theta_{0})=O_{\mathbb{P}}\left(  1\right)  $.
Furthermore, $\sqrt{NT}\widehat{\lambda}=O_{\mathbb{P}}(1)$ as implied by
(\ref{Eq.Lemma.ConstrainedTheta.1}) since $\sqrt{NT}\triangledown_{\theta
}L(\theta_{0})=O_{\mathbb{P}}(1)$ and $J(\theta_{0})$ is of full row rank. By
Theorem \ref{Theorem.CorrectedLike}, a mean value expansion of $\triangledown
_{\theta}L(\widehat{\theta}_{R})$ around $\theta_{0}$ is
\begin{align*}
\triangledown_{\theta}L(\widehat{\theta}_{R})  &  =\triangledown_{\theta
}L(\theta_{0})+\triangledown_{\theta\theta^{\prime}}L(\overline{\theta
})(\widehat{\theta}_{R}-\theta_{0})\\[3pt]
&  =\triangledown_{\theta}l(\theta_{0})+\mathbb{E}[\triangledown_{\theta
\theta^{\prime}}l(\theta_{0})](\widehat{\theta}_{R}-\theta_{0})+o_{\mathbb{P}%
}(1/\sqrt{NT}),
\end{align*}
where each component of $\overline{\theta}$ lies between the corresponding
components of $\widehat{\theta}_{R}$ and $\theta_{0}.$ Combining this with
(\ref{Eq.Lemma.ConstrainedTheta.1}) yields that
\begin{equation}
\mathbb{E}[\triangledown_{\theta\theta^{\prime}}l(\theta_{0})]\sqrt
{NT}(\widehat{\theta}_{R}-\theta_{0})+[J(\theta_{0})]^{\prime}\sqrt
{NT}\widehat{\lambda}=-\sqrt{NT}\triangledown_{\theta}l(\theta_{0}%
)+o_{\mathbb{P}}(1). \label{Eq.Lemma.ConstrainedTheta.4}%
\end{equation}
Under $H_{0}$ and from (\ref{Eq.Lemma.ConstrainedTheta.2}), we also obtain
\begin{align}
0  &  =\sqrt{NT}R(\widehat{\theta}_{R})\nonumber\\[3pt]
&  =\sqrt{NT}R(\theta_{0})+J(\theta_{0})\sqrt{NT}(\widehat{\theta}_{R}%
-\theta_{0})+o_{\mathbb{P}}(1)\nonumber\\[3pt]
&  =J(\theta_{0})\sqrt{NT}(\widehat{\theta}_{R}-\theta_{0})+o_{\mathbb{P}}(1).
\label{Eq.Lemma.ConstrainedTheta.5}%
\end{align}
Now, combining (\ref{Eq.Lemma.ConstrainedTheta.4}) and
(\ref{Eq.Lemma.ConstrainedTheta.5}), we have
\[%
\begin{pmatrix}
\mathbb{E}[\triangledown_{\theta\theta^{\prime}}l(\theta_{0})] & \left[
J(\theta_{0})\right]  ^{\prime}\\
J(\theta_{0}) & 0
\end{pmatrix}%
\begin{pmatrix}
\sqrt{NT}(\widehat{\theta}_{R}-\theta_{0})\\
\sqrt{NT}\widehat{\lambda}%
\end{pmatrix}
=%
\begin{pmatrix}
-\sqrt{NT}\triangledown_{\theta}l(\theta_{0})\\
0
\end{pmatrix}
+o_{\mathbb{P}}(1)
\]
which has the solution
\begin{equation}
\sqrt{NT}(\widehat{\theta}_{R}-\theta_{0})=-\left(  \{\mathbb{E}%
[\triangledown_{\theta\theta^{\prime}}l(\theta_{0})]\}^{-1}-\Xi\left(
\theta_{0}\right)  \right)  \sqrt{NT}\triangledown_{\theta}l(\theta
_{0})+o_{\mathbb{P}}\left(  1\right)  . \label{Eq.Lemma.ConstrainedTheta.8}%
\end{equation}
The conclusion follows by subtracting (\ref{Eq.Theorem.Theta.2}) in Corollary
\ref{Theorem.Theta} from both sides of (\ref{Eq.Lemma.ConstrainedTheta.8}).%
\end{proof}%

Below, we proof the asymptotic distribution of $\widehat{\xi}_{LR}%
,\widehat{\xi}_{LM},\widehat{\xi}_{Wald}$ below.%

\begin{proof}[Proof of Corollary \ref{Theorem.Inference}]%
Let $\widehat{\xi}:=-NT\triangledown_{\theta^{\prime}}l(\theta_{0})\Xi\left(
\theta_{0}\right)  \triangledown_{\theta}l(\theta_{0})$ for $\Xi\left(
\theta_{0}\right)  $ defined in Lemma \ref{Lemma.ConstrainedTheta}. We first
show that
\begin{equation}
\widehat{\xi}_{LR}=\widehat{\xi}+o_{\mathbb{P}}(1),\qquad\widehat{\xi}%
_{LM}=\widehat{\xi}+o_{\mathbb{P}}(1),\qquad\widehat{\xi}_{Wald}=\widehat{\xi
}+o_{\mathbb{P}}(1). \label{Eq.Theorem.Inference.1}%
\end{equation}
A mean value expansion of $L(\widehat{\theta}_{R})$ around $\widehat{\theta
}_{L}$ gives
\[
{L}(\widehat{\theta}_{R})={L}(\widehat{\theta}_{L})+\triangledown
_{\theta^{\prime}}{L}(\widehat{\theta}_{L})(\widehat{\theta}_{R}%
-\widehat{\theta}_{L})+\frac{1}{2}(\widehat{\theta}_{R}-\widehat{\theta}%
_{L})^{\prime}\triangledown_{\theta\theta^{\prime}}L(\dot{\theta
})(\widehat{\theta}_{R}-\widehat{\theta}_{L})
\]
where each component of $\dot{\theta}$ lies between the corresponding
components of $\widehat{\theta}_{R}$ and $\widehat{\theta}_{L}$. Noting that
$\triangledown_{\theta}{L}(\widehat{\theta}_{L})=0$, we obtain
\[
\widehat{\xi}_{LR}=-NT(\widehat{\theta}_{R}-\widehat{\theta}_{L})^{\prime
}\triangledown_{\theta\theta^{\prime}}L(\dot{\theta})(\widehat{\theta}%
_{R}-\widehat{\theta}_{L}).
\]
Since $\triangledown_{\theta\theta^{\prime}}L(\dot{\theta})-\mathbb{E}%
[\triangledown_{\theta\theta^{\prime}}L(\theta_{0})]=o_{\mathbb{P}}(1)$, by
Lemma \ref{Lemma.ConstrainedTheta}, we further conclude that
\[
\widehat{\xi}_{LR}=\widehat{\xi}+o_{\mathbb{P}}(1).
\]
For $\widehat{\xi}_{LM},$ note that
\begin{align*}
\triangledown_{\theta}L(\widehat{\theta}_{R})  &  =\triangledown_{\theta
}L(\widehat{\theta}_{L})+\triangledown_{\theta\theta^{\prime}}L(\ddot{\theta
})(\widehat{\theta}_{R}-\widehat{\theta}_{L})\\[3pt]
&  =\mathbb{E}[\triangledown_{\theta\theta^{\prime}}L(\theta_{0}%
)](\widehat{\theta}_{R}-\widehat{\theta}_{L})+o_{\mathbb{P}}(1/\sqrt{NT}),
\end{align*}
where each component of $\ddot{\theta}$ lies between the corresponding
components of $\widehat{\theta}_{R}$ and $\widehat{\theta}_{L}$. Thus,
\begin{align*}
\widehat{\xi}_{LM}  &  =-NT(\widehat{\theta}_{R}-\widehat{\theta}_{L}%
)^{\prime}\mathbb{E}[\triangledown_{\theta\theta^{\prime}}L(\theta
_{0})][\triangledown_{\theta\theta^{\prime}}L(\widehat{\theta}_{R}%
)]^{-1}\mathbb{E}[\triangledown_{\theta\theta^{\prime}}L(\theta_{0}%
)](\widehat{\theta}_{R}-\widehat{\theta}_{L})+o_{\mathbb{P}}\left(  1\right)
\\[3pt]
&  =-NT(\widehat{\theta}_{R}-\widehat{\theta}_{L})^{\prime}\mathbb{E}%
[\triangledown_{\theta\theta^{\prime}}L(\theta_{0})](\widehat{\theta}%
_{R}-\widehat{\theta}_{L})+o_{\mathbb{P}}\left(  1\right) \\[3pt]
&  =\widehat{\xi}+o_{\mathbb{P}}\left(  1\right)  .
\end{align*}
For $\widehat{\xi}_{Wald}$, a mean value expansion of $R(\widehat{\theta}%
_{L})$ gives
\[
R(\widehat{\theta}_{L})=R(\widehat{\theta}_{R})+J(\dddot{\theta}%
)(\widehat{\theta}_{L}-\widehat{\theta}_{R})=J(\dddot{\theta})(\widehat{\theta
}_{L}-\widehat{\theta}_{R})
\]
where each component of $\dddot{\theta}$ lies between the corresponding
components of $\widehat{\theta}_{L}$ and $\widehat{\theta}_{R}$. Since both
$\widehat{\theta}_{L}$ and $\widehat{\theta}_{R}$(under $H_{0}$) are
consistent estimators for $\theta_{0},$ we obtain that
\begin{align*}
\widehat{\xi}_{Wald}  &  =-NT(\widehat{\theta}_{L}-\widehat{\theta}%
_{R})^{\prime}J(\dddot{\theta})^{\prime}\left\{  J(\widehat{\theta}%
_{L})[\triangledown_{\theta\theta^{\prime}}L(\widehat{\theta}_{L}%
)]^{-1}J(\widehat{\theta}_{L})^{\prime}\right\}  ^{-1}J(\dddot{\theta
})(\widehat{\theta}_{L}-\widehat{\theta}_{R})\\[3pt]
&  =\widehat{\xi}+o_{\mathbb{P}}\left(  1\right)  .
\end{align*}

Next, since $\Sigma^{-1/2}\sqrt{NT}\triangledown_{\theta}l(\theta
_{0})\overset{\mathbb{D}}{\longrightarrow}\mathcal{N}(0,\mathbb{I}_{K})$,
applying the delta method delivers $\widehat{\xi}\overset{\mathbb{D}%
}{\longrightarrow}\chi^{2}\left(  r\right)  $. The desired results follows
from here.%
\end{proof}%

\subsection{Additional Simulation Results}

\label{Section.AdditionalSim}

\subsubsection{Additional Results under Design 1}

\label{Section.AdditionalSim.Design1}%

\begin{table}[H]
\begin{centering}%
\caption{Comparisons of Uncorrected and Corrected Probit Estimates, Design 1}%
\setstretch{1.2}%
%

\begin{tabular*}
{\linewidth}[c]{@{\extracolsep{\fill}}llrrcrrcrr}\hline\hline
$(N,T)$ &  & \multicolumn{2}{c}{$(30,30)$} &  & \multicolumn{2}{c}{$(60,60)$}
&  & \multicolumn{2}{c}{$(90,90)$}\\
&  & Bias & RMSE & \multicolumn{1}{r}{} & Bias & RMSE & \multicolumn{1}{r}{} &
Bias & RMSE\\\hline
&  & \multicolumn{8}{c}{Static Probit Model}\\
$\widehat{\beta}$ &  & $0.175$ & $0.208$ & \multicolumn{1}{r}{} & $0.069$ &
$0.076$ & \multicolumn{1}{r}{} & $0.044$ & $0.048$\\
$\widehat{\beta}_{L}$ &  & $0.067$ & $0.101$ & \multicolumn{1}{r}{} & $0.016$
& $0.034$ & \multicolumn{1}{r}{} & $0.008$ & $0.019$\\
$\widehat{\beta}_{J}$ &  & $-1.966$ & $3.942$ & \multicolumn{1}{r}{} &
$-0.040$ & $0.101$ & \multicolumn{1}{r}{} & $-0.010$ & $0.022$\\
$\widehat{\beta}_{B}$ &  & $-0.114$ & $0.153$ & \multicolumn{1}{r}{} &
$-0.013$ & $0.030$ & \multicolumn{1}{r}{} & $-0.005$ & $0.018$\\\hline
&  & \multicolumn{8}{c}{Dynamic Probit Model}\\
$\widehat{\beta}$ &  & $0.209$ & $0.335$ & \multicolumn{1}{r}{} & $0.072$ &
$0.080$ & \multicolumn{1}{r}{} & $0.045$ & $0.049$\\
$\widehat{\beta}_{L}^{(1)}$ &  & $0.079$ & $0.116$ & \multicolumn{1}{r}{} &
$0.019$ & $0.036$ & \multicolumn{1}{r}{} & $0.008$ & $0.021$\\
$\widehat{\beta}_{L}^{(2)}$ &  & $0.082$ & $0.118$ & \multicolumn{1}{r}{} &
$0.020$ & $0.037$ & \multicolumn{1}{r}{} & $0.009$ & $0.021$\\
$\widehat{\beta}_{J}$ &  & $-2.161$ & $3.956$ & \multicolumn{1}{r}{} &
$-0.077$ & $0.218$ & \multicolumn{1}{r}{} & $-0.013$ & $0.025$\\
$\widehat{\beta}_{B}$ &  & $-0.155$ & $0.482$ & \multicolumn{1}{r}{} &
$-0.015$ & $0.032$ & \multicolumn{1}{r}{} & $-0.006$ & $0.019$\\
$\widehat{\rho}$ &  & $-0.044$ & $0.126$ & \multicolumn{1}{r}{} & $-0.021$ &
$0.059$ & \multicolumn{1}{r}{} & $-0.015$ & $0.036$\\
$\widehat{\rho}_{L}^{(1)}$ &  & $-0.024$ & $0.109$ & \multicolumn{1}{r}{} &
$-0.009$ & $0.053$ & \multicolumn{1}{r}{} & $-0.007$ & $0.032$\\
$\widehat{\rho}_{L}^{(2)}$ &  & $-0.016$ & $0.109$ & \multicolumn{1}{r}{} &
$-0.004$ & $0.052$ & \multicolumn{1}{r}{} & $-0.003$ & $0.032$\\
$\widehat{\rho}_{J}$ &  & $-0.008$ & $0.116$ & \multicolumn{1}{r}{} & $0.001$
& $0.055$ & \multicolumn{1}{r}{} & $0.000$ & $0.032$\\
$\widehat{\rho}_{B}$ &  & $-0.004$ & $0.100$ & \multicolumn{1}{r}{} & $0.000$
& $0.052$ & \multicolumn{1}{r}{} & $-0.001$ & $0.031$\\\hline\hline
\end{tabular*}
%

\begin{footnotesize}
\justify
\end{footnotesize}%
%

\end{centering}
\end{table}%

\bigskip%

\begin{table}[H]
\begin{centering}%
\caption{Empirical Size and Power from LR Test for Probit Model, Design 1}%
\setstretch{1.2}%
%

\begin{subtable}{\textwidth}\centering
\caption{Size}%
%

\begin{tabular*}
{\linewidth}[c]{@{\extracolsep{\fill}}cccc}\hline\hline
& \multicolumn{3}{c}{Size}\\
$(N,T)$ & Uncorrected & Analytical & Bootstrap\\\hline
& \multicolumn{3}{c}{Static Probit Model}\\
$(30,30)$ & $66$ & $26$ & $2$\\
$(60,60)$ & $64$ & $13$ & $2$\\
$(90,90)$ & $66$ & $8$ & $3$\\\hline
& \multicolumn{3}{c}{Dynamic Probit Model}\\
$(30,30)$ & $61$ & $25$ & $2$\\
$(60,60)$ & $60$ & $12$ & $3$\\
$(90,90)$ & $57$ & $9$ & $2$\\\hline\hline
\end{tabular*}
%

\end{subtable}%

\bigskip%

\begin{subtable}{\textwidth}\centering
\caption{Power}%
%

\begin{tabular*}
{\linewidth}[c]{@{\extracolsep{\fill}}ccccccccc}\hline\hline
& \multicolumn{4}{c}{Power at $\delta$ (Analytical)} &
\multicolumn{4}{c}{Power at $\delta$ (Bootstrap)}\\
$(N,T)$ & $-0.2$ & $-0.1$ & $0.1$ & $0.2$ & $-0.2$ & $-0.1$ & $0.1$ &
$0.2$\\\hline
& \multicolumn{8}{c}{Static Probit Model}\\
$(30,30)$ & $99$ & $80$ & $16$ & $58$ & $85$ & $24$ & $0$ & $0$\\
$(60,60)$ & $100$ & $99$ & $83$ & $100$ & $100$ & $96$ & $0$ & $56$\\
$(90,90)$ & $100$ & $100$ & $100$ & $100$ & $100$ & $100$ & $15$ &
$100$\\\hline
& \multicolumn{8}{c}{Dynamic Probit Model}\\
$(30,30)$ & $99$ & $74$ & $28$ & $73$ & $78$ & $18$ & $1$ & $2$\\
$(60,60)$ & $100$ & $99$ & $88$ & $100$ & $100$ & $94$ & $4$ & $86$\\
$(90,90)$ & $100$ & $100$ & $100$ & $100$ & $100$ & $100$ & $50$ &
$100$\\\hline\hline
\end{tabular*}
%

\end{subtable}%

\bigskip%

\begin{footnotesize}
\justify

\noindent\textit{Notes}: See also Table
\ref{Table.Simulation.Design1TestLargeNT} where we show that the sizes for the
probit model with $\left(  N,T\right)  =\left(  120,120\right)  $ and $\left(
150,150\right)  $ are close to $5\%$.%

\end{footnotesize}%
%

\end{centering}
\end{table}%

\bigskip%

\begin{table}[H]
\begin{centering}%
\caption
{Empirical Size and Power from LM Test for Logit and Probit Model, Design 1}%
\setstretch{1.2}%
%

\begin{tabular*}
{\linewidth}[c]{@{\extracolsep{\fill}}cccccccccc}\hline\hline
&  & \multicolumn{2}{c}{Size} &  &  & \multicolumn{4}{c}{Power at $\delta$}\\
$(N,T)$ &  & Uncorrected & Analytical &  &  & $-0.2$ & $-0.1$ & $0.1$ &
$0.2$\\\hline
&  & \multicolumn{8}{c}{Static Logit Model}\\
$(30,30)$ &  & $37$ & $13$ &  &  & $87$ & $47$ & $13$ & $46$\\
$(60,60)$ &  & $40$ & $8$ &  &  & $100$ & $85$ & $59$ & $99$\\
$(90,90)$ &  & $38$ & $5$ &  &  & $100$ & $98$ & $95$ & $100$\\\hline
&  & \multicolumn{8}{c}{Static Probit Model}\\
$(30,30)$ &  & $64$ & $25$ &  &  & $99$ & $79$ & $17$ & $60$\\
$(60,60)$ &  & $63$ & $13$ &  &  & $100$ & $99$ & $84$ & $100$\\
$(90,90)$ &  & $66$ & $8$ &  &  & $100$ & $100$ & $100$ & $100$\\\hline
&  & \multicolumn{8}{c}{Dynamic Logit Model}\\
$(30,30)$ &  & $35$ & $13$ &  &  & $83$ & $41$ & $20$ & $56$\\
$(60,60)$ &  & $35$ & $9$ &  &  & $100$ & $79$ & $64$ & $99$\\
$(90,90)$ &  & $33$ & $6$ &  &  & $100$ & $98$ & $95$ & $100$\\\hline
&  & \multicolumn{8}{c}{Dynamic Probit Model}\\
$(30,30)$ &  & $58$ & $22$ &  &  & $99$ & $73$ & $28$ & $75$\\
$(60,60)$ &  & $58$ & $12$ &  &  & $100$ & $99$ & $89$ & $99$\\
$(90,90)$ &  & $56$ & $9$ &  &  & $100$ & $100$ & $100$ & $99$\\\hline\hline
\end{tabular*}
%

\begin{footnotesize}
\justify

\noindent\textit{Notes}: See also Table
\ref{Table.Simulation.Design1TestLargeNT} where we show that the sizes for the
probit model with $\left(  N,T\right)  =\left(  120,120\right)  $ and $\left(
150,150\right)  $ are close to $5\%$.%

\end{footnotesize}%
%

\end{centering}
\end{table}%

\bigskip%

\begin{table}[H]
\begin{centering}%
\caption
{Empirical Size and Power from Wald Test for Logit and Probit Model, Design 1}%
\setstretch{1.2}%
%

\begin{tabular*}
{\linewidth}[c]{@{\extracolsep{\fill}}cccccccccc}\hline\hline
&  & \multicolumn{2}{c}{Size} &  &  & \multicolumn{4}{c}{Power at $\delta$}\\
$(N,T)$ &  & Uncorrected & Analytical &  &  & $-0.2$ & $-0.1$ & $0.1$ &
$0.2$\\\hline
&  & \multicolumn{8}{c}{Static Logit Model}\\
$(30,30)$ &  & $37$ & $13$ &  &  & $87$ & $46$ & $12$ & $45$\\
$(60,60)$ &  & $40$ & $8$ &  &  & $100$ & $85$ & $59$ & $99$\\
$(90,90)$ &  & $38$ & $5$ &  &  & $100$ & $98$ & $95$ & $100$\\\hline
&  & \multicolumn{8}{c}{Static Probit Model}\\
$(30,30)$ &  & $63$ & $23$ &  &  & $98$ & $78$ & $17$ & $59$\\
$(60,60)$ &  & $63$ & $13$ &  &  & $100$ & $99$ & $84$ & $100$\\
$(90,90)$ &  & $66$ & $8$ &  &  & $100$ & $100$ & $100$ & $100$\\\hline
&  & \multicolumn{8}{c}{Dynamic Logit Model}\\
$(30,30)$ &  & $34$ & $12$ &  &  & $82$ & $40$ & $20$ & $56$\\
$(60,60)$ &  & $35$ & $9$ &  &  & $100$ & $79$ & $64$ & $99$\\
$(90,90)$ &  & $32$ & $6$ &  &  & $100$ & $98$ & $95$ & $100$\\\hline
&  & \multicolumn{8}{c}{Dynamic Probit Model}\\
$(30,30)$ &  & $55$ & $21$ &  &  & $98$ & $72$ & $29$ & $75$\\
$(60,60)$ &  & $58$ & $12$ &  &  & $99$ & $98$ & $89$ & $99$\\
$(90,90)$ &  & $56$ & $9$ &  &  & $100$ & $100$ & $100$ & $100$\\\hline\hline
\end{tabular*}
%

\begin{footnotesize}
\justify

\noindent\textit{Notes}: See also Table
\ref{Table.Simulation.Design1TestLargeNT} where we show that the sizes for the
probit model with $\left(  N,T\right)  =\left(  120,120\right)  $ and $\left(
150,150\right)  $ are close to $5\%$.%

\end{footnotesize}%
%

\end{centering}
\end{table}%

\bigskip%

\begin{table}[H]
\begin{centering}%
\caption
{Empirical Size from LR, LM, and Wald Test for Probit Model, Design 1 with $N,T=(120,120)$ and $(150,150)$}%
\label{Table.Simulation.Design1TestLargeNT}%
\setstretch{1.2}%
\scriptsize
%

\begin{tabular*}
{\linewidth}[c]{@{\extracolsep{\fill}}cccccccc}\hline\hline
&  & \multicolumn{2}{c}{LR Test} & \multicolumn{2}{c}{LM Test} &
\multicolumn{2}{c}{Wald Test}\\
$(N,T)$ &  & Uncorrected & Analytical & Uncorrected & Analytical &
Uncorrected & Analytical\\\hline
&  & \multicolumn{6}{c}{Static Probit Model}\\
$(120,120)$ &  & $63$ & $6$ & $63$ & $6$ & $63$ & $6$\\
$(150,150)$ &  & $64$ & $5$ & $63$ & $5$ & $63$ & $5$\\\hline
&  & \multicolumn{6}{c}{Dynamic Probit Model}\\
$(120,120)$ &  & $59$ & $8$ & $59$ & $8$ & $59$ & $8$\\
$(150,150)$ &  & $56$ & $7$ & $56$ & $7$ & $56$ & $7$\\\hline\hline
\end{tabular*}
%

\begin{footnotesize}
\justify
\end{footnotesize}%
%

\end{centering}
\end{table}%

\bigskip%

\begin{table}[H]
\begin{centering}%
\caption
{Comparisons of Uncorrected and Corrected Logit and Probit Estimates, Design 1 with $N=90$ and $T=10$}%
\label{Table.Simulation.Design1UnequalNT}%
\setstretch{1.2}%
\footnotesize
%

\begin{tabular*}
{\linewidth}[c]{@{\extracolsep{\fill}}ccccccccccc}\hline\hline
&  & \multicolumn{2}{c}{Probit} & \multicolumn{2}{c}{Logit} &  &
\multicolumn{2}{c}{Probit} & \multicolumn{2}{c}{Logit}\\
&  & Bias & RMSE & Bias & RMSE &  & Bias & RMSE & Bias & RMSE\\\hline
&  & \multicolumn{4}{c}{Static Model, $\theta_{0}=\beta_{0}=0$} &  &
\multicolumn{4}{c}{Static Model, $\theta_{0}=\beta_{0}=0.5$}\\
$\widehat{\beta}$ &  & \multicolumn{1}{r}{$0.056$} &
\multicolumn{1}{r}{$2.821$} & \multicolumn{1}{r}{$0.074$} &
\multicolumn{1}{r}{$1.029$} &  & \multicolumn{1}{r}{$4.048$} &
\multicolumn{1}{r}{$6.310$} & \multicolumn{1}{r}{$1.742$} &
\multicolumn{1}{r}{$2.523$}\\
$\widehat{\beta}_{L}$ &  & \multicolumn{1}{r}{$0.039$} &
\multicolumn{1}{r}{$1.048$} & \multicolumn{1}{r}{$0.084$} &
\multicolumn{1}{r}{$1.030$} &  & \multicolumn{1}{r}{$0.623$} &
\multicolumn{1}{r}{$1.577$} & \multicolumn{1}{r}{$0.299$} &
\multicolumn{1}{r}{$1.029$}\\\hline
&  & \multicolumn{4}{c}{Dynamic Model, $\theta_{0}=\left(  0,0\right)  $} &  &
\multicolumn{4}{c}{Dynamic Model, $\theta_{0}=\left(  0.5,0.5\right)  $}\\
$\widehat{\beta}$ &  & \multicolumn{1}{r}{$0.104$} &
\multicolumn{1}{r}{$2.501$} & \multicolumn{1}{r}{$0.081$} &
\multicolumn{1}{r}{$1.342$} &  & \multicolumn{1}{r}{$3.931$} &
\multicolumn{1}{r}{$6.008$} & \multicolumn{1}{r}{$1.953$} &
\multicolumn{1}{r}{$2.982$}\\
$\widehat{\beta}_{L}$ &  & \multicolumn{1}{r}{$0.045$} &
\multicolumn{1}{r}{$0.733$} & \multicolumn{1}{r}{$0.039$} &
\multicolumn{1}{r}{$0.467$} &  & \multicolumn{1}{r}{$0.520$} &
\multicolumn{1}{r}{$1.324$} & \multicolumn{1}{r}{$0.326$} &
\multicolumn{1}{r}{$1.348$}\\
$\widehat{\rho}$ &  & \multicolumn{1}{r}{$-0.331$} &
\multicolumn{1}{r}{$0.354$} & \multicolumn{1}{r}{$-0.537$} &
\multicolumn{1}{r}{$0.572$} &  & \multicolumn{1}{r}{$-0.266$} &
\multicolumn{1}{r}{$0.298$} & \multicolumn{1}{r}{$-0.465$} &
\multicolumn{1}{r}{$0.507$}\\
$\widehat{\rho}_{L}$ &  & \multicolumn{1}{r}{$-0.161$} &
\multicolumn{1}{r}{$0.197$} & \multicolumn{1}{r}{$-0.230$} &
\multicolumn{1}{r}{$0.284$} &  & \multicolumn{1}{r}{$-0.161$} &
\multicolumn{1}{r}{$0.222$} & \multicolumn{1}{r}{$-0.240$} &
\multicolumn{1}{r}{$0.305$}\\\hline\hline
\end{tabular*}

\bigskip%

\begin{footnotesize}
\justify

\noindent\textit{Notes}: $\left(  N,T\right)  =\left(  90,10\right)  $. For
the dynamic model we set $\tau=1$, producing corrected estimators
$\widehat{\rho}_{L}$ and $\widehat{\beta}_{L}$ for $\rho_{0}$ and $\beta_{0}$ respectively.%

\end{footnotesize}%
%

\end{centering}
\end{table}%

\subsubsection{Heterogeneous Autoregressive Parameters}

\label{Section.AdditionalSim.Poisson}%

In this section, we present simulation studies for models with two-way
heterogeneous autoregressive coefficient. We consider three models here: a
Poission model, a logit model, and a probit model. For the Poisson, we
consider a model specified according to the data-generating process
$Y_{it}\sim\operatorname*{Poisson}\left(  \lambda_{it}\right)  $ for%
\[
\lambda_{it}=\exp\left\{  \left(  \rho_{0}+\alpha_{1,i}^{0}+\gamma_{1,t}%
^{0}\right)  Y_{it-1}+(\beta_{0}+\alpha_{2,i}^{0}+\gamma_{2,t}^{0}%
)Z_{it}+\alpha_{3,i}^{0}+\gamma_{3,t}^{0}\right\}  ,
\]
with $Y_{i0}\sim\operatorname*{Poisson}\left(  \exp\left\{  (\beta_{0}%
+\alpha_{1,i}^{0})Z_{i0}+\alpha_{2,i}^{0}\right\}  \right)  $, where the
parameter of interests is $\theta_{0}=(\rho_{0},\beta_{0})^{\prime
}=(-0.5,0.5)^{\prime}$. We set $\rho_{0}=-0.5$ to keep $Y_{it}$ from exploding
as $t$ progresses. Similar to Section \ref{Section.Simulation}, we generate
$\{\alpha_{k,i}^{0},\gamma_{k,t}^{0}\}\sim\mathcal{N}(0,0.04)$ for
$k=1,\ldots,3$ and demean them afterwards\footnote{As in Section
\ref{Section.Simulation}, the individual effects are not normalized when the
model is being estimated.}. The rest of the experiment is the same as Design 1
in Section \ref{Section.Simulation}.

Table \ref{Table.Simulation.PoissonEst} presents the simulation results. We
see that the bias of the MLE $\widehat{\beta}$ of a static Poisson model is
relative small, compared to the probit and logit models above. The biases in
$\widehat{\beta}_{L}$, $\widehat{\beta}_{J}$, and $\widehat{\beta}_{B}$ are
all very small, with $\widehat{\beta}_{L}$ and $\widehat{\beta}_{B}$ very
similar and outperforming $\widehat{\beta}_{J}$. For the dynamic Poisson
model, however, the MLE $\widehat{\rho}$ is severely biased, showing a bias of
$-0.204$ for $\left(  N,T\right)  =\left(  30,30\right)  $. On the other hand,
our bias correction procedure reduces the bias considerably, delivering
$\widehat{\rho}_{L}^{(1)}$ which contains a bias of $-0.077$, roughly $1/3$ of
the bias of the MLE. Our procedure also outperforms the jackknife and the
bootstrap on $\rho$ for small samples. As the sample size increases, the
biases of all estimators decrease, as expected, with the performance of our
procedure and that of the bootstrap similar, outperforming the jackknife. For
$\left(  N,T\right)  =\left(  30,30\right)  $, we notice that the bias of
$\widehat{\rho}_{B}$ is $1.323$, which is very severe. We have double checked
and this seems to be a valid result. We think this might be a coincidence and
would like to emphasize that, as the sample size increases, the bootstrap is
an effective bias correction device. For the hypothesis tests, we only report
the results from the LR test for the dynamic model in Table
\ref{Table.Simulation.PoissonTest}. For the static model, the sizes of the
uncorrected, corrected, and the bootstrap LR test are all very close to the
nominal level of $5\%$. The patterns of the LR test are similar to those from
the probit and logit models in Section \ref{Section.Simulation}. For the LM
and Wald test, the results are comparable to the LR test, and are, therefore,
omitted here.%

\begin{table}[H]
\begin{centering}%
\caption{Comparisons of Uncorrected and Corrected Poisson Estimates}%
\label{Table.Simulation.PoissonEst}%
\setstretch{1.2}%
%

\begin{tabular*}
{\linewidth}[c]{@{\extracolsep{\fill}}lrrrrrr}\hline\hline
$(N,T)$ & \multicolumn{2}{c}{$(30,30)$} & \multicolumn{2}{c}{$(60,60)$} &
\multicolumn{2}{c}{$(90,90)$}\\
& Bias & RMSE & Bias & RMSE & Bias & RMSE\\\hline
& \multicolumn{6}{c}{Static Poisson Model}\\
$\widehat{\beta}$ & $0.016$ & $0.043$ & $0.005$ & $0.018$ & $0.003$ &
$0.012$\\
$\widehat{\beta}_{L}$ & $-0.002$ & $0.039$ & $-0.002$ & $0.018$ & $-0.001$ &
$0.011$\\
$\widehat{\beta}_{J}$ & $-0.014$ & $0.051$ & $-0.004$ & $0.019$ & $-0.002$ &
$0.012$\\
$\widehat{\beta}_{B}$ & $0.001$ & $0.039$ & $0.000$ & $0.017$ & $0.000$ &
$0.011$\\\hline
& \multicolumn{6}{c}{Dynamic Poisson Model}\\
$\widehat{\beta}$ & $0.025$ & $0.058$ & $0.008$ & $0.022$ & $0.005$ &
$0.014$\\
$\widehat{\beta}_{L}^{(1)}$ & $-0.005$ & $0.051$ & $-0.004$ & $0.021$ &
$-0.002$ & $0.014$\\
$\widehat{\beta}_{L}^{(2)}$ & $-0.005$ & $0.051$ & $-0.004$ & $0.021$ &
$-0.002$ & $0.014$\\
$\widehat{\beta}_{J}$ & $-0.060$ & $0.218$ & $-0.010$ & $0.025$ & $-0.004$ &
$0.015$\\
$\widehat{\beta}_{B}$ & $-0.003$ & $0.054$ & $0.000$ & $0.021$ & $0.000$ &
$0.014$\\
$\widehat{\rho}$ & $-0.204$ & $0.250$ & $-0.067$ & $0.075$ & $-0.041$ &
$0.045$\\
$\widehat{\rho}_{L}^{(1)}$ & $-0.077$ & $0.110$ & $-0.017$ & $0.036$ &
$-0.006$ & $0.023$\\
$\widehat{\rho}_{L}^{(2)}$ & $-0.086$ & $0.118$ & $-0.022$ & $0.038$ &
$-0.010$ & $0.022$\\
$\widehat{\rho}_{J}$ & $0.588$ & $0.821$ & $0.060$ & $0.083$ & $0.014$ &
$0.027$\\
$\widehat{\rho}_{B}$ & $1.323$ & $1.548$ & $0.024$ & $0.056$ & $0.006$ &
$0.020$\\\hline\hline
\end{tabular*}
%

\begin{footnotesize}
\justify
\end{footnotesize}%
%

\end{centering}
\end{table}%

\bigskip%

\begin{table}[H]
\begin{centering}%
\caption
{Empirical Size and Power from LR Test for Two-Way Fixed-Effect Probit Estimates}%
\label{Table.Simulation.PoissonTest}%
\setstretch{1.2}%
%

\begin{subtable}{\textwidth}\centering
\caption{Empirical Size}%
%

\begin{tabular*}
{\linewidth}[c]{@{\extracolsep{\fill}}ccccc}\hline\hline
&  & \multicolumn{3}{c}{Size}\\
$(N,T)$ &  & Uncorrected & Analytical & Bootstrap\\\hline
$(30,30)$ &  & $53$ & $25$ & $12$\\
$(60,60)$ &  & $48$ & $12$ & $5$\\
$(90,90)$ &  & $46$ & $8$ & $4$\\\hline\hline
\end{tabular*}
%

\end{subtable}%

\bigskip%

\begin{subtable}{\textwidth}\centering
\caption{Empirical Power}%
%

\begin{tabular*}
{\linewidth}[c]{@{\extracolsep{\fill}}cccccccccc}\hline\hline
&  & \multicolumn{4}{c}{Power at $\delta$ (Analytical)} &
\multicolumn{4}{c}{Power at $\delta$ (Bootstrap)}\\\hline
$(N,T)$ &  & $-0.2$ & $-0.1$ & $0.1$ & $0.2$ & $-0.2$ & $-0.1$ & $0.1$ &
$0.2$\\
$(30,30)$ &  & $99$ & $57$ & $93$ & $100$ & $76$ & $19$ & $56$ & $98$\\
$(60,60)$ &  & $99$ & $99$ & $99$ & $99$ & $100$ & $89$ & $100$ & $100$\\
$(90,90)$ &  & $99$ & $99$ & $99$ & $99$ & $99$ & $99$ & $99$ & $99$%
\\\hline\hline
\end{tabular*}
%

\end{subtable}%
%

\begin{footnotesize}
\justify
\end{footnotesize}%
%

\end{centering}
\end{table}%
%

For the two binary response models, we find that adding heterogeneities to the
autoregressive coefficient may induce some numerical difficulty. To deal with
it, we apply certain restrictions to the data-generating process. In Remark
\ref{Remark.BinRespHeteroAutoCoeff} below, we first explain the numerical
issue, followed by introducing the data-generating process and the
restrictions we apply.

\begin{remark}
[binary response model with heterogeneous autoregressive coefficient]%
\label{Remark.BinRespHeteroAutoCoeff}For binary response models with
heterogeneous autogressive coefficient, numerical difficulty may arise if the
(lagged) dependent variable contains insufficient variations. Consider a
simple binary response model%
\[
Y_{it}=%
\mathds{1}%
\{\left(  \rho_{0}+\alpha_{1,i}+\gamma_{1,t}\right)  Y_{it-1}+\alpha
_{2,i}+\gamma_{2,t}+\varepsilon_{it}>0\}.
\]
Suppose, for an individual $i$, there are too many time periods $t$ such that
$Y_{it-1}=1$. Then, the regressor vector $\left(  Y_{i1},\ldots,Y_{iT-1}%
\right)  ^{\prime}$ could be nearly collinear to the constant of $1$. This
renders the separate identification of $\alpha_{1,i}$ and $\alpha_{2,i}$
numerically difficult. Similar situations also happen if, for some $t$, there
are too many individuals with $Y_{it-1}=1$. When there are excessive zeros in
$Y_{it-1}$ for some $i$ or $t$, the identification of $\alpha_{1,i}\ $or
$\gamma_{1,t}$ is also very difficult.
\end{remark}

For the simulation, we consider a binary response model specified according to
the data-generating process%
\[
Y_{it}=%
\mathds{1}%
\{\left(  \rho_{0}+\alpha_{1,i}^{0}+\gamma_{1,t}^{0}\right)  Y_{it-1}%
+(\beta_{0}+\alpha_{2,i}^{0}+\gamma_{2,t}^{0})Z_{it}+\delta_{0}+\alpha
_{2,i}^{0}+\gamma_{2,t}^{0}+\varepsilon_{it}>0\},
\]
where, for $k=1,2,3$, $\{\alpha_{k,i}^{0},\gamma_{k,t}^{0}\}\sim
\mathcal{N}(0,0.04)$ and are demeaned; $\varepsilon_{it}$ are either standard
normal (probit) or standard logistic (logit); and $\delta_{0}$ is explained
below. All other settings are the same as in Section \ref{Section.Simulation},
except that we restrict both $0.1\leq T^{-1}\sum_{t=1}^{T}Y_{it}\leq0.6$ for
all $i$ and $0.1\leq N^{-1}\sum_{i=1}^{N}Y_{it}\leq0.6$ for all $t$. An
additional common parameter $\delta_{0}=-1$ is included in the intercept to
make it easier to generate a data set which satisfies the two
restrictions\footnote{The response $Y_{it}$ tends to contain too many ones
without $\delta_{0}=-1$. This is because $\left(  \rho_{0}+\alpha_{1,i}%
^{0}+\gamma_{1,t}^{0}\right)  Y_{it-1}$ is positive on average.}. Note that
$\delta_{0}$ is considered nuisance and is not a part of $\theta_{0}$. These
measures seem to alleviate the numerical difficulty considerably, although
they are, admittedly, somewhat ad-hoc.

Table \ref{Table.Simulation.BinRespHeteroLagEst} presents the simulation
results for both the probit and logit models with heterogeneous autoregressive
coefficient. For dynamic models, we see that $\widehat{\rho}$ exhibits larger
biases, and our bias correction procedure reduces the bias considerably. Under
the current setting, neither the jackknife nor the bootstrap perform better
than our bias correction procedure. For static models, the conclusion is the
same as in Section \ref{Section.Simulation}, except that the biases of all
estimators are slightly larger (which is likely caused by having $\delta
_{0}=-1$). Table \ref{Table.Simulation.BinRespHeteroLagTest} presents the
empirical sizes and powers from the LR tests. We see that the LR test based on
the uncorrected likelihood is fatally size-distorted. This is anticipated
because the IPP is much more severe with an additional set of heterogeneities.
Our bias correction procedure reduces the empirical type-I error risk
considerably. Similar to the results in Section \ref{Section.Simulation}, the
bootstrap reduces the empirical type-I error risk at the cost of reducing the
power of the test as well.%

\begin{table}[H]
\begin{centering}%
\caption
{Comparisons of Uncorrected and Corrected Probit and Logit Estimates with Heterogeneous Autoregressive Coefficient, Design 1}%
\label{Table.Simulation.BinRespHeteroLagEst}%
\setstretch{1.2}%
%

\begin{tabular*}
{\linewidth}[c]{@{\extracolsep{\fill}}llrrcrrcrr}\hline\hline
$(N,T)$ &  & \multicolumn{2}{c}{$(30,30)$} &  & \multicolumn{2}{c}{$(60,60)$}
&  & \multicolumn{2}{c}{$(90,90)$}\\
&  & Bias & RMSE & \multicolumn{1}{r}{} & Bias & RMSE & \multicolumn{1}{r}{} &
Bias & RMSE\\\hline
&  & \multicolumn{8}{c}{Static Logit Model}\\
$\widehat{\beta}$ &  & $0.196$ & $0.255$ & \multicolumn{1}{r}{} & $0.081$ &
$0.097$ & \multicolumn{1}{r}{} & $0.050$ & $0.059$\\
$\widehat{\beta}_{L}$ &  & $0.065$ & $0.128$ & \multicolumn{1}{r}{} & $0.018$
& $0.051$ & \multicolumn{1}{r}{} & $0.007$ & $0.030$\\
$\widehat{\beta}_{J}$ &  & $-1.344$ & $1.939$ & \multicolumn{1}{r}{} &
$-0.049$ & $0.108$ & \multicolumn{1}{r}{} & $-0.013$ & $0.033$\\
$\widehat{\beta}_{B}$ &  & $-0.143$ & $0.179$ & \multicolumn{1}{r}{} &
$-0.014$ & $0.046$ & \multicolumn{1}{r}{} & $-0.006$ & $0.029$\\\hline
& \multicolumn{1}{c}{} & \multicolumn{8}{c}{Static Probit Model}\\
$\widehat{\beta}$ &  & $0.374$ & $0.566$ & \multicolumn{1}{r}{} & $0.105$ &
$0.113$ & \multicolumn{1}{r}{} & $0.062$ & $0.066$\\
$\widehat{\beta}_{L}$ &  & $0.125$ & $0.164$ & \multicolumn{1}{r}{} & $0.038$
& $0.053$ & \multicolumn{1}{r}{} & $0.017$ & $0.027$\\
$\widehat{\beta}_{J}$ &  & $-5.362$ & $7.774$ & \multicolumn{1}{r}{} &
$-0.350$ & $1.101$ & \multicolumn{1}{r}{} & $-0.037$ & $0.059$\\
$\widehat{\beta}_{B}$ &  & $-0.704$ & $0.923$ & \multicolumn{1}{r}{} &
$-0.032$ & $0.043$ & \multicolumn{1}{r}{} & $-0.009$ & $0.022$\\\hline
& \multicolumn{1}{c}{} & \multicolumn{8}{c}{Dynamic Logit Model}\\
$\widehat{\beta}$ &  & $0.279$ & $0.332$ & \multicolumn{1}{r}{} & $0.106$ &
$0.119$ & \multicolumn{1}{r}{} & $0.065$ & $0.073$\\
$\widehat{\beta}_{L}^{(1)}$ &  & $0.101$ & $0.160$ & \multicolumn{1}{r}{} &
$0.026$ & $0.054$ & \multicolumn{1}{r}{} & $0.011$ & $0.032$\\
$\widehat{\beta}_{L}^{(2)}$ &  & $0.105$ & $0.164$ & \multicolumn{1}{r}{} &
$0.027$ & $0.055$ & \multicolumn{1}{r}{} & $0.012$ & $0.032$\\
$\widehat{\beta}_{J}$ &  & $-2.345$ & $3.017$ & \multicolumn{1}{r}{} &
$-0.085$ & $0.138$ & \multicolumn{1}{r}{} & $-0.020$ & $0.037$\\
$\widehat{\beta}_{B}$ &  & $-0.369$ & $0.407$ &  & $-0.029$ & $0.051$ &  &
$-0.011$ & $0.030$\\
$\widehat{\rho}$ &  & $-0.546$ & $0.725$ &  & $-0.120$ & $0.169$ &  & $-0.059$
& $0.089$\\
$\widehat{\rho}_{L}^{(1)}$ &  & $-0.194$ & $0.396$ &  & $-0.050$ & $0.127$ &
& $-0.022$ & $0.063$\\
$\widehat{\rho}_{L}^{(2)}$ &  & $-0.193$ & $0.404$ &  & $-0.047$ & $0.127$ &
& $-0.019$ & $0.063$\\
$\widehat{\rho}_{J}$ &  & $-0.363$ & $6.295$ &  & $0.305$ & $0.577$ &  &
$0.099$ & $0.267$\\
$\widehat{\rho}_{B}$ &  & $0.551$ & $0.734$ &  & $0.160$ & $0.186$ &  &
$0.043$ & $0.071$\\\hline
& \multicolumn{1}{c}{} & \multicolumn{8}{c}{Dynamic Probit Model}\\
$\widehat{\beta}$ &  & $0.500$ & $0.700$ & \multicolumn{1}{r}{} & $0.130$ &
$0.137$ & \multicolumn{1}{r}{} & $0.079$ & $0.083$\\
$\widehat{\beta}_{L}^{(1)}$ &  & $0.179$ & $0.215$ & \multicolumn{1}{r}{} &
$0.046$ & $0.059$ & \multicolumn{1}{r}{} & $0.023$ & $0.032$\\
$\widehat{\beta}_{L}^{(2)}$ &  & $0.189$ & $0.261$ & \multicolumn{1}{r}{} &
$0.046$ & $0.077$ & \multicolumn{1}{r}{} & $0.023$ & $0.032$\\
$\widehat{\beta}_{J}$ &  & $-7.236$ & $9.499$ & \multicolumn{1}{r}{} &
$-0.499$ & $0.932$ & \multicolumn{1}{r}{} & $-0.057$ & $0.104$\\
$\widehat{\beta}_{B}$ &  & $-2.386$ & $4.786$ & \multicolumn{1}{r}{} &
$-0.062$ & $0.069$ & \multicolumn{1}{r}{} & $-0.018$ & $0.027$\\
$\widehat{\rho}$ &  & $-0.518$ & $0.781$ & \multicolumn{1}{r}{} & $-0.173$ &
$0.208$ & \multicolumn{1}{r}{} & $-0.083$ & $0.106$\\
$\widehat{\rho}_{L}^{(1)}$ &  & $-0.226$ & $0.485$ & \multicolumn{1}{r}{} &
$-0.039$ & $0.259$ & \multicolumn{1}{r}{} & $-0.024$ & $0.155$\\
$\widehat{\rho}_{L}^{(2)}$ &  & $-0.213$ & $0.533$ & \multicolumn{1}{r}{} &
$-0.037$ & $0.261$ & \multicolumn{1}{r}{} & $-0.023$ & $0.167$\\
$\widehat{\rho}_{J}$ &  & $-4.299$ & $12.866$ & \multicolumn{1}{r}{} &
$-0.066$ & $2.459$ & \multicolumn{1}{r}{} & $0.116$ & $1.033$\\
$\widehat{\rho}_{B}$ &  & $-1.621$ & $3.122$ &  & $0.126$ & $0.186$ &  &
$0.091$ & $0.105$\\\hline\hline
\end{tabular*}
%

\begin{footnotesize}
\justify
\end{footnotesize}%
%

\end{centering}
\end{table}%

\bigskip%

\begin{table}[H]
\begin{centering}%
\caption
{Empirical Size and Power from LR Test for Probit and Logit Models with Heterogeneous Autoregressive Coefficient, Design 1}%
\label{Table.Simulation.BinRespHeteroLagTest}%

\begin{subtable}{\textwidth}\centering
\caption{Empirical Size}%
\setstretch{1.2}%
%

\begin{tabular*}
{\linewidth}[c]{@{\extracolsep{\fill}}ccccc}\hline\hline
&  & \multicolumn{3}{c}{Size}\\
$(N,T)$ &  & Uncorrected & Analytical & Bootstrap\\\hline
&  & \multicolumn{3}{c}{Static Logit Model}\\
$(30,30)$ &  & $38$ & $16$ & $1$\\
$(60,60)$ &  & $41$ & $11$ & $3$\\
$(90,90)$ &  & $39$ & $7$ & $2$\\\hline
&  & \multicolumn{3}{c}{Static Probit Model}\\
$(30,30)$ &  & $78$ & $41$ & $9$\\
$(60,60)$ &  & $81$ & $26$ & $3$\\
$(90,90)$ &  & $79$ & $15$ & $3$\\\hline
&  & \multicolumn{3}{c}{Dynamic Logit Model}\\
$(30,30)$ &  & $62$ & $37$ & $8$\\
$(60,60)$ &  & $61$ & $16$ & $2$\\
$(90,90)$ &  & $55$ & $10$ & $2$\\\hline
&  & \multicolumn{3}{c}{Dynamic Probit Model}\\
$(30,30)$ &  & $93$ & $78$ & $52$\\
$(60,60)$ &  & $93$ & $45$ & $7$\\
$(90,90)$ &  & $94$ & $30$ & $2$\\\hline\hline
\end{tabular*}
%

\end{subtable}%

\bigskip%

\begin{subtable}{\textwidth}\centering
\caption{Empirical Power}%
\setstretch{1.2}%
\footnotesize
%

\begin{tabular*}
{\linewidth}[c]{@{\extracolsep{\fill}}cccccccccc}\hline\hline
&  & \multicolumn{4}{c}{Power at $\delta$ (Analytical)} &
\multicolumn{4}{c}{Power at $\delta$ (Bootstrap)}\\
$(N,T)$ &  & $-0.2$ & $-0.1$ & $0.1$ & $0.2$ & $-0.2$ & $-0.1$ & $0.1$ &
$0.2$\\\hline
&  & \multicolumn{8}{c}{Static Logit Model}\\
$(30,30)$ &  & $83$ & $44$ & $13$ & $37$ & $36$ & $8$ & $1$ & $1$\\
$(60,60)$ &  & $100$ & $77$ & $49$ & $98$ & $100$ & $62$ & $1$ & $21$\\
$(90,90)$ &  & $100$ & $97$ & $91$ & $100$ & $100$ & $95$ & $6$ & $91$\\\hline
&  & \multicolumn{8}{c}{Static Probit Model}\\
$(30,30)$ &  & $99$ & $85$ & $12$ & $27$ & $61$ & $19$ & $5$ & $4$\\
$(60,60)$ &  & $100$ & $99$ & $53$ & $99$ & $100$ & $87$ & $0$ & $5$\\
$(90,90)$ &  & $100$ & $100$ & $97$ & $100$ & $100$ & $100$ & $1$ &
$88$\\\hline
&  & \multicolumn{8}{c}{Dynamic Logit Model}\\
$(30,30)$ &  & $80$ & $54$ & $39$ & $59$ & $33$ & $14$ & $5$ & $6$\\
$(60,60)$ &  & $100$ & $77$ & $59$ & $99$ & $98$ & $42$ & $1$ & $20$\\
$(90,90)$ &  & $100$ & $97$ & $93$ & $100$ & $100$ & $87$ & $6$ & $90$\\\hline
&  & \multicolumn{8}{c}{Dynamic Probit Model}\\
$(30,30)$ &  & $99$ & $91$ & $67$ & $71$ & $81$ & $62$ & $44$ & $40$\\
$(60,60)$ &  & $100$ & $98$ & $76$ & $100$ & $100$ & $81$ & $2$ & $11$\\
$(90,90)$ &  & $100$ & $100$ & $98$ & $100$ & $100$ & $99$ & $2$ &
$84$\\\hline\hline
\end{tabular*}
%

\end{subtable}%
%

\begin{footnotesize}
\justify
\end{footnotesize}%
%

\end{centering}
\end{table}%

\subsubsection{Probit and Logit under Design 2}

\label{Section.AdditionalSim.Design2}

In this section, we provide the simulation results under the same settings as
in Section \ref{Section.Simulation}, except that $\{\alpha_{k,i}^{0}%
,\gamma_{k,t}^{0}\}\sim\mathcal{N}(0,0.25)$, $Z_{it}\sim\mathcal{N}%
(\alpha_{1,i}^{0}+\alpha_{2,i}^{0}+\gamma_{1,t}^{0}+\gamma_{2,t}^{0},1)$, and
$Z_{i0}\sim\mathcal{N}(\alpha_{1,i}^{0}+\alpha_{2,i}^{0},1)$.%

\begin{table}[H]
\begin{centering}%
\caption
{Comparisons of Uncorrected and Corrected Logit and Probit Estimates, Design 2}%
\label{Table.Simulation.Design2Est}%
\setstretch{1.2}%
\tiny
%

\begin{tabular*}
{\linewidth}[c]{@{\extracolsep{\fill}}crrrrrrrrrr}\hline\hline
& \multicolumn{2}{c}{$(30,30)$} & \multicolumn{2}{c}{$(60,60)$} &
\multicolumn{2}{c}{$(90,90)$} & \multicolumn{2}{c}{$(120,120)$} &
\multicolumn{2}{c}{$(150,150)$}\\
\multicolumn{1}{l}{} & Bias & RMSE & Bias & RMSE & Bias & RMSE & Bias & RMSE &
Bias & RMSE\\\hline
\multicolumn{1}{l}{} & \multicolumn{10}{c}{Static Logit Model}\\
\multicolumn{1}{l}{$\widehat{\beta}$} & $0.45$ & $1.20$ & $0.10$ & $0.13$ &
$0.06$ & $0.07$ & $0.08$ & $0.05$ & $0.07$ & $0.04$\\
\multicolumn{1}{l}{$\widehat{\beta}_{L}$} & $0.10$ & $0.20$ & $0.02$ & $0.06$
& $0.01$ & $0.04$ & $0.01$ & $0.03$ & $0.01$ & $0.02$\\\hline
\multicolumn{1}{l}{} & \multicolumn{10}{c}{Static Probit Model}\\
\multicolumn{1}{l}{$\widehat{\beta}$} & $0.94$ & $2.28$ & $0.15$ & $0.25$ &
$0.07$ & $0.08$ & $0.10$ & $0.05$ & $0.08$ & $0.04$\\
\multicolumn{1}{l}{$\widehat{\beta}_{L}$} & $0.33$ & $5.17$ & $0.04$ & $0.06$
& $0.02$ & $0.04$ & $0.02$ & $0.03$ & $0.01$ & $0.02$\\\hline
\multicolumn{1}{l}{} & \multicolumn{10}{c}{Dynamic Logit Model}\\
\multicolumn{1}{l}{$\widehat{\beta}$} & $0.53$ & $1.58$ & $0.11$ & $0.21$ &
$0.06$ & $0.07$ & $0.08$ & $0.05$ & $0.06$ & $0.04$\\
\multicolumn{1}{l}{$\widehat{\beta}_{L}^{(1)}$} & $0.12$ & $0.20$ & $0.02$ &
$0.06$ & $0.01$ & $0.04$ & $0.01$ & $0.03$ & $0.01$ & $0.02$\\
\multicolumn{1}{l}{$\widehat{\beta}_{L}^{(2)}$} & $0.13$ & $0.28$ & $0.03$ &
$0.06$ & $0.01$ & $0.04$ & $0.01$ & $0.03$ & $0.01$ & $0.02$\\
\multicolumn{1}{l}{$\widehat{\rho}$} & $-0.11$ & $0.24$ & $-0.06$ & $0.11$ &
$-0.04$ & $0.07$ & $-0.05$ & $0.05$ & $-0.04$ & $0.04$\\
\multicolumn{1}{l}{$\widehat{\rho}_{L}^{(1)}$} & $-0.04$ & $0.19$ & $-0.02$ &
$0.09$ & $-0.01$ & $0.06$ & $-0.01$ & $0.04$ & $-0.01$ & $0.03$\\
\multicolumn{1}{l}{$\widehat{\rho}_{L}^{(2)}$} & $-0.04$ & $0.20$ & $-0.01$ &
$0.09$ & $-0.01$ & $0.06$ & $0.00$ & $0.04$ & $0.00$ & $0.03$\\\hline
\multicolumn{1}{l}{} & \multicolumn{10}{c}{Dynamic Probit Model}\\
\multicolumn{1}{l}{$\widehat{\beta}$} & $0.91$ & $1.84$ & $0.17$ & $0.25$ &
$0.08$ & $0.10$ & $0.11$ & $0.06$ & $0.08$ & $0.04$\\
\multicolumn{1}{l}{$\widehat{\beta}_{L}^{(1)}$} & $0.21$ & $0.40$ & $0.05$ &
$0.08$ & $0.02$ & $0.04$ & $0.02$ & $0.03$ & $0.01$ & $0.03$\\
\multicolumn{1}{l}{$\widehat{\beta}_{L}^{(2)}$} & $0.20$ & $0.28$ & $0.05$ &
$0.08$ & $0.02$ & $0.04$ & $0.03$ & $0.03$ & $0.02$ & $0.04$\\
\multicolumn{1}{l}{$\widehat{\rho}$} & $-0.03$ & $0.16$ & $-0.02$ & $0.07$ &
$-0.01$ & $0.05$ & $-0.02$ & $0.03$ & $-0.02$ & $0.03$\\
\multicolumn{1}{l}{$\widehat{\rho}_{L}^{(1)}$} & $-0.02$ & $0.14$ & $-0.01$ &
$0.07$ & $-0.00$ & $0.04$ & $-0.01$ & $0.04$ & $-0.01$ & $0.03$\\
\multicolumn{1}{l}{$\widehat{\rho}_{L}^{(2)}$} & $-0.01$ & $0.14$ & $-0.00$ &
$0.07$ & $-0.00$ & $0.05$ & $0.00$ & $0.04$ & $-0.01$ & $0.04$\\\hline\hline
\end{tabular*}
%

\begin{footnotesize}
\justify
\end{footnotesize}%
%

\end{centering}
\end{table}%

\bigskip%

\begin{table}[H]
\begin{centering}%
\caption
{Empirical Size and Power from LR Test for Logit and Probit Model, Design 2}%
\setstretch{1.2}%
%

\begin{tabular*}
{\linewidth}[c]{@{\extracolsep{\fill}}cccccccccc}\hline\hline
&  & \multicolumn{2}{c}{Size} &  &  & \multicolumn{4}{c}{Power at $\delta$}\\
$(N,T)$ &  & Uncorrected & Analytical &  &  & $-0.2$ & $-0.1$ & $0.1$ &
$0.2$\\\hline
&  & \multicolumn{8}{c}{Static Logit Model}\\
$(30,30)$ &  & $47$ & $20$ &  &  & $79$ & $45$ & $12$ & $26$\\
$(60,60)$ &  & $43$ & $10$ &  &  & $100$ & $73$ & $40$ & $93$\\
$(90,90)$ &  & $42$ & $9$ &  &  & $100$ & $94$ & $81$ & $100$\\
$(120,120)$ &  & $40$ & $7$ &  &  & $100$ & $99$ & $97$ & $99$\\
$(150,150)$ &  & $39$ & $5$ &  &  & $100$ & $100$ & $100$ & $100$\\\hline
&  & \multicolumn{8}{c}{Static Probit Model}\\
$(30,30)$ &  & $77$ & $39$ &  &  & $97$ & $77$ & $14$ & $15$\\
$(60,60)$ &  & $76$ & $20$ &  &  & $100$ & $96$ & $34$ & $83$\\
$(90,90)$ &  & $72$ & $14$ &  &  & $100$ & $99$ & $85$ & $96$\\
$(120,120)$ &  & $72$ & $10$ &  &  & $99$ & $99$ & $98$ & $98$\\
$(150,150)$ &  & $72$ & $9$ &  &  & $99$ & $99$ & $99$ & $99$\\\hline
&  & \multicolumn{8}{c}{Dynamic Logit Model}\\
$(30,30)$ &  & $43$ & $18$ &  &  & $75$ & $41$ & $19$ & $35$\\
$(60,60)$ &  & $38$ & $9$ &  &  & $100$ & $69$ & $44$ & $96$\\
$(90,90)$ &  & $34$ & $7$ &  &  & $100$ & $93$ & $86$ & $100$\\
$(120,120)$ &  & $33$ & $6$ &  &  & $100$ & $99$ & $98$ & $100$\\
$(150,150)$ &  & $31$ & $5$ &  &  & $100$ & $100$ & $100$ & $100$\\\hline
&  & \multicolumn{8}{c}{Dynamic Probit Model}\\
$(30,30)$ &  & $71$ & $37$ &  &  & $95$ & $74$ & $25$ & $38$\\
$(60,60)$ &  & $66$ & $19$ &  &  & $100$ & $93$ & $47$ & $96$\\
$(90,90)$ &  & $63$ & $14$ &  &  & $100$ & $100$ & $90$ & $100$\\
$(120,120)$ &  & $64$ & $11$ &  &  & $99$ & $99$ & $99$ & $99$\\
$(150,150)$ &  & $62$ & $8$ &  &  & $99$ & $99$ & $99$ & $99$\\\hline\hline
\end{tabular*}
%

\begin{footnotesize}
\justify
\end{footnotesize}%
%

\end{centering}
\end{table}%

\bigskip%

\begin{table}[H]
\begin{centering}%
\caption
{Empirical Size and Power from LM Test for Logit and Probit Model, Design 2}%
\setstretch{1.2}%
%

\begin{tabular*}
{\linewidth}[c]{@{\extracolsep{\fill}}cccccccccc}\hline\hline
&  & \multicolumn{2}{c}{Size} &  &  & \multicolumn{4}{c}{Power at $\delta$}\\
$(N,T)$ &  & Uncorrected & Analytical &  &  & $-0.2$ & $-0.1$ & $0.1$ &
$0.2$\\\hline
&  & \multicolumn{8}{c}{Static Logit Model}\\
$(30,30)$ &  & $42$ & $18$ &  &  & $77$ & $42$ & $12$ & $28$\\
$(60,60)$ &  & $42$ & $9$ &  &  & $99$ & $72$ & $41$ & $93$\\
$(90,90)$ &  & $41$ & $9$ &  &  & $100$ & $94$ & $82$ & $100$\\
$(120,120)$ &  & $40$ & $7$ &  &  & $100$ & $99$ & $97$ & $100$\\
$(150,150)$ &  & $39$ & $5$ &  &  & $100$ & $99$ & $100$ & $100$\\\hline
&  & \multicolumn{8}{c}{Static Probit Model}\\
$(30,30)$ &  & $69$ & $34$ &  &  & $96$ & $73$ & $12$ & $17$\\
$(60,60)$ &  & $74$ & $18$ &  &  & $100$ & $96$ & $37$ & $76$\\
$(90,90)$ &  & $71$ & $14$ &  &  & $100$ & $99$ & $86$ & $88$\\
$(120,120)$ &  & $72$ & $10$ &  &  & $99$ & $99$ & $96$ & $91$\\
$(150,150)$ &  & $72$ & $9$ &  &  & $99$ & $100$ & $98$ & $94$\\\hline
&  & \multicolumn{8}{c}{Dynamic Logit Model}\\
$(30,30)$ &  & $38$ & $16$ &  &  & $73$ & $38$ & $19$ & $36$\\
$(60,60)$ &  & $37$ & $9$ &  &  & $100$ & $68$ & $45$ & $94$\\
$(90,90)$ &  & $34$ & $7$ &  &  & $100$ & $93$ & $86$ & $99$\\
$(120,120)$ &  & $33$ & $6$ &  &  & $100$ & $99$ & $98$ & $99$\\
$(150,150)$ &  & $31$ & $5$ &  &  & $100$ & $100$ & $100$ & $99$\\\hline
&  & \multicolumn{8}{c}{Dynamic Probit Model}\\
$(30,30)$ &  & $59$ & $30$ &  &  & $94$ & $70$ & $21$ & $38$\\
$(60,60)$ &  & $62$ & $17$ &  &  & $100$ & $92$ & $48$ & $91$\\
$(90,90)$ &  & $60$ & $13$ &  &  & $100$ & $99$ & $90$ & $94$\\
$(120,120)$ &  & $62$ & $10$ &  &  & $99$ & $99$ & $97$ & $96$\\
$(150,150)$ &  & $62$ & $7$ &  &  & $99$ & $99$ & $99$ & $96$\\\hline\hline
\end{tabular*}
%

\begin{footnotesize}
\justify
\end{footnotesize}%
%

\end{centering}
\end{table}%

\bigskip%

\begin{table}[H]
\begin{centering}%
\caption
{Empirical Size and Power from Wald Test for Logit and Probit Model, Design 2}%
\setstretch{1.2}%
%

\begin{tabular*}
{\linewidth}[c]{@{\extracolsep{\fill}}cccccccccc}\hline\hline
&  & \multicolumn{2}{c}{Size} &  &  & \multicolumn{4}{c}{Power at $\delta$}\\
$(N,T)$ &  & Uncorrected & Analytical &  &  & $-0.2$ & $-0.1$ & $0.1$ &
$0.2$\\\hline
&  & \multicolumn{8}{c}{Static Logit Model}\\
$(30,30)$ &  & $32$ & $15$ &  &  & $75$ & $39$ & $10$ & $29$\\
$(60,60)$ &  & $41$ & $9$ &  &  & $100$ & $72$ & $41$ & $95$\\
$(90,90)$ &  & $42$ & $9$ &  &  & $100$ & $94$ & $81$ & $100$\\
$(120,120)$ &  & $40$ & $7$ &  &  & $100$ & $99$ & $97$ & $100$\\
$(150,150)$ &  & $39$ & $5$ &  &  & $100$ & $100$ & $100$ & $100$\\\hline
&  & \multicolumn{8}{c}{Static Probit Model}\\
$(30,30)$ &  & $37$ & $29$ &  &  & $91$ & $68$ & $10$ & $19$\\
$(60,60)$ &  & $71$ & $18$ &  &  & $99$ & $95$ & $39$ & $92$\\
$(90,90)$ &  & $72$ & $14$ &  &  & $99$ & $99$ & $88$ & $99$\\
$(120,120)$ &  & $72$ & $10$ &  &  & $99$ & $99$ & $98$ & $99$\\
$(150,150)$ &  & $72$ & $9$ &  &  & $99$ & $99$ & $99$ & $99$\\\hline
&  & \multicolumn{8}{c}{Dynamic Logit Model}\\
$(30,30)$ &  & $27$ & $15$ &  &  & $69$ & $34$ & $18$ & $36$\\
$(60,60)$ &  & $36$ & $8$ &  &  & $100$ & $68$ & $45$ & $97$\\
$(90,90)$ &  & $34$ & $7$ &  &  & $100$ & $92$ & $86$ & $100$\\
$(120,120)$ &  & $33$ & $6$ &  &  & $100$ & $99$ & $98$ & $100$\\
$(150,150)$ &  & $31$ & $5$ &  &  & $100$ & $100$ & $100$ & $100$\\\hline
&  & \multicolumn{8}{c}{Dynamic Probit Model}\\
$(30,30)$ &  & $25$ & $26$ &  &  & $88$ & $60$ & $21$ & $39$\\
$(60,60)$ &  & $56$ & $16$ &  &  & $100$ & $91$ & $49$ & $98$\\
$(90,90)$ &  & $60$ & $13$ &  &  & $99$ & $99$ & $91$ & $100$\\
$(120,120)$ &  & $64$ & $10$ &  &  & $99$ & $99$ & $98$ & $98$\\
$(150,150)$ &  & $62$ & $7$ &  &  & $99$ & $99$ & $99$ & $99$\\\hline\hline
\end{tabular*}
%

\begin{footnotesize}
\justify
\end{footnotesize}%
%

\end{centering}
\end{table}%

\bigskip%

\begin{table}[H]
\begin{centering}%
\caption
{Comparisons of Uncorrected, Corrected, and Jackknife Estimates for Logit and Probit, Design 2}%
\label{Table.Simulation.Design2Jack}%
\setstretch{1.2}%
%

\begin{tabular*}
{\linewidth}[c]{@{\extracolsep{\fill}}llrrcrrcrr}\hline\hline
$(N,T)$ &  & \multicolumn{2}{c}{$(30,30)$} &  & \multicolumn{2}{c}{$(60,60)$}
&  & \multicolumn{2}{c}{$(90,90)$}\\
&  & Bias & RMSE & \multicolumn{1}{r}{} & Bias & RMSE & \multicolumn{1}{r}{} &
Bias & RMSE\\\hline
&  & \multicolumn{8}{c}{Static Logit Model}\\
$\widehat{\beta}$ &  & $0.400$ & $1.435$ & \multicolumn{1}{r}{} & $0.101$ &
$0.125$ & \multicolumn{1}{r}{} & $0.060$ & $0.072$\\
$\widehat{\beta}_{L}$ &  & $0.093$ & $0.188$ & \multicolumn{1}{r}{} & $0.022$
& $0.056$ & \multicolumn{1}{r}{} & $0.010$ & $0.036$\\
$\widehat{\beta}_{J}$ &  & $-2.314$ & $4.328$ & \multicolumn{1}{r}{} &
$-0.172$ & $0.406$ & \multicolumn{1}{r}{} & $-0.031$ & $0.070$\\\hline
&  & \multicolumn{8}{c}{Static Probit Model}\\
$\widehat{\beta}$ &  & $0.846$ & $2.168$ & \multicolumn{1}{r}{} & $0.151$ &
$0.261$ & \multicolumn{1}{r}{} & $0.074$ & $0.082$\\
$\widehat{\beta}_{L}$ &  & $0.182$ & $0.339$ & \multicolumn{1}{r}{} & $0.041$
& $0.068$ & \multicolumn{1}{r}{} & $0.019$ & $0.033$\\
$\widehat{\beta}_{J}$ &  & $-4.929$ & $14.383$ & \multicolumn{1}{r}{} &
$-0.504$ & $0.981$ & \multicolumn{1}{r}{} & $-0.105$ & $0.214$\\\hline
& \multicolumn{1}{c}{} & \multicolumn{8}{c}{Dynamic Logit Model}\\
$\widehat{\beta}$ &  & $0.415$ & $0.858$ & \multicolumn{1}{r}{} & $0.104$ &
$0.137$ & \multicolumn{1}{r}{} & $0.061$ & $0.074$\\
$\widehat{\beta}_{L}$ &  & $0.104$ & $0.201$ & \multicolumn{1}{r}{} & $0.023$
& $0.060$ & \multicolumn{1}{r}{} & $0.012$ & $0.043$\\
$\widehat{\beta}_{J}$ &  & $-2.442$ & $4.223$ & \multicolumn{1}{r}{} &
$-0.231$ & $0.481$ & \multicolumn{1}{r}{} & $-0.043$ & $0.105$\\
$\widehat{\rho}$ &  & $-0.120$ & $0.257$ & \multicolumn{1}{r}{} & $-0.055$ &
$0.108$ & \multicolumn{1}{r}{} & $-0.036$ & $0.071$\\
$\widehat{\rho}_{L}$ &  & $-0.047$ & $0.205$ & \multicolumn{1}{r}{} & $-0.016$
& $0.089$ & \multicolumn{1}{r}{} & $-0.010$ & $0.059$\\
$\widehat{\rho}_{J}$ &  & $0.020$ & $0.218$ & \multicolumn{1}{r}{} & $0.003$ &
$0.092$ & \multicolumn{1}{r}{} & $0.002$ & $0.060$\\\hline
&  & \multicolumn{8}{c}{Dynamic Probit Model}\\
$\widehat{\beta}$ &  & $1.040$ & $2.343$ & \multicolumn{1}{r}{} & $0.184$ &
$0.287$ & \multicolumn{1}{r}{} & $0.080$ & $0.090$\\
$\widehat{\beta}_{L}$ &  & $0.209$ & $0.332$ & \multicolumn{1}{r}{} & $0.052$
& $0.082$ & \multicolumn{1}{r}{} & $0.020$ & $0.049$\\
$\widehat{\beta}_{J}$ &  & $-3.352$ & $7.708$ & \multicolumn{1}{r}{} &
$-0.558$ & $1.125$ & \multicolumn{1}{r}{} & $-0.118$ & $0.217$\\
$\widehat{\rho}$ &  & $-0.046$ & $0.169$ & \multicolumn{1}{r}{} & $-0.022$ &
$0.072$ & \multicolumn{1}{r}{} & $-0.014$ & $0.046$\\
$\widehat{\rho}_{L}$ &  & $-0.028$ & $0.145$ & \multicolumn{1}{r}{} & $-0.011$
& $0.067$ & \multicolumn{1}{r}{} & $-0.009$ & $0.053$\\
$\widehat{\rho}_{J}$ &  & $-0.029$ & $0.169$ & \multicolumn{1}{r}{} & $-0.006$
& $0.068$ & \multicolumn{1}{r}{} & $-0.002$ & $0.044$\\\hline\hline
\end{tabular*}

\bigskip%

\begin{footnotesize}
\justify

\noindent\textit{Notes}: Simulated separately from Table
\ref{Table.Simulation.Design2Est}. For dynamic models, We set $\tau=1$ in
$L\left(  \theta\right)  $, producing corrected estimators $\widehat{\rho}%
_{L}$ and $\widehat{\beta}_{L}$ for $\rho_{0}$ and $\beta_{0}$ respectively.%

\end{footnotesize}%
%

\end{centering}
\end{table}%

\subsubsection{Two-way Fixed-Effect Probit}

\label{Section.AdditionalSim.FVW2016}%

In this section, we consider a dynamic probit model with two-way scalar fixed
effects appearing only in the intercept. This is a special case of the two-way
DN-HP model. We take Design 1 in \cite{fw2016} and simulate our method, the
jackknife, and the bootstrap here. We set $N=56$ and $T\in\left\{
14,28,56\right\}  $, and consider%
\[
Y_{it}=%
\mathds{1}%
\left\{  \rho_{0}Y_{it-1}+\beta_{0}Z_{it}+\alpha_{i}^{0}+\gamma_{t}%
^{0}+\varepsilon_{it}>0\right\}
\]
with $Y_{i0}=%
\mathds{1}%
\left\{  \beta_{0}Z_{i0}+\alpha_{i}^{0}+\gamma_{0}^{0}+\varepsilon
_{i0}>0\right\}  $, where $\rho_{0}=0.5$, $\beta_{0}=1$, $\varepsilon_{it}$
and $\varepsilon_{i0}$ are standard normal, $\left\{  \alpha_{i}^{0}%
,\gamma_{t}^{0},\gamma_{0}^{0}\right\}  \sim\mathcal{N}\left(
0,0.0625\right)  $, and $Z_{it}=Z_{it-1}/2+\alpha_{i0}+\gamma_{t0}+\nu_{it}$
with $\nu_{it}\sim\mathcal{N}\left(  0,1/2\right)  $ and $Z_{i0}%
\sim\mathcal{N}\left(  0,1\right)  $. We set $\tau\in\left\{  1,2\right\}  $.
For the hypothesis tests, we only present the results from the LR tests, using
the uncorrected likelihood $\widehat{l}\left(  \theta\right)  $, the corrected
likelihood $L\left(  \theta\right)  $ (setting $\tau=1$), and the bootstrap.
We set the number of repetitions to $499$ for the bootstrap and run $1000$
replications. The rest are the same as in Section \ref{Section.Simulation}.

Table \ref{Table.Simulation.FVW2016Est} presents the simulation results.
First, the biases and RMSEs of $\widehat{\beta}$, $\widehat{\beta}_{J}$,
$\widehat{\rho}$, and $\widehat{\rho}_{J}$ are in line with the original
simulations of \cite{fw2016}. For instance for $\left(  N,T\right)  =\left(
56,14\right)  $, the bias of $\widehat{\beta}$ is $0.182$, roughly $18\%$
relative to the true value of $1$, and its RMSE is $0.221$, roughly $22\%$.
For $\widehat{\beta}_{J}$, the bias is $-8\%$ and the RMSE is $11\%$. These
are very close to the results of \cite{fw2016}. As for the corrected
estimators, we find that the bootstrap estimators $\widehat{\beta}_{B}$ and
$\widehat{\rho}_{B}$ are generally the best, outperforming the jackknife
estimators $\widehat{\beta}_{J}$ and $\widehat{\rho}_{J}$ and our corrected
estimators $\widehat{\beta}_{L}^{\left(  \tau\right)  }$ and $\widehat{\rho
}_{L}^{\left(  \tau\right)  }$. The biases of $\widehat{\beta}_{B}$ and
$\widehat{\rho}_{B}$ are also comparable to the parameter-based procedure of
\cite{fw2016}. Our procedure outperforms the jackknife in estimating
$\beta_{0}$ (but not $\rho_{0}$) when the sample size is small. However, the
differences between all estimators (except the MLE) become small as the sample
size increases. While all being analytical corrections, our corrected
estimator performs slightly worse than the parameter-based corrected estimator
of \cite{fw2016}. This can be anticipated, because our procedure produces the
bias correction effect only indirectly via the likelihood, whereas the
parameter-based approach applies directly to the estimator. However, we would
like to point out that the procedure of \cite{fw2016} does not apply to the
DN-HP model without modifications.%

\begin{table}[H]
\begin{centering}%
\caption
{Comparisons of Uncorrected and Corrected Two-Way Fixed-Effect Probit Estimates}%
\label{Table.Simulation.FVW2016Est}%
\setstretch{1.2}%
%

\begin{tabular*}
{\linewidth}[c]{@{\extracolsep{\fill}}llrrcrrcrr}\hline\hline
$(N,T)$ &  & \multicolumn{2}{c}{$(56,14)$} &  & \multicolumn{2}{c}{$(56,28)$}
&  & \multicolumn{2}{c}{$(56,56)$}\\
&  & Bias & RMSE & \multicolumn{1}{r}{} & Bias & RMSE & \multicolumn{1}{r}{} &
Bias & RMSE\\\hline
$\widehat{\beta}$ &  & $0.182$ & $0.221$ & \multicolumn{1}{r}{} & $0.093$ &
$0.120$ & \multicolumn{1}{r}{} & $0.062$ & $0.078$\\
$\widehat{\beta}_{L}^{(1)}$ &  & $0.063$ & $0.156$ & \multicolumn{1}{r}{} &
$0.020$ & $0.073$ & \multicolumn{1}{r}{} & $0.012$ & $0.047$\\
$\widehat{\beta}_{L}^{(2)}$ &  & $0.076$ & $0.159$ & \multicolumn{1}{r}{} &
$0.022$ & $0.075$ & \multicolumn{1}{r}{} & $0.012$ & $0.047$\\
$\widehat{\beta}_{J}$ &  & $-0.083$ & $0.151$ & \multicolumn{1}{r}{} &
$-0.028$ & $0.081$ & \multicolumn{1}{r}{} & $-0.007$ & $0.047$\\
$\widehat{\beta}_{B}$ &  & $-0.045$ & $0.105$ & \multicolumn{1}{r}{} &
$-0.017$ & $0.069$ & \multicolumn{1}{r}{} & $-0.002$ & $0.045$\\\hline
$\widehat{\rho}$ &  & $-0.226$ & $0.266$ & \multicolumn{1}{r}{} & $-0.098$ &
$0.134$ & \multicolumn{1}{r}{} & $-0.047$ & $0.081$\\
$\widehat{\rho}_{L}^{(1)}$ &  & $-0.100$ & $0.166$ & \multicolumn{1}{r}{} &
$-0.030$ & $0.090$ & \multicolumn{1}{r}{} & $-0.016$ & $0.065$\\
$\widehat{\rho}_{L}^{(2)}$ &  & $-0.102$ & $0.171$ & \multicolumn{1}{r}{} &
$-0.024$ & $0.091$ & \multicolumn{1}{r}{} & $-0.011$ & $0.065$\\
$\widehat{\rho}_{J}$ &  & $0.046$ & $0.161$ & \multicolumn{1}{r}{} & $0.015$ &
$0.095$ & \multicolumn{1}{r}{} & $0.000$ & $0.067$\\
$\widehat{\rho}_{B}$ &  & $0.015$ & $0.131$ & \multicolumn{1}{r}{} & $0.010$ &
$0.088$ & \multicolumn{1}{r}{} & $-0.001$ & $0.064$\\\hline\hline
\end{tabular*}
%

\begin{footnotesize}
\justify
\end{footnotesize}%
%

\end{centering}
\end{table}%

Table \ref{Table.Simulation.FVW2016Size} presents the empirical sizes and
powers of the LR test using the uncorrected likelihood (size only), our
corrected likelihood, and the bootstrap. Briefly speaking, our procedure
delivers an empirical size close to the nominal level of $5\%$ and provide a
reasonable empirical power, starting from $T=28$. On the opposite, the
empirical size from the bootstrap is close to $5\%$ for all $T\in\left\{
14,28,56\right\}  $, but its power is relatively low, especially near the true hypothesis.%

\begin{table}[H]
\begin{centering}%
\caption
{Empirical Size and Power from LR Test for Two-Way Fixed-Effect Probit Estimates}%
\label{Table.Simulation.FVW2016Size}%
\setstretch{1.2}%
%

\begin{subtable}{\textwidth}\centering
\caption{Empirical Size}%
%

\begin{tabular*}
{\linewidth}[c]{@{\extracolsep{\fill}}ccccc}\hline\hline
&  & \multicolumn{3}{c}{Size}\\
$(N,T)$ &  & Uncorrected & Analytical & Bootstrap\\\hline
$(56,14)$ &  & $53$ & $17$ & $4$\\
$(56,28)$ &  & $30$ & $8$ & $3$\\
$(56,56)$ &  & $25$ & $7$ & $5$\\\hline\hline
\end{tabular*}
%

\end{subtable}%

\bigskip%

\begin{subtable}{\textwidth}\centering
\caption{Empirical Power}%
%

\begin{tabular*}
{\linewidth}[c]{@{\extracolsep{\fill}}ccccccccccc}\hline\hline
&  & \multicolumn{4}{c}{Power at $\delta$ (Analytical)} &  &
\multicolumn{4}{c}{Power at $\delta$ (Bootstrap)}\\\hline
$(N,T)$ &  & $-0.2$ & $-0.1$ & $0.1$ & $0.2$ &  & $-0.2$ & $-0.1$ & $0.1$ &
$0.2$\\
$(56,14)$ &  & $79$ & $37$ & $33$ & $70$ &  & $35$ & $10$ & $6$ & $19$\\
$(56,28)$ &  & $97$ & $46$ & $45$ & $95$ &  & $87$ & $25$ & $10$ & $56$\\
$(56,56)$ &  & $100$ & $79$ & $72$ & $100$ &  & $100$ & $70$ & $27$ &
$94$\\\hline\hline
\end{tabular*}
%

\end{subtable}%
%

\begin{footnotesize}
\justify
\end{footnotesize}%
%

\end{centering}
\end{table}%

\subsection{Additional Empirical Results}

\label{sec:APP}

\subsubsection{Sample}

The data set is constructed from waves 20--30 of PSID, covering the survey
years 1987 to 1997. We selected this time span because 1997 was the last year
PSID conducted annual surveys, switching to a biennial schedule in 1999. Each
survey wave includes questions about employment status in the prior calendar
year, such as, \textquotedblleft About how many weeks did (he/she) work on
that job last year?\textquotedblright\ and \textquotedblleft During the weeks
that (he/she) worked, about how many hours did (he/she) usually work per
week?\textquotedblright\ We use responses to these questions to construct
individual labor force participation (LFP) status for the prior year. Thus,
wave 30 (1997) provides LFP data for 1996, wave 29 (1996) for 1995, and so on.
In our final sample, labor force participation status is observed from 1987 to
1996, with lagged participation status observed from 1986 to 1995.

We study labor force participation among female household heads with children
who are not legally married. In the PSID, marital status is reported as
married, widowed, divorced, separated, or single; women reporting any of the
latter four categories are classified as unmarried. Because cohabitation
status is not systematically identified in the PSID, the term
\textquotedblleft single mothers\textquotedblright\ in this paper refers to
unmarried female household heads with children and may include a small number
of cohabiting women.

\begin{center}
\begin{figure}[H]
\caption{Labor Force Participation Contingency Table}
\label{fig:lfpct}
\centering\includegraphics[width=0.6\columnwidth]{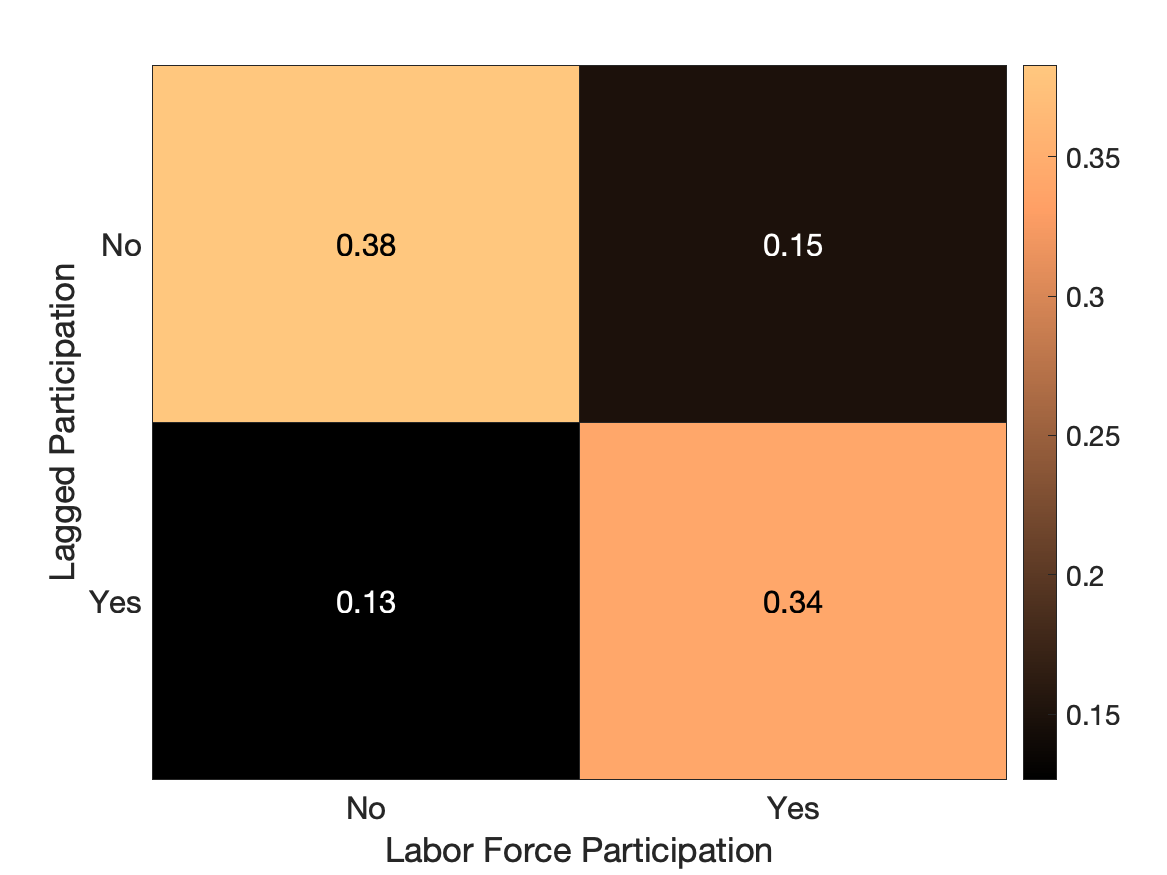}
\vspace{1mm}
\begin{flushleft}
\begin{justify}
\begin{footnotesize}
\noindent\textit{Notes}%
: Each cell contains the percentage of individuals with given current and past labor force participation status.
\end{footnotesize}
\end{justify}
\end{flushleft}
\end{figure}
\end{center}

\begin{center}
\begin{figure}[H]
\caption{Labor Force Participation Status Transition Rates By Year}
\label{fig:lfpflow}
\centering\includegraphics[width=0.7\columnwidth]{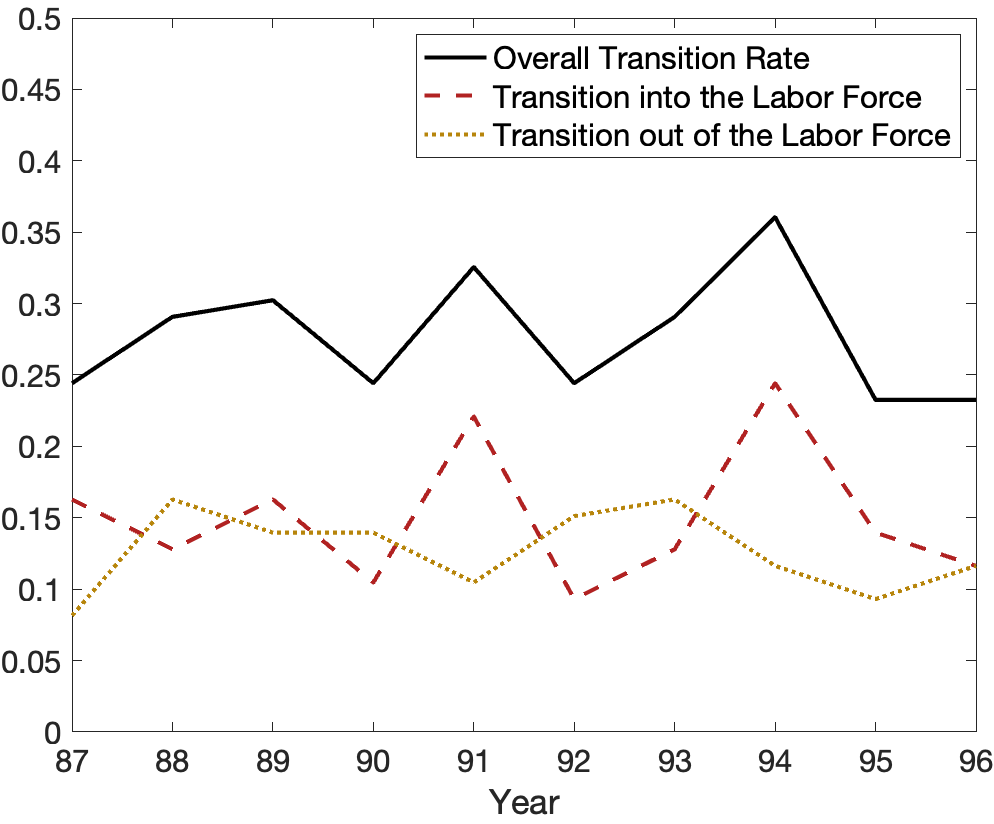}
\end{figure}
\end{center}

We further restrict the data to an informative sample of individuals whose
current and lagged LFP status each varied at least once during the sample
period. This ensures within-person variation in both variables, which is
necessary for identification of the dynamic model. The final sample contains
$N=86$ single mothers observed for $T=10$ consecutive years. Table
\ref{tab:DS} reports the summary statistics. Figure \ref{fig:lfpct} displays
the joint distribution of current and lagged LFP status, and Figure
\ref{fig:lfpflow} plots the year-specific transition rates between employment
and non-employment.%

\setlength{\tabcolsep}{10pt}
\begin{table}[H]
\centering\begin{threeparttable}
\caption{Descriptive Statistics\protect\label{tab:DS}}
\begin{tabular}{lcccc}
\toprule\toprule Variable & Mean & Std. Dev. & Min & Max\tabularnewline
\midrule
Labor Force Participation & 0.491 & 0.500 & 0.000 & 1.000\tabularnewline
Lagged Participation & 0.467 & 0.499 & 0.000 & 1.000\tabularnewline
Number of Children & 2.25 & 1.104 & 1.000 & 6.000\tabularnewline
\bottomrule\end{tabular}
\vspace{1mm}
\begin{tablenotes}
\footnotesize\item\emph{Notes:}
Summary statistics are computed based on a  sample of 86 individuals observed from 1987 to 1996. All variables have 860 observations. Data source: PSID 1987--1997.
\end{tablenotes}
\end{threeparttable}
\end{table}

\subsubsection{Extra Empirical Results}%

\begin{table}[H]
\centering\begin{threeparttable}
\caption{Estimated Heterogeneous Effects on Labor Force Participation}
\protect\label{tab:M2d}
\begin{tabular}{lccccccc}
\hline\hline&  &  &  & 10th & 25th & 75th & 90th\tabularnewline
& Mean & Median & Std. Dev. & Percentile & Percentile & Percentile & Percentile\tabularnewline
\hline$\alpha_{i}%
$ & 0.000 & 0.012 & 2.329 & -2.477 & -1.858 & 1.883 & 3.521\tabularnewline
$\gamma_{t}%
$ & 0.000 & -0.125 & 0.450 & -0.435 & -0.361 & 0.248 & 0.692\tabularnewline
$\beta_{it}%
$ & 0.436 & 0.388 & 2.372 & -2.186 & -1.487 & 2.216 & 3.768\tabularnewline
\hline\end{tabular}
\end{threeparttable}
\end{table}
\begin{table}[H]
\centering\begin{threeparttable}
\caption
{Estimated Heterogeneous State Dependence in Labor Force Participation}
\begin{tabular}{lccccccc}
\hline\hline&  &  &  & 10th & 25th & 75th & 90th\tabularnewline
& Mean & Median & Std. Dev. & Percentile & Percentile & Percentile & Percentile\tabularnewline
\hline$\zeta_{i}%
$ & 0.000 & -0.021 & 2.359 & -2.831 & -1.766 & 1.730 & 3.353\tabularnewline
$\eta_{t}%
$ & 0.000 & -0.149 & 0.812 & -0.926 & -0.708 & 0.245 & 1.358\tabularnewline
$\rho_{it}%
$ & 0.253 & 0.083 & 2.495 & -2.812 & -1.635 & 2.090 & 3.725\tabularnewline
\hline\end{tabular}
\end{threeparttable}
\end{table}%

\subsection{Average Partial Effects}

\label{Section.APE}%

We discuss the average partial effects (APEs) in this section. Denote%
\[
\Delta\left(  \theta,\phi\right)  :=\frac{1}{NT}\sum_{i=1}^{N}\sum_{t=1}%
^{T}\Delta_{i,t}\left(  \theta,\alpha_{i},\gamma_{t}\right)
\]
where $\Delta_{i,t}\left(  \theta,\alpha_{i},\gamma_{t}\right)  $ is, for
simplicity, a scalar partial effect function of interests. For the probit
model in Section \ref{Section.Simulation}, $\Delta_{i,t}\left(  \theta
,\alpha_{i},\gamma_{t}\right)  $ could be, e.g.,%
\[
\Delta_{i,t}\left(  \theta,\alpha_{i},\gamma_{t}\right)  :=F\left(
\rho+Z_{it}\left(  \beta+\alpha_{1,i}+\gamma_{1,t}\right)  +\alpha
_{2,i}+\gamma_{2,t}\right)  -F\left(  Z_{it}\left(  \beta+\alpha_{1,i}%
+\gamma_{1,t}\right)  +\alpha_{2,i}+\gamma_{2,t}\right)  ,
\]
which is the partial effect of the binary variable $Y_{it-1}$, where
$\theta=\left(  \rho,\beta\right)  $ and $F\left(  \cdot\right)  $ is the
standard normal cumulative distribution function. Let $\overline{\mathbb{E}}$
denote the unconditional expectation w.r.t. the joint density of $\left(
Y_{it},X_{it}^{\prime},\phi^{0\prime}\right)  ^{\prime}$, $\Delta
:=\Delta(\theta_{0},\phi^{0})$, and $\overline{\Delta}:=\overline{\mathbb{E}%
}\Delta$. For nonlinear models with two-way fixed effects in the intercept
only, \cite{fw2016} show that the APE estimator $\widehat{\Delta}%
:=\Delta(\widehat{\theta},\widehat{\phi}(\widehat{\theta}))$ is
\emph{asymptotically} unbiased. We briefly re-visit this result here and
justify that it also applies to the DN-HP models.

Consider the decomposition%
\[
\widehat{\Delta}-\overline{\Delta}=(\widehat{\Delta}-\Delta)+(\Delta
-\overline{\Delta}).
\]
Here $\widehat{\Delta}-\Delta$ is due to the estimation error of the
parameters. As demonstrated by \cite{fw2016}, $\widehat{\Delta}$ is affected
by the IPP in that it possesses a bias of order $1/\sqrt{NT}$ away from
$\Delta$. On the other hand, $\Delta-\overline{\Delta}$ is due to the
estimation of the population mean $\overline{\Delta}$ by the sample mean
$\Delta$. The component $\Delta$ does not suffer from the IPP. However, its
convergence rate depends on the data-generating process and could potentially
be slower than $1/\sqrt{NT}$. To see this, consider $\left\{  \alpha_{i}%
^{0},\gamma_{t}^{0}\right\}  $ being i.i.d. across both $i$ and $t$.
Calculating the variance of $(\Delta-\overline{\Delta})$, we have%
\[
\overline{\mathbb{E}}(\Delta-\overline{\Delta})^{2}=\frac{1}{N^{2}T^{2}}%
\sum_{i=1}^{N}\sum_{j=1}^{N}\sum_{t=1}^{T}\sum_{s=1}^{T}\overline{\mathbb{E}%
}\left[  \widetilde{\Delta}_{i,t}(\theta_{0},\alpha_{i}^{0},\gamma_{t}%
^{0})\widetilde{\Delta}_{j,s}(\theta_{0},\alpha_{j}^{0},\gamma_{s}%
^{0})\right]  ,
\]
where $\widetilde{\Delta}_{i,t}\left(  \theta,\alpha_{i},\gamma_{t}\right)
:=\Delta_{i,t}\left(  \theta,\alpha_{i},\gamma_{t}\right)  -\overline
{\mathbb{E}}\Delta_{i,t}\left(  \theta,\alpha_{i},\gamma_{t}\right)  $. Here,
$\widetilde{\Delta}_{i,t}(\theta_{0},\alpha_{i}^{0},\gamma_{t}^{0})$ and
$\widetilde{\Delta}_{j,t}(\theta_{0},\alpha_{j}^{0},\gamma_{t}^{0})$, for
$i\neq j$, depend on the same $\gamma_{t}^{0}$; while $\widetilde{\Delta
}_{i,t}(\theta_{0},\alpha_{i}^{0},\gamma_{t}^{0})$ and $\widetilde{\Delta
}_{i,s}(\theta_{0},\alpha_{i}^{0},\gamma_{s}^{0})$, for $t\neq s$, depend on
the same $\alpha_{i}^{0}$. Since $\alpha_{i}^{0}$ and $\gamma_{t}^{0}$ are
held as random variables in the APE context, $\widetilde{\Delta}_{i,t}%
(\theta_{0},\alpha_{i}^{0},\gamma_{t}^{0})$ exhibits dependencies across both
$i$ or $t$. As a consequence, the rate of the variance $\overline{\mathbb{E}%
}(\Delta-\overline{\Delta})^{2}$ may be slower than $1/NT$. In other words,
the bias $\widehat{\Delta}-\Delta$ is negligible relative to the variance of
$\Delta-\overline{\Delta}$, implying $\widehat{\Delta}$ being asymptotically
unbiased towards $\overline{\Delta}$. Of course, there still might be some
finite-sample bias. For the two-way DN-HP model here, applying the same
expansion as in Section \ref{Section.IPP} to $\Delta\left(  \theta
,\phi\right)  $\ would establish that the bias $\widehat{\Delta}-\Delta$ is
also of order $1/\sqrt{NT}$. Therefore, the same arguments above still apply.

Next, we\ conduct simulation studies to demonstrate that estimating the APEs
using $\widehat{\theta}_{L}$ already leads to a considerable reduction of its
finite-sample bias. Under the same settings in Section
\ref{Section.Simulation}, Table \ref{Table.Simulation.Design1APE} compares the
biases and RMSEs of $\widehat{\Delta}^{Z}:=\Delta^{Z}(\widehat{\theta
},\widehat{\phi}(\widehat{\theta}))$ v.s. $\widehat{\Delta}_{L}^{Z}%
:=\Delta^{Z}(\widehat{\theta}_{L},\widehat{\phi}(\widehat{\theta}_{L}))$ and
$\widehat{\Delta}^{Y}:=\Delta^{Y}(\widehat{\theta},\widehat{\phi
}(\widehat{\theta}))$ v.s. $\widehat{\Delta}_{L}^{Y}:=\Delta^{Y}%
(\widehat{\theta}_{L},\widehat{\phi}(\widehat{\theta}_{L}))$, where%
\begin{align*}
\Delta^{Z}(\theta,\phi)  &  :=\frac{1}{NT}\sum_{i=1}^{N}\sum_{t=1}^{T}%
\Delta_{i,t}^{Z}\left(  \theta,\alpha_{i},\gamma_{t}\right)  ,\qquad\Delta
^{Y}(\theta,\phi):=\frac{1}{NT}\sum_{i=1}^{N}\sum_{t=1}^{T}\Delta_{i,t}%
^{Y}\left(  \theta,\alpha_{i},\gamma_{t}\right)  ,\\
\Delta_{i,t}^{Z}\left(  \theta,\alpha_{i},\gamma_{t}\right)   &  :=f\left(
\rho Y_{it-1}+Z_{it}\left(  \beta+\alpha_{1,i}+\gamma_{1,t}\right)
+\alpha_{2,i}+\gamma_{2,t}\right)  \left(  \beta+\alpha_{1,i}+\gamma
_{1,t}\right)  ,\\
\Delta_{i,t}^{Y}\left(  \theta,\alpha_{i},\gamma_{t}\right)   &  :=F\left(
\rho+Z_{it}\left(  \beta+\alpha_{1,i}+\gamma_{1,t}\right)  +\alpha
_{2,i}+\gamma_{2,t}\right)  -F\left(  Z_{it}\left(  \beta+\alpha_{1,i}%
+\gamma_{1,t}\right)  +\alpha_{2,i}+\gamma_{2,t}\right)  ,
\end{align*}
$F\left(  \cdot\right)  $ is the cumulative distribution function of a
standard normal (for the probit model) or a standard logistic (for the logit
model) distribution, and $f\left(  \cdot\right)  $ is the probability density
function corresponding to $F\left(  \cdot\right)  $. The biases and RMSEs are
calculated relative to the Monte-Carlo means of the infeasible APEs
$\Delta^{Z}:=\Delta^{Z}(\theta_{0},\phi^{0})$ and $\Delta^{Y}:=\Delta
^{Y}(\theta_{0},\phi^{0})$, respectively, and the biases are reported in
percentages relative to the Monte-Carlo means of $\Delta^{Z}$ and $\Delta^{Y}$
(for better comparison). Finally, we only report the APE estimators calculated
using $\widehat{\beta}_{L}^{(1)}$ and $\widehat{\rho}_{L}^{(1)}$ (omitting
those using $\widehat{\beta}_{L}^{(2)}$ and $\widehat{\rho}_{L}^{(2)}$ because
they are similar). For the static logit model with $(N,T)=(30,30)$, we find
that the bias of the plug-in estimator $\widehat{\Delta}_{L}^{Z}$ is only
$-2\%$, while the uncorrected APE estimator $\widehat{\Delta}^{Z}$ shows a
$16\%$ bias. The RMSE of $\widehat{\Delta}_{L}^{Z}$ is also smaller than that
of $\widehat{\Delta}^{Z}$. As the sample size increases, the RMSEs of both
estimators decrease, with $\widehat{\Delta}_{L}^{Z}$ outperforming
$\widehat{\Delta}^{Z}$ constantly. The results here suggest that estimating
the APEs using $\widehat{\theta}_{L}$ already reduces the finite-sample bias
without inflating the RMSE, compared to using $\widehat{\theta}$ directly,
even that $\widehat{\Delta}_{L}^{Z}$ and $\widehat{\Delta}_{L}^{Y}$ are
intrinsically not bias-corrected estimators.%

\begin{table}[H]
\begin{centering}%
\caption
{Comparisons of Average Partial Effects for Logit and Probit Models, Design 1}%
\label{Table.Simulation.Design1APE}%
\setstretch{1.2}%
%

\begin{tabular*}
{\linewidth}[c]{@{\extracolsep{\fill}}lrrrrrr}\hline\hline
$(N,T)$ & \multicolumn{2}{c}{$(30,30)$} & \multicolumn{2}{c}{$(60,60)$} &
\multicolumn{2}{c}{$(90,90)$}\\
& Bias & RMSE & Bias & RMSE & Bias & RMSE\\\hline
& \multicolumn{6}{c}{Static Logit Model}\\
$\widehat{\Delta}^{Z}$ & $16$ & $0.028$ & $7$ & $0.012$ & $4$ & $0.008$\\
$\widehat{\Delta}_{L}^{Z}$ & $-2$ & $0.018$ & $-3$ & $0.009$ & $-3$ &
$0.006$\\\hline
& \multicolumn{6}{c}{Static Probit Model}\\
$\widehat{\Delta}^{Z}$ & $18$ & $0.057$ & $7$ & $0.014$ & $4$ & $0.009$\\
$\widehat{\Delta}_{L}^{Z}$ & $1$ & $0.019$ & $-2$ & $0.008$ & $-2$ &
$0.006$\\\hline
& \multicolumn{6}{c}{Dynamic Logit Model}\\
$\widehat{\Delta}^{Z}$ & $15$ & $0.027$ & $6$ & $0.011$ & $5$ & $0.008$\\
$\widehat{\Delta}_{L}^{Z}$ & $-2$ & $0.018$ & $-3$ & $0.009$ & $-2$ &
$0.006$\\
$\widehat{\Delta}^{Y}$ & $-32$ & $0.049$ & $-16$ & $0.024$ & $-10$ & $0.016$\\
$\widehat{\Delta}_{L}^{Y}$ & $-19$ & $0.038$ & $-9$ & $0.019$ & $-5$ &
$0.012$\\\hline
& \multicolumn{6}{c}{Dynamic Probit Model}\\
$\widehat{\Delta}^{Z}$ & $26$ & $0.121$ & $8$ & $0.015$ & $5$ & $0.010$\\
$\widehat{\Delta}_{L}^{Z}$ & $4$ & $0.027$ & $0$ & $0.008$ & $-1$ & $0.005$\\
$\widehat{\Delta}^{Y}$ & $-21$ & $0.048$ & $-10$ & $0.022$ & $-6$ & $0.015$\\
$\widehat{\Delta}_{L}^{Y}$ & $-15$ & $0.039$ & $-6$ & $0.018$ & $-3$ &
$0.012$\\\hline\hline
\end{tabular*}
%

\begin{footnotesize}
\justify

\noindent\textit{Notes}: Bias is reported in percentage relative to the
Monte-Carlo mean of the infeasible APEs $\Delta^{Z}$ or $\Delta^{Y}$. For the
dynamic models, $\widehat{\Delta}_{L}^{Z}$ or $\widehat{\Delta}_{L}^{Y}$ are
calculated using $\widehat{\beta}_{L}^{(1)}$ and $\widehat{\rho}_{L}^{(1)}$.%

\end{footnotesize}%
%

\end{centering}
\end{table}%

\subsection{The Failure of the Jackknife}

\label{Section.Jackknife}

In our simulation studies, we find that the jackknife does not perform well in
reducing the bias of structural parameters associated with two-way
heterogeneities for small $N,T$. This could be due to the incompatibility of
the jackknife with the normalization for the identification. Note that the
normalization is crucial for the identification under the DN-HP model. In this
section, we discuss the possible cause of the failure of the jackknife.

Consider even $N,T$ and a simple probit model%
\[
Y_{it}=%
\mathds{1}%
\left\{  \left(  \beta+\alpha_{i}+\gamma_{t}\right)  X_{it}+\varepsilon
_{it}>0\right\}  ,
\]
where $X_{it}$ is a scalar regressor and $\left\{  \beta,\alpha_{i},\gamma
_{t}\right\}  $ are all scalar parameters. For simplicity, we assume that
there is no intercept, which does not affect our discussion here. Recall that
we set $N^{-1}\sum_{i=1}^{N}\alpha_{i}^{0}=T^{-1}\sum_{t=1}^{T}\gamma_{t}%
^{0}=0$ in the data-generating process and normalize $N^{-1}\sum_{i=1}%
^{N}\alpha_{i}=T^{-1}\sum_{t=1}^{T}\gamma_{t}=0$ during the estimation. The
normalization causes no problem on the full data set. However, this is not the
case for half panels. To see the problem, define%
\begin{align*}
\mathcal{P}_{N,T/2}^{\left(  1\right)  }  &  :=\left\{  \left(  i,t\right)
:i=1,\ldots,N;t=1,\ldots,T/2\right\}  ,\\
\mathcal{P}_{N,T/2}^{\left(  2\right)  }  &  :=\left\{  \left(  i,t\right)
:i=1,\ldots,N;t=T/2+1,\ldots,T\right\}  ,\\
\mathcal{P}_{N/2,T}^{\left(  1\right)  }  &  :=\left\{  \left(  i,t\right)
:i=1,\ldots,N/2;t=1,\ldots,T\right\}  ,\\
\mathcal{P}_{N/2,T}^{\left(  2\right)  }  &  :=\left\{  \left(  i,t\right)
:i=N/2+1,\ldots,N;t=1,\ldots,T\right\}  ,
\end{align*}
where $\mathcal{P}_{N,T/2}^{\left(  1\right)  }$ and $\mathcal{P}%
_{N,T/2}^{\left(  2\right)  }$ are the two half panels obtained by splitting
along the time dimension, and $\mathcal{P}_{N/2,T}^{\left(  1\right)  }$ and
$\mathcal{P}_{N/2,T}^{\left(  2\right)  }$ are constructed by splitting along
the individual dimension. Focus on the two half panels $\mathcal{P}%
_{N,T/2}^{\left(  1\right)  }$ and $\mathcal{P}_{N,T/2}^{\left(  2\right)  }$,
and define $g_{T/2}^{\left(  1\right)  }:=\left(  T/2\right)  ^{-1}\sum
_{t=1}^{T/2}\gamma_{t}^{0}$ and $g_{T/2}^{\left(  2\right)  }:=\left(
T/2\right)  ^{-1}\sum_{T/2+1}^{T}\gamma_{t}^{0}$. Here $g_{T/2}^{\left(
1\right)  }$ is the summation of $\gamma_{t}^{0}$ over the time periods in
$\mathcal{P}_{N,T/2}^{\left(  1\right)  }$ and $g_{T/2}^{\left(  2\right)  }$
similarly over $\mathcal{P}_{N,T/2}^{\left(  2\right)  }$. Note that
$g_{T/2}^{\left(  1\right)  }+g_{T/2}^{\left(  2\right)  }=0$ but they are
generally nonzero themselves. When the model is estimated on $\mathcal{P}%
_{N,T/2}^{\left(  1\right)  }$, the normalization $\sum_{t=1}^{T/2}\gamma
_{t}=0$ is imposed. However, this normalization does not agree with
$g_{T/2}^{\left(  1\right)  }\neq0$. Consequently, the effect from
$g_{T/2}^{\left(  1\right)  }$ is absorbed into the common parameter. In other
words, the MLE computed on $\mathcal{P}_{N,T/2}^{\left(  1\right)  }$, say
$\widehat{\beta}_{N,T/2}^{\left(  1\right)  }$, is essentially estimating
$\beta_{N,T/2}^{\left(  1\right)  }=\beta_{0}+g_{T/2}^{\left(  1\right)  }$.
The same issue happens on $\mathcal{P}_{N,T/2}^{\left(  2\right)  }$, causing
the MLE obtained on $\mathcal{P}_{N,T/2}^{\left(  2\right)  }$, say
$\widehat{\beta}_{N,T/2}^{\left(  2\right)  }$, to be estimating
$\beta_{N,T/2}^{\left(  2\right)  }=\beta_{0}+g_{T/2}^{\left(  2\right)  }$.
Note that $\beta_{N,T/2}^{\left(  1\right)  }\neq\beta_{N,T/2}^{\left(
2\right)  }$ generally but $(\beta_{N,T/2}^{\left(  1\right)  }+\beta
_{N,T/2}^{\left(  2\right)  })/2=\beta_{0}$. The same happens on
$\mathcal{P}_{N/2,T}^{\left(  1\right)  }$ and $\mathcal{P}_{N/2,T}^{\left(
2\right)  }$, causing the corresponding MLEs, say $\widehat{\beta}%
_{N/2,T}^{\left(  1\right)  }$ and $\widehat{\beta}_{N/2,T}^{\left(  2\right)
}$, to be estimating some $\beta_{N/2,T}^{\left(  1\right)  }$ and
$\beta_{N/2,T}^{\left(  2\right)  }$ generally different from $\beta_{0}$.

Now, suppose that the MLE admits the expansion%
\begin{equation}
\widehat{\beta}-\beta_{0}=\frac{b_{\alpha}\left(  \beta_{0}\right)  }{T}%
+\frac{b_{\gamma}\left(  \beta_{0}\right)  }{N}+o_{\mathbb{P}}\left(
N^{-1}\vee T^{-1}\right)  \label{Eq.Jackknife.MLEExpansion}%
\end{equation}
for some bias terms $b_{\alpha}\left(  \beta\right)  $ and $b_{\gamma}\left(
\beta\right)  $, as in \cite{fw2016}. On half panels, because the estimated
targets are generally different, we would have%
\begin{align*}
\widehat{\beta}_{N,T/2}^{\left(  1\right)  }-\beta_{N,T/2}^{\left(  1\right)
}  &  =\frac{b_{\alpha}(\beta_{N,T/2}^{\left(  1\right)  })}{T/2}%
+\frac{b_{\gamma}(\beta_{N,T/2}^{\left(  1\right)  })}{N}+o_{\mathbb{P}%
}\left(  N^{-1}\vee T^{-1}\right)  ,\\
\widehat{\beta}_{N,T/2}^{\left(  2\right)  }-\beta_{N,T/2}^{\left(  2\right)
}  &  =\frac{b_{\alpha}(\beta_{N,T/2}^{\left(  2\right)  })}{T/2}%
+\frac{b_{\gamma}(\beta_{N,T/2}^{\left(  2\right)  })}{N}+o_{\mathbb{P}%
}\left(  N^{-1}\vee T^{-1}\right)  ,\\
\widehat{\beta}_{N/2,T}^{\left(  1\right)  }-\beta_{N/2,T}^{\left(  1\right)
}  &  =\frac{b_{\alpha}(\beta_{N/2,T}^{\left(  1\right)  })}{T}+\frac
{b_{\gamma}(\beta_{N/2,T}^{\left(  1\right)  })}{N/2}+o_{\mathbb{P}}\left(
N^{-1}\vee T^{-1}\right)  ,\\
\widehat{\beta}_{N/2,T}^{\left(  2\right)  }-\beta_{N/2,T}^{\left(  2\right)
}  &  =\frac{b_{\alpha}(\beta_{N/2,T}^{\left(  2\right)  })}{T}+\frac
{b_{\gamma}(\beta_{N/2,T}^{\left(  2\right)  })}{N/2}+o_{\mathbb{P}}\left(
N^{-1}\vee T^{-1}\right)  .
\end{align*}
Since $\beta_{0}$, $\beta_{N,T/2}^{\left(  1\right)  }$, $\beta_{N,T/2}%
^{\left(  2\right)  }$, $\beta_{N/2,T}^{\left(  1\right)  }$, and
$\beta_{N/2,T}^{\left(  2\right)  }$ generally differ from each other, the
five $b_{\alpha}\left(  \cdot\right)  $ terms would also be different. The
same happens to $b_{\gamma}\left(  \cdot\right)  $. Next, define%
\begin{align*}
\widehat{\beta}_{N,T/2}  &  :=\frac{1}{2}\left(  \widehat{\beta}%
_{N,T/2}^{\left(  1\right)  }+\widehat{\beta}_{N,T/2}^{\left(  2\right)
}\right) \\
&  =\beta_{0}+\frac{b_{\alpha}(\beta_{N,T/2}^{\left(  1\right)  })+b_{\alpha
}(\beta_{N,T/2}^{\left(  2\right)  })}{T}+\frac{b_{\gamma}(\beta
_{N,T/2}^{\left(  1\right)  })+b_{\gamma}(\beta_{N,T/2}^{\left(  2\right)  }%
)}{2N}+o_{\mathbb{P}}\left(  N^{-1}\vee T^{-1}\right)  ,\\
\widehat{\beta}_{N/2,T}  &  :=\frac{1}{2}\left(  \widehat{\beta}%
_{N/2,T}^{\left(  1\right)  }+\widehat{\beta}_{N/2,T}^{\left(  2\right)
}\right) \\
&  =\beta_{0}+\frac{b_{\alpha}(\beta_{N/2,T}^{\left(  1\right)  })+b_{\alpha
}(\beta_{N/2,T}^{\left(  2\right)  })}{2T}+\frac{b_{\gamma}(\beta
_{N/2,T}^{\left(  1\right)  })+b_{\gamma}(\beta_{N/2,T}^{\left(  2\right)  }%
)}{N}+o_{\mathbb{P}}\left(  N^{-1}\vee T^{-1}\right)  .
\end{align*}
Noting (\ref{Eq.Jackknife.MLEExpansion}) we have%
\begin{align*}
\widehat{\beta}_{N,T/2}-\widehat{\beta}  &  =\frac{b_{\alpha}(\beta
_{N,T/2}^{\left(  1\right)  })+b_{\alpha}(\beta_{N,T/2}^{\left(  2\right)
})-b_{\alpha}\left(  \beta_{0}\right)  }{T}+\frac{b_{\gamma}(\beta
_{N,T/2}^{\left(  1\right)  })+b_{\gamma}(\beta_{N,T/2}^{\left(  2\right)
})-2b_{\gamma}\left(  \beta_{0}\right)  }{2N}\\
&  \quad\quad+o_{\mathbb{P}}\left(  N^{-1}\vee T^{-1}\right)  ,\\
\widehat{\beta}_{N/2,T}-\widehat{\beta}  &  =\frac{b_{\alpha}(\beta
_{N/2,T}^{\left(  1\right)  })+b_{\alpha}(\beta_{N/2,T}^{\left(  2\right)
})-2b_{\alpha}\left(  \beta_{0}\right)  }{2T}+\frac{b_{\gamma}(\beta
_{N/2,T}^{\left(  1\right)  })+b_{\gamma}(\beta_{N/2,T}^{\left(  2\right)
})-b_{\gamma}\left(  \beta_{0}\right)  }{N}\\
&  \quad\quad+o_{\mathbb{P}}\left(  N^{-1}\vee T^{-1}\right)  .
\end{align*}
Given the jackknife estimator $\widehat{\beta}_{J}:=3\widehat{\beta
}-\widehat{\beta}_{N,T/2}-\widehat{\beta}_{N/2,T}$, we have%
\begin{align}
\widehat{\beta}_{J}-\beta^{0}  &  =\widehat{\beta}-\beta_{0}-\left(
\widehat{\beta}_{N,T/2}-\widehat{\beta}\right)  -\left(  \widehat{\beta
}_{N/2,T}-\widehat{\beta}\right) \nonumber\\
&  =\frac{6b_{\alpha}(\beta_{0})-2b_{\alpha}(\beta_{N,T/2}^{\left(  1\right)
})-2b_{\alpha}(\beta_{N,T/2}^{\left(  2\right)  })-b_{\alpha}(\beta
_{N/2,T}^{\left(  1\right)  })-b_{\alpha}(\beta_{N/2,T}^{\left(  2\right)  }%
)}{2T}\nonumber\\
&  \quad\quad+\frac{6b_{\gamma}(\beta_{0})-b_{\gamma}(\beta_{N,T/2}^{\left(
1\right)  })-b_{\gamma}(\beta_{N,T/2}^{\left(  2\right)  })-2b_{\gamma}%
(\beta_{N/2,T}^{\left(  1\right)  })-2b_{\gamma}(\beta_{N/2,T}^{\left(
2\right)  })}{2N}\nonumber\\
&  \quad\quad+o_{\mathbb{P}}\left(  N^{-1}\vee T^{-1}\right)  .
\label{Eq.Jackknife.JackknifeFailed}%
\end{align}
If $\beta_{0}$, $\beta_{N,T/2}^{\left(  1\right)  }$, $\beta_{N,T/2}^{\left(
2\right)  }$, $\beta_{N/2,T}^{\left(  1\right)  }$, and $\beta_{N/2,T}%
^{\left(  2\right)  }$ were the same, the terms in the numerator would cancel
out; and $\widehat{\beta}_{J}-\beta^{0}$ would be $o_{\mathbb{P}}\left(
N^{-1}\vee T^{-1}\right)  $. For two-way DN-HP models, $\beta_{N,T/2}^{\left(
1\right)  }$, $\beta_{N,T/2}^{\left(  2\right)  }$, $\beta_{N/2,T}^{\left(
1\right)  }$, and $\beta_{N/2,T}^{\left(  2\right)  }$ only approach
$\beta_{0}$, as $N,T$ increase, because $g_{T/2}^{\left(  1\right)  }\ $and
$g_{T/2}^{\left(  2\right)  }$ (and similar terms from subpanels
$\mathcal{P}_{N/2,T}^{\left(  1\right)  }$ and $\mathcal{P}_{N/2,T}^{\left(
2\right)  }$) approach zero. When $N$ and $T$ are small, they remain
different. This causes $\widehat{\beta}_{J}$ to perform less well under small
$N,T$.%

\renewcommand{\thesubsection}{\arabic{section}.\arabic{subsection}}
\renewcommand{\thesubsubsection}{\arabic{section}.\arabic{subsection}
.\arabic{subsubsection}}
\numberwithin{equation}{section}
\numberwithin{lemma}{section}
\numberwithin{theorem}{section}%

\bibliographystyle{chicago}
\bibliography{DoubleEffects}

\end{document}